\documentclass[twocolumn,aps,prl,notitlepage]{revtex4-2}
\pdfoutput=1
\usepackage{bm}
\usepackage{graphicx}
\usepackage{amssymb}
\usepackage{amsmath}
\usepackage{tensor}
\usepackage{dcolumn}
\usepackage{amsfonts}
\usepackage{upgreek}
\usepackage{comment}
\usepackage[margin=.7in]{geometry}
\usepackage[euler]{textgreek}
\usepackage{lineno}
\usepackage{braket}
\usepackage[caption=false]{subfig}
\usepackage{floatrow}
\usepackage{gensymb}
\usepackage{multirow}
\usepackage{wasysym}
\usepackage{mathtools}
\usepackage[usenames,dvipsnames]{xcolor}
\usepackage[normalem]{ulem}
\usepackage[colorlinks]{hyperref}
\hypersetup{
    colorlinks=true,
    citecolor=blue,
    linkcolor=blue,
    filecolor=magenta,      
    urlcolor=blue,
    }
\usepackage{lineno}
\usepackage{tikz}
\usepackage{upgreek}
\usepackage{slashed}
\allowdisplaybreaks
\newcommand{\stkout}[1]{\ifmmode\text{\sout{\ensuremath{#1}}}\else\sout{#1}\fi}
\newcommand{\bs}{\boldsymbol}
\newcommand{\pd}{\partial}

\newcommand{\eps}{\epsilon}
\newcommand{\veps}{\varepsilon}

\begin{document}

\title{Quantum geometry of gravitons}
\author{M. Mehraeen}
\email{mandela.mehraeen@manchester.ac.uk}
\affiliation{Department of Physics and Astronomy, University of Manchester, Oxford Road, Manchester M13 9PL, UK}
\date{\today}
\begin{abstract}
We uncover the quantum geometric structure of graviton Wigner functions and stress-energy tensors via quantum field theory in curved spacetime, revealing the phase-space geometry of spacetime and relativistic quantum-state manifolds. We show that the polarization mode expansion of the underlying graviton field operator is fully captured by quantum geometry, as encoded in the spinor-helicity formalism, thereby establishing the geometric framework from the outset. Applying this to a weak gravitational background, we demonstrate the roles of the quantum metric and connection in describing graviton transport beyond the chiral vortical effect. We also clarify the role of the quantum metric in normalizing the perfect-fluid graviton stress tensor within this approach. This work paves the path for explorations of multistate Hilbert-space geometry in high-energy and gravitational physics. In addition, this framework naturally encompasses lower-spin excitations, allowing for a unified quantum geometric treatment of bosonic and fermionic many-body systems in condensed matter and particle physics.
\end{abstract}
\maketitle

\textit{Introduction.}$-$The detection of gravitational waves~\cite{abbott2016observation} has greatly renewed interest in probing fundamental aspects of gravity through gravitational-wave astronomy. A notable area of exploration in this regard is the study of spin-dependent phenomena, with a prominent example being the family of gravitational spin Hall effects. These predict the chiral splitting of gravitational waves and other perturbations in curved spacetime. The spin Hall effect of gravitational waves has been derived through various approaches, ranging from classical to quantum field-theoretic methods~\cite{yamamoto2018spin, andersson2021propagation, andersson2023spin, oancea2024frequency, kubota2024spin, kubota2024spin, frolov2025spinoptics, ito2026spin}. In close analogy with the original electronic counterpart in condensed matter physics~\cite{dyakonov1971current, dyakonov1971possibility, hirsch1999spin, zhang2000spin, murakami2003dissipationless, sinova2004universal, sinova2015spin}, the effect arises from spin-orbit coupling and can be attributed to the Berry curvature of gravitons~\cite{yamamoto2018spin, andersson2021propagation, andersson2023spin, oancea2024frequency, ito2026spin}.

Another intriguing manifestation of the Berry curvature is the chiral vortical effect (CVE), which is the generation of a current in a chirally imbalanced medium in the presence of rotation or local vorticity. Initially, this was predicted for fermion and electric currents in rotating thermal radiation~\cite{vilenkin1978parity}, examined for neutrino emission from rotating black holes~\cite{vilenkin1979macroscopic} and further developed within quantum field theory in rotating systems~\cite{vilenkin1980quantum}. Later on, partly driven by possible connections with quantum anomalies, the CVE reemerged in various contexts and was recovered within anomalous hydrodynamics~\cite{son2009hydrodynamics}, holography~\cite{erdmenger2009fluid}, Kubo formulas~\cite{landsteiner2011gravitational} and kinetic theory~\cite{stephanov2012chiral}. In particular, the connection to the Berry curvature monopole was first understood in the latter approach. Since then, Berry curvature-induced chiral transport has been widely explored and generalized to fermionic and bosonic systems in various contexts in both flat and curved spacetime~\cite{landsteiner2016notes, kharzeev2016chiral, yan2017topological, armitage2018weyl, gorbar2018anomalous, nagaosa2020transport, kamada2023chiral}. Recently, the CVE of polarized gravitons  was also proposed~\cite{ito2026spin}, which, among other things, can contribute to frame-dragging effects induced by rotating celestial bodies.

While the quantum geometric exploration of transport in high-energy and gravitational physics has largely been limited to the Berry curvature~\footnote{Two recent exceptions to this with quantum metric effects are Refs.~\cite{oancea2026quantum} and~\cite{akiba2026quantum}. The former focusses on fermion propagation via a subbundle geometry formalism, while the latter explores photon transport using a semiclassical action.}, in condensed matter physics, this area of research has been flourishing in recent years~\cite{torma2023essay, liu2024quantum, chen2024quantum, shim2025spin, jiang2025revealing, yu2025quantum, verma2026quantum, gao2025quantum}. Historically, the Berry curvature--the imaginary part of the quantum geometric tensor~\cite{provost1980riemannian}, has been the central geometric structure of interest for several decades in studies of materials responses~\cite{xiao2010berry}. Until recently, the quantum metric, the real part of the quantum geometric tensor, had received relatively little attention~\cite{marzari1997maximally, souza2000polarization, venuti2007quantum, ma2010abelian, resta2011insulating, degrandi2011universal, neupert2013measuring, claassen2015position}. Furthermore, driven by attempts to understand nonlinear responses and multiband physics, other quantum geometric structures have recently emerged, including the quantum Levi-Civita connection~\cite{gao2014field}, as well as torsion~\cite{ahn2022riemannian, hsu2023nonlinear}, nonmetricty~\cite{jain2026nonlinear, jain2026topological}, and higher-state quantum geometric tensors~\cite{mehraeen2025quantum, guo2025bicircular}. These geometric structures have deepened our understanding of materials responses, and highlight the increasingly important role of geometry in classifying and probing quantum matter.

Motivated by these developments and insights from condensed matter physics, a natural question that arises is whether this growing inventory of gauge-invariant geometric structures also manifest in and help classify transport phenomena in high-energy and gravitational physics. This would constitute a notable expansion in the scope of applications of quantum geometry, as these transport phenomena exist on all scales in physics, ranging from relativistic heavy-ion collisions in quark-gluon plasmas, to cosmological electromagnetic and gravitational fields~\cite{kharzeev2008effects, fukushima2008chiral, kharzeev2011testing, tashiro2012chiral, boyarsky2012self, long2014leptogenesis, abbaslu2019contribution, huang2021vorticity, abbaslu2021generation}.

Here, we address this question by exploring graviton Wigner functions and stress-energy tensors, constructed from graviton quantum fields in curved spacetime, beyond the linear approximation in the gradient expansion. In doing so, we find that, in addition to the Berry curvature--which describes the graviton spin Hall effect and CVE, the quantum metric, Levi-Civita connection and nonmetricity all appear at higher orders, and help capture transport effects driven by spacetime inhomogeneities. More importantly, perhaps, we find that the phase-space Wigner functions and stress tensors can be fully described by quantum geometric quantities. We attribute this to the quantum geometric nature that underpins the spinor-helicity formalism developed in particle physics~\cite{kleiss1985spinor, gunion1985improved, xu1987helicity, peskin2011simplifying}, namely that spin-1 polarization vectors constituting graviton polarization tensors~\cite{gupta1952quantization, weinberg1964photons} are expressible as off-diagonal Berry connections weighted by graviton momenta. As a result, in this approach, the quantum geometric representation is manifest from the outset, with the gradient expansion capturing higher-order geometric quantities.

We apply this approach to evaluate graviton transport in a weak-field gravitational background, and demonstrate both chiral and nonchiral contributions beyond the standard CVE. We show that these contributions, which can arise in the presence of local variations in spacetime vorticity, are precisely captured by the combined effects of the graviton quantum geometric tensor and quantum connection in the phase-space picture. We conclude by presenting an outlook on future directions, including the applicability of these results to high-energy and condensed matter physics.

\textit{Graviton Wigner functions in curved spacetime.}$-$We consider the spacetime metric
$g_{\mu\nu}
=
\bar{g}_{\mu\nu}
+
h_{\mu\nu}$, which represents an arbitrarily curved background $\bar{g}_{\mu\nu}$ perturbed by metric fluctuations $h_{\mu\nu}$. Following the standard quantization procedure in curved geometry, one can promote $h_{\mu\nu}$ to a field operator, which admits a plane-wave mode expansion arising from solving the equations of motion for the Einstein-Hilbert action~\footnote{For further details, see the Supplemental Material at [url], which includes Refs.~\cite{ahn2022riemannian, provost1980riemannian, mehraeen2025quantum, hehl1995metric, blauGR, jain2026topological, gao2014field, hsu2023nonlinear, mehraeen2025quantum, jain2026nonlinear, ito2026spin, bellac2011thermal, gupta1952quantization, weinberg1964photons, kleiss1985spinor, gunion1985improved, xu1987helicity, hattori2021wigner, fonarev1994wigner, nakahara2018geometry, liu2019chiral, altas2019second, mameda2022photonic} and the following sections: S1.~Graviton field operator in linearized gravity; 
S2.~Quantum state geometry;
S3.~Graviton Wigner functions in flat spacetime;
S4.~Phase-space horizontal lift and Wigner functions in curved spacetime;
S5.~Dressed distribution function;
S6.~Graviton stress-energy tensor in phase space;
S7.~Graviton transport in linearized background.}.

For metric perturbations in an arbitrary background, the helicity-dependent lesser propagator in the Wigner function formalism is given by~\cite{bellac2011thermal, ito2026spin}
\begin{equation}
\label{eq_Wigner_curved}
W^{h}_{\phantom{h}\mu\nu\rho\sigma}(x,q)
=
\int_{y}e^{-i\frac{q\cdot y}{\hbar}}
\left\langle
h^{h}_{\rho\sigma}\left(x,\frac{y}{2}\right)
h^{h}_{\mu\nu}\left(x,-\frac{y}{2}\right)
\right\rangle,
\end{equation}
with 
$\int_{y} = \int d^{4}y\sqrt{-\bar{g}}$
,
$\bar{g} = \det(\bar{g}_{\mu\nu})$, $q$ the graviton four-momentum, and $h=\text{R}, \text{L}$ denoting the helicity of the graviton state~\footnote{Regarding notation and conventions, throughout this work, we assume natural units $\hbar=c=k_B=1$, unless to show explicit $\hbar$ dependence. We work in the metric signature $\text{diag}(1,-\bs{1})$, and the Levi-Civita tensor $\veps^{\mu\nu\rho\sigma}$ is defined as 
$\veps^{0123}
=
(-g)^{-1/2}$. Tensor (anti-)symmetrizations are defined as
$\mathcal{O}_{(\mu\nu)}
\equiv
(\mathcal{O}_{\mu\nu}
+
\mathcal{O}_{\nu\mu})/2
$
and
$\mathcal{O}_{[\mu\nu]}
\equiv
(\mathcal{O}_{\mu\nu}
-
\mathcal{O}_{\nu\mu})/2
$.
}.
$h_{\mu\nu}(x,y)$ defines the graviton field at covariant separation $y$ from the base point $x$ in the spacetime manifold, where the covariant Taylor expansion of a field about $x$ is exponentiated into a translation generated by a suitable horizontal lift~\cite{nakahara2018geometry}, $h_{\mu\nu}(x,y)
\equiv
\left(1+y^{\lambda}\nabla_{\lambda}
+\frac{1}{2}y^{\lambda}y^{\kappa}\nabla_{\lambda}\nabla_{\kappa}
+\cdots\right)h_{\mu\nu}(x)
=
e^{y\cdot D}h_{\mu\nu}(x)$.

Accordingly, a natural choice of covariant derivative in phase space is the horizontal lift $D$ of the spacetime covariant derivative $\nabla$ to the tangent and cotangent bundles. For an arbitrary phase-space function $\Phi(x,q)$, the latter is defined as~\cite{fonarev1994wigner, nakahara2018geometry, liu2019chiral}
$D_{\mu} \Phi(x,q)
=
( \nabla_{\mu} + \Gamma^{\lambda}_{\mu\nu} q_{\lambda} \partial_{q}^{\nu} ) \Phi(x,q)$,
with 
$\Gamma^{\lambda}_{\mu\nu}$ a component of the spacetime Levi-Civita connection and
$\partial^\nu_q
=
\pd / \pd q_\nu$.
This simplifies the analysis in curved spacetime, as
$[ D_{\mu},y^{\nu} ] = 0$,
$[ D_{\mu},q_{\nu} ] = 0$, and 
$D_{\mu}\bar{g}_{\nu\lambda} = 0$.

Specifically, inserting the graviton field operator
\begin{equation}
\label{eq_field_exp}
h^h_{\mu\nu}(x)
=
\int_{\bs{p}}
\left[
a^h_{\bs{p}} 
e^h_{\mu\nu}(p)\text{e}^{-\text{i}p\cdot x}
+
a^{h\dagger}_{\bs{p}}
e^{h*}_{\mu\nu}(p)\text{e}^{\text{i}p\cdot x}
\right],
\end{equation}
with
$\int_{\bs{p}}
=
\int \frac{\text{d}^3 \bs{p}}{(2\pi)^3} 
\sqrt{\frac{16\pi G}{|\bs{p}|}}$ and $G$ Newton's constant, into Eq.~(\ref{eq_Wigner_curved}), for
$\bar{g}_{\mu\nu} = \eta_{\mu\nu}$, yields the flat spacetime Wigner function gradient expansion in terms of that of the graviton polarization tensor $e_{\mu\nu}^h$. The properties of the horizontal lift then imply that the generalization to curved space readily follows from the flat-space results through the substitution $\pd \to D$, provided one also imposes parallel transport conditions on any constituent frame vectors, $\nabla_{\mu}n_{\nu}=0$, to fix frame ambiguities.

\textit{Quantum geometric structure.}$-$To retrieve the quantum geometry of gravitons in a systematic manner, we first reexpress graviton polarization tensors as local tensor products of spin-1 polarization vectors in the momentum-space of gravitons~\cite{gupta1952quantization, weinberg1964photons},
$e^h_{\mu\nu}(q) = \epsilon^h_\mu(q) \epsilon^h_\nu(q)$. This then sets up the implementation of the spinor-helicity formalism~\cite{kleiss1985spinor, gunion1985improved, xu1987helicity, peskin2011simplifying}, which allows one to reexpress spin-1 polarization vectors as spinor products. 

\begin{table*}
\begin{ruledtabular}
\normalsize
\begin{tabular}{l c c c c}
{} & $n=0$ & $n=1$ & $n=2$  & \multicolumn{1}{c}{$n=3$}
\\
\hline
\\[-.3em]
Description & Perfect Fluid & CVE & QGME & ICVE
\vspace{.2cm}
\\
Stress tensor 
& 
$\frac{\rho_{h}}{3}\left(4u_{\mu}u_{\nu} -\bar g_{\mu\nu}\right)$ 
& 
$\propto \iota^{h} \hbar T^3 u_{(\mu}\omega_{\nu)}$ 
&
$\propto \hbar^2 T^{2} u_{(\mu}R_{\nu)\lambda}u^{\lambda}$
&
$\propto \iota^h \hbar^3 T u_{(\mu} \nabla^2 \omega_{\nu)}$
\vspace{.2cm}
\\
Energy current
&
$\frac{\rho_h}{3} \bs{u}$ 
&
$\propto \iota^h \hbar T^{3} \bs{\omega}$
&
$\propto \hbar^2 T^{2} \bs{R}$
&
$\propto \iota^h \hbar^3 T \bs{\pd}^2 \bs{\omega}$
\vspace{.2cm}
\\
Geometric description & Tr\,$\mathcal{G}$  & $\Omega$ 
&
$\mathcal{G}$, $\Omega$, $\Gamma_\mathcal{G}$ 
&
$\mathcal{G}$, $\Omega$, $\Gamma_\mathcal{G}$, 
$\mathcal{N}$
\vspace{.2cm}  
\\
Helicity & + & -- & + & --
\end{tabular}
\end{ruledtabular}
\caption{Stress tensor contributions in a weak-field gravitational background at different orders in the gradient expansion, and the quantum geometric structures that give rise to them in the quantum phase-space picture. Here, 
$\rho_{h}=\frac{\pi^{2}T^{4}}{30}$ is the energy density per helicity channel of the thermal graviton gas, and
$\omega^{\mu}
=
(\omega^0, \bs{\omega})
=
\frac{1}{2}\epsilon^{\mu\nu\rho\sigma}u_{\nu}
\nabla_{\rho}u_{\sigma}$ is the covariant spacetime vorticity.}
\label{tab1}
\end{table*}

Specifically, for massless photons, one can express the polarization vector as a product of Weyl spinors,
$\epsilon_\mu^{h}(q)
=
\frac{1}{\sqrt{2}} c_h^{- \dagger}(q) \sigma_\mu^{h} c_h^+(q)$.
Here, $c_h^\pm$ are the positive- and negative-energy eigenstates of the helicity-$h$ Weyl Hamiltonian,
$\bs{\sigma}\cdot \hat{\bs{q}} c_h^\pm
=
\pm \iota^h c_h^\pm$, 
with 
$\iota^{R,L} = \pm 1$. 
And we have introduced helicity-dependent Pauli matrix 4-vectors as
$\sigma^{\mathrm{R}} = \sigma$, $\sigma^{\mathrm{L}} = \bar{\sigma}$
and correspondingly
$\bar{\sigma}^{\mathrm{R}} = \bar{\sigma}$,
$\bar{\sigma}^{\mathrm{L}} = \sigma$,
such that
$\bar{\sigma}^{\mathrm{R,L}} = \sigma^{\mathrm{L,R}}$,
with
$\sigma^{\mu}
=
\left(1,\bs{\sigma}\right)$
and
$\bar{\sigma}^{\mu}
=
\left(1,-\bs{\sigma}\right)$.

Taking the derivative of the Weyl equation then reveals the quantum geometric relation
\begin{equation}
\label{eq_eps_berry}
\epsilon_\mu^h
=
i\sqrt{2}\vert{}\boldsymbol{q}\vert{}\mathcal{A}^{\prime - +}_{\mu h},
\end{equation}
where $\mathcal{A}^{\prime ab}_{\mu h}$ is the off-diagonal component of the helicity-dependent Berry connection between energy levels $a$ and $b$ for a general $N$-level system, 
$\mathcal{A}^{ba}_{\alpha h}
=
i c^{b\dagger}_{h} \partial^{q}_{\alpha} c^{a}_{h}$,
as
$\mathcal{A}^{ba}_{\mu h}
=
a^{a}_{\mu h} \delta^{ba} 
+
\mathcal{A}^{\prime ba}_{\mu h}$,
which is induced by the momentum-dependence of the eigenstates,
$\partial^{q}_{\alpha} c^{a}_{h}
=
-i \sum_{b} \mathcal{A}^{ba}_{\alpha h} c^{b}_{h}$.

The off-diagonal Berry connections can be regarded as elements of helicity-dependent complex-valued veilbeins in the basis of states~\cite{ahn2022riemannian}, 
$\hat{\theta}^{ab}_{\mu h}
=
\mathcal{A}^{\prime ab}_{\mu h} c^{a}_{h} c^{b\dagger}_{h}$,
with the Hilbert-Schmidt inner product $\left(A, B\right) =
\mathrm{Tr}\left(A^{\dagger} B\right)$ of tangent basis vectors and their derivatives inducing gauge-invariant geometric structures on the quantum-state manifold. In particular, the quantum geometric tensor and quantum connection between states $a$ and $b$ are defined as
\begin{subequations}
\begin{align}
\mathcal{Q}^{ab}_{\mu\nu,h}
\equiv&
\left(\hat{\theta}^{ab}_{\nu h}, \hat{\theta}^{ab}_{\mu h}\right)
=
\mathcal{A}^{\prime ab}_{\mu h} 
\mathcal{A}^{\prime ba}_{\nu h}
\\
\mathcal{C}^{ba}_{\mu\nu\rho,h}
\equiv&
\left(\hat{\theta}^{ba}_{\mu h},
\partial^{q}_{\nu} \hat{\theta}^{ba}_{\rho h}\right)
= \mathcal{A}^{\prime ab}_{\mu h}
\left(\mathcal{D}^{h}_{\nu} \mathcal{A}^{\prime}_{\rho h}\right)^{ba},
\end{align}
\end{subequations}
where
$(\mathcal{D}^{h}_{\alpha} O_{\beta})^{ab}
=
\left(\partial^{q}_{\alpha} - i \Delta^{ab}_{\alpha h}\right)
O^{ab}_{\beta}$
is the Berry covariant derivative~\cite{sipe1993nonlinear, aversa1995nonlinear, sipe2000second}, with
$\Delta^{ab}_{\alpha h} \equiv a^{a}_{\alpha h} -
a^{b}_{\alpha h}$ the diagonal Berry connection mismatch. The real and imaginary parts of the quantum geometric tensor  are the quantum metric and Berry curvature~\cite{provost1980riemannian}, respectively,
$\mathcal{Q}
=
\mathcal{G}
-
\frac{i}{2} \Omega$.
And the quantum connection admits~\cite{jain2026nonlinear, jain2026topological} the standard decomposition of affine geometry into the quantum Levi-Civita connection $\Gamma_{\mathcal{G}}$, disformation $\mathcal{L}$ (nonmetricity $\mathcal{N}$), and(con-)torsion. The multistate quantum torsion tensor requires at least a three-level quantum system to exist~\cite{ahn2022riemannian}, which is not the case for massless gravitons. Thus, one can write
$\mathcal{C}
=
\Gamma_{\mathcal{G}}
+
\mathcal{L}$, with the quantum Levi-Civita connection purely real and the quantum nonmetricity (disformation) purely imaginary.

These geometric structures manifest in the gradient expansion of the graviton polarization modes in Eq.~(\ref{eq_Wigner_curved}), which are, in turn, inherited from the photon polarization expansion as the tensor product 
$\mathcal{P}_{\mu\nu\rho\sigma}^h
=
\tilde{\mathcal{P}}_{\mu\rho}^h
\tilde{\mathcal{P}}_{\nu\sigma}^h$,
where
$\tilde{\mathcal{P}}_{\mu\rho}^h (q ; p)
=
\eps^h_{\mu} \left(q + \frac{p}{2} \right)
\eps^{h*}_{\rho} \left(q - \frac{p}{2} \right)
=
\tilde{\Pi} _{\mu\rho}^h
+
p^{\alpha}
\tilde{\Sigma} _{\mu\rho\alpha}^h
+
O(p^2)$,
with $p \sim i \hbar \pd$ in the Wigner transformation.
From Eq.~(\ref{eq_eps_berry}), the leading and subleading terms read
\begin{subequations}
\label{eq_tilde_Pi}
\begin{align}
\label{eq_tilde_Pi_a}
\tilde{\Pi}^{h}_{\mu\nu}
=&
\eps^h_{\mu} \eps^{h*}_{\nu}
=
2 |\boldsymbol{q} |^{2} \mathcal{Q}^{-+}_{\mu\nu,h},
\\
\tilde{\Sigma}^{h}_{\mu\nu\alpha}
=&
\frac{1}{2}\left[(\partial^{q}_{\alpha}\epsilon^{h}_{\mu})\epsilon^{h*}_{\nu}-\epsilon^{h}_{\mu}(\partial^{q}_{\alpha}\epsilon^{h*}_{\nu})\right]
\nonumber
\\
=&
|\boldsymbol{q}|^{2}\left[2i\Delta^{-+}_{\alpha h}\mathcal{Q}^{-+}_{\mu\nu,h}
+\mathcal{C}^{-+}_{\nu\alpha\mu,h}
-\mathcal{C}^{+-}_{\mu\alpha\nu,h}\right],
\end{align}
\end{subequations}
with the lengthier higher-order expansion terms presented in the Supplemental Material. 

The appearance of the quantum geometric tensor at $O(\hbar^0)$ in Eq.~(\ref{eq_tilde_Pi_a}), which is the geometrical optics limit of the analysis, merits elaboration. While we discuss this further in the Discussion, here we point out that this can be regarded as reflecting a dual geometric description that underpins the spinor-helicity formalism, namely between the Bloch sphere of Weyl spinors and the Poincaré sphere of polarization states. Indeed, the right- and left-handed photon polarization states can be identified with antipodal points on the Poincaré sphere~\cite{collett2008visualization}, as can Weyl eigenstates on the Bloch sphere. 

Moreover, this geometric duality establishes quantum geometric interpretations for the real and imaginary parts of 
$\tilde{\Pi}^{h}_{\mu\nu} (q)$. The real part, which is symmetric, is precisely the standard polarization tensor in the Coulomb gauge~\cite{hattori2021wigner}. This projects four-vectors onto the 2d polarization hypersurface transverse to $q$, and is mapped to the quantum metric tensor through the spinor-helicity formalism. And the antisymmetric part is the spin tensor, which generates rotations within the transverse subspace of polarizations. This is mapped to the Berry curvature in the quantum geometric language.

\textit{Graviton stress tensor.}$-$The stress tensor of metric perturbations is identified from the expansion of the Einstein tensor in powers of $h_{\mu\nu}$ around the background,
$G_{\mu\nu}
=
G^{[0]}_{\mu\nu}
+
G^{[1]}_{\mu\nu}
+
G^{[2]}_{\mu\nu}$.
One can reorganize the Einstein equation as
$G^{[0]}_{\mu\nu} 
=
- (G^{[1]}_{\mu\nu} + G^{[2]}_{\mu\nu})$, 
so that the perturbation sources the gravitational field, with stress tensor
$T_{\mu\nu}
=
-\frac{1}{8\pi G} (G^{[1]}_{\mu\nu}
+
G^{[2]}_{\mu\nu} )$.
Within the transverse-traceless gauge, the linear term vanishes and the quadratic term reduces to the form
$G^{[2]}_{\mu\nu}
=
R^{[2]}_{\mu\nu}
-
\frac{1}{2}\bar{g}_{\mu\nu}R^{[2]}$~\cite{altas2019second}.

To derive the phase-space representation of the stress tensor, as discussed above, the spacetime covariant derivative is promoted to the phase-space horizontal lift and the Wigner transformation prescription~\cite{ito2026spin} is applied. This consists of symmetrizing the arguments of the gravitational field pairs appearing in the stress tensor, as
\begin{align}
h_{\rho\sigma}\nabla_{\mu}\nabla_{\nu}h_{\lambda\eta}
\to&
\frac{1}{2}\left[\left\{D_{\mu}D_{\nu}h_{\lambda\eta}
\left(x, \frac{y}{2}\right)\right\}h_{\rho\sigma}
\left(x, -\frac{y}{2}\right)
\right.
\nonumber
\\
&+
\left.
h_{\rho\sigma}\left(x, \frac{y}{2}\right)
\left\{D_{\mu}D_{\nu}h_{\lambda\eta}
\left(x, -\frac{y}{2}\right)\right\}\right],
\end{align}
and similarly for other terms. As a result, one obtains the phase-space stress tensor 
$\mathcal{T}_{\mu\nu}^h(x,q)$ 
in terms of horizontal lifts of Wigner function contractions. Integrating out graviton momenta, the stress-energy tensor including gradient expansion corrections is then obtained as
$T^{h}_{\mu\nu} (x)
=
\int_{q} \mathcal{T}^{h}_{\mu\nu}(x,q)$,
with
$\int_{q} = \int\frac{d^{4}q}{(2\pi)^{4}\sqrt{-\bar g}}$.

\textit{Application to linearized background.}$-$We now discuss the explicit application of the quantum geometric formalism to the evaluation of the graviton stress tensor in the weak-field limit, the results of which are summarized in Table~\ref{tab1}. We consider the background metric
$\bar g_{\mu\nu}=\eta_{\mu\nu}+\gamma_{\mu\nu}$,
with
$\gamma_{0i}=\gamma_{0i}(\boldsymbol{x})$,
$\gamma_{00},\gamma_{ij}=0$,
and work to linear order in $\gamma_{\mu\nu}$ by assuming
$|\gamma_{\mu\nu}|\ll 1$.
In the presence of spin-vorticity coupling, the equilibrium distribution function reads~\cite{liu2019chiral, mameda2022photonic}
$f^{h}(x,q)
=
N (g^{h})$,
with
$g^{h}
=
q\cdot U + \frac{\hbar}{2} S^{\alpha\beta}_h
\nabla_{\alpha}U_{\beta}$,
where
$N(x)=(e^{x}-1)^{-1}$, $U^{\mu}=\beta u^{\mu}$ with $\beta=1/T$ and $u^{\mu}$ the spacetime fluid velocity. The helicity-odd spin tensor is given by 
$S_{\mu\nu}^h
=
2 \iota^h \epsilon_{\mu\nu\alpha\beta} \frac{q^\alpha n^\beta}{q \cdot n}$ in the frame $n$.
We take
$u^{\mu}=n^{\mu}=(1,\boldsymbol{0})$, 
so that 
$u_{\mu} = (u_0, - \bs{u}) =(1,\gamma_{0i})$ 
and $q\cdot U=\beta q_{0}$. This corresponds to the Killing condition
$\nabla_{(\mu}U_{\nu)}=0$.

At zeroth order in $\hbar$, the phase-space stress tensor contribution is proportional to the second momentum-space cumulant of the classical Wigner function contraction,
$q_\mu q_\nu 
{W_0^h}_{\alpha\beta}^{\phantom{\alpha\beta}\alpha\beta}$, which is in turn, captured by the trace of the quantum metric in the geometric formalism,
${W^{h}_{0}}_{\alpha\beta}^{\phantom{\alpha\beta}\alpha\beta}
\propto
(\bar g^{\alpha\beta}\mathcal{G}^{-+}_{\alpha\beta} )^{2}$. This yields the well-known perfect-fluid stress tensor of a graviton gas in the geometrical optics limit. Thus, it is interesting to conclude that, in the dual quantum geometric picture, this well-known classical result can be viewed as being due to a normalization condition arising from the trace of the quantum metric.

Beyond the helicity-odd graviton CVE at linear order, at $O(\hbar^2)$, we find a helicity-even current that, in the fluid rest frame, is along
$\bs{R} = \bs{\pd} \times \bs{\omega}$, where 
$R_i \equiv R_{0i}$ is a spatiotemporal component of the spacetime Ricci tensor.  Thus, this constitutes a nonchiral quantum gravitomagnetic effect (QGME) for gravitons that arises from the concerted actions of the Berry curvature, quantum metric and quantum Levi-Civita connection in the quantum phase-space picture. We note that a fermionic analog of this graviton current was also recently predicted~\cite{hayata2021second}, which we expect to have an analogous quantum geometric description.

And at third order, we find a helicity-odd contribution along 
$\bs{\pd} \times \bs{R} = - \bs{\pd}^2 \bs{\omega}$, which can be regarded as a correction to the graviton chiral vortical current from local variations in the spacetime vorticity. In addition to the quantum geometric structures describing the second-order response, this inhomogeneous chiral vortical effect (ICVE) is also captured by the recently proposed quantum nonmetricity~\cite{jain2026nonlinear, jain2026topological} in the geometric description.

\textit{Discussion and outlook.}$-$In summary, this work uncovers the quantum geometric structure of graviton Wigner functions and stress-energy tensors beyond the Berry curvature, and reveals the roles of the quantum metric and quantum connection in describing both chiral and nonchiral graviton transport in an arbitrary spacetime geometry. A central result in this work is that graviton transport can be regarded as being entirely quantum geometrical. This stems from the observation that the spinor-helicity formalism that underpins this analysis can be essentially viewed as establishing a dual relation between the geometry of quantum state manifolds and that of polarization tensors.

This dual geometric description also helps to clarify why, at the classical level, the perfect-fluid stress tensor of gravitons obtained within the geometrical optics limit can be understood from the trace of the quantum metric. That is, one can regard the $O(\hbar^0)$ result not as a quantum mechanical statement, but rather, as a manifestation of the dual description between classical and quantum geometry encoded in the spinor-helicity formalism. Thus, this interpretation would constitute a geometric confirmation of a recent argument that not all quantum geometric responses are quantum~\cite{stern2026how}.

We envision several directions for future investigations arising from these results. First, the present analysis is directly applicable to chiral photon and fermion transport in the relativistic framework and in various contexts, as it naturally captures lower-spin excitations. In addition, as our analysis has been limited to collisionless gravitons, a next step would be to include the effect of collisions and to derive the full quantum geometric kinetic theory for graviton transport. This would likely give rise to a rich interplay between quantum geometry and scattering events, as has similarly been demonstrated recently in the condensed matter context for electrons in disordered crystalline solids~\cite{atencia2022semiclassical, atencia2023disorder, mehraeen2024quantum, huang2025scaling, gong2026scaling}. Related to this, recent theoretical advances in quantum geometry in condensed matter systems have explored geometries of density matrices~\cite{ji2025density, guan2026exploring, bradlyn2026multistate}, inspired by quantum information methods. It would therefore be interesting to explore what role this could play in quantum geometric chiral transport.

On the condensed matter side, the approach explored in the present work suggests several fruitful directions for future exploration as well. First, it may play a role in generalizations of quantum geometric viscous and elastic responses~\cite{bradlyn2012kubo, shapourian2015viscoelastic, jain2026nonlinear, jain2026topological, osborne2026geometry} beyond global distortions of flat crystalline systems. This is particularly relevant given recent advances in the spatiotemporal control of strain in quantum materials, leading to programmable local strainscapes~\cite{liu2025programmable}
and more general curved nanostructures for straintronic applications~\cite{bukharaev2018straintronics}. Second, we anticipate applications in electron hydrodynamics of topological quantum materials~\cite{berdyugin2019measuring, aharon-steinberg2022direct}, through quantum-corrected Navier-Stokes equations obtained from quantum geometric Wigner functions. In addition, it would also be interesting to leverage the present approach to explore orbital Wigner functions~\cite{mitscherling2026orbital} from a quantum geometric perspective, potentially providing a theoretical toolkit for the emerging field of orbitronics.

Finally, in the present work, we have used the simplest form of the spinor-helicity formalism, namely that for Weyl spinors capturing correlators of massless photons and gravitons. Given recent advances in the understanding of on-shell methods for scattering amplitudes in modern high-energy physics~\cite{arkani-hamed2021scattering}, it would be quite interesting to explore whether this can lead to new insights in quantum geometric descriptions of massive or interacting systems in condensed matter physics and beyond. We leave this for future works.


\bibliography{qg_grav}

\begin{thebibliography}{106}%
\makeatletter
\providecommand \@ifxundefined [1]{%
 \@ifx{#1\undefined}
}%
\providecommand \@ifnum [1]{%
 \ifnum #1\expandafter \@firstoftwo
 \else \expandafter \@secondoftwo
 \fi
}%
\providecommand \@ifx [1]{%
 \ifx #1\expandafter \@firstoftwo
 \else \expandafter \@secondoftwo
 \fi
}%
\providecommand \natexlab [1]{#1}%
\providecommand \enquote  [1]{``#1''}%
\providecommand \bibnamefont  [1]{#1}%
\providecommand \bibfnamefont [1]{#1}%
\providecommand \citenamefont [1]{#1}%
\providecommand \href@noop [0]{\@secondoftwo}%
\providecommand \href [0]{\begingroup \@sanitize@url \@href}%
\providecommand \@href[1]{\@@startlink{#1}\@@href}%
\providecommand \@@href[1]{\endgroup#1\@@endlink}%
\providecommand \@sanitize@url [0]{\catcode `\\12\catcode `\$12\catcode
  `\&12\catcode `\#12\catcode `\^12\catcode `\_12\catcode `\%12\relax}%
\providecommand \@@startlink[1]{}%
\providecommand \@@endlink[0]{}%
\providecommand \url  [0]{\begingroup\@sanitize@url \@url }%
\providecommand \@url [1]{\endgroup\@href {#1}{\urlprefix }}%
\providecommand \urlprefix  [0]{URL }%
\providecommand \Eprint [0]{\href }%
\providecommand \doibase [0]{https://doi.org/}%
\providecommand \selectlanguage [0]{\@gobble}%
\providecommand \bibinfo  [0]{\@secondoftwo}%
\providecommand \bibfield  [0]{\@secondoftwo}%
\providecommand \translation [1]{[#1]}%
\providecommand \BibitemOpen [0]{}%
\providecommand \bibitemStop [0]{}%
\providecommand \bibitemNoStop [0]{.\EOS\space}%
\providecommand \EOS [0]{\spacefactor3000\relax}%
\providecommand \BibitemShut  [1]{\csname bibitem#1\endcsname}%
\let\auto@bib@innerbib\@empty
\bibitem [{\citenamefont {Abbott}\ \emph {et~al.}(2016)\citenamefont {Abbott}
  \emph {et~al.}}]{abbott2016observation}%
  \BibitemOpen
  \bibfield  {author} {\bibinfo {author} {\bibfnamefont {B.~P.}\ \bibnamefont
  {Abbott}} \emph {et~al.},\ }\bibfield  {title} {\bibinfo {title}
  {{Observation of Gravitational Waves from a Binary Black Hole Merger}},\
  }\href {https://doi.org/10.1103/PhysRevLett.116.061102} {\bibfield  {journal}
  {\bibinfo  {journal} {Phys. Rev. Lett.}\ }\textbf {\bibinfo {volume} {116}},\
  \bibinfo {pages} {061102} (\bibinfo {year} {2016})}\BibitemShut {NoStop}%
\bibitem [{\citenamefont {Yamamoto}(2018)}]{yamamoto2018spin}%
  \BibitemOpen
  \bibfield  {author} {\bibinfo {author} {\bibfnamefont {N.}~\bibnamefont
  {Yamamoto}},\ }\bibfield  {title} {\bibinfo {title} {{Spin Hall effect of
  gravitational waves}},\ }\href {https://doi.org/10.1103/PhysRevD.98.061701}
  {\bibfield  {journal} {\bibinfo  {journal} {Phys. Rev. D}\ }\textbf {\bibinfo
  {volume} {98}},\ \bibinfo {pages} {061701} (\bibinfo {year}
  {2018})}\BibitemShut {NoStop}%
\bibitem [{\citenamefont {Andersson}\ \emph {et~al.}(2021)\citenamefont
  {Andersson}, \citenamefont {Joudioux}, \citenamefont {Oancea},\ and\
  \citenamefont {Raj}}]{andersson2021propagation}%
  \BibitemOpen
  \bibfield  {author} {\bibinfo {author} {\bibfnamefont {L.}~\bibnamefont
  {Andersson}}, \bibinfo {author} {\bibfnamefont {J.}~\bibnamefont {Joudioux}},
  \bibinfo {author} {\bibfnamefont {M.~A.}\ \bibnamefont {Oancea}},\ and\
  \bibinfo {author} {\bibfnamefont {A.}~\bibnamefont {Raj}},\ }\bibfield
  {title} {\bibinfo {title} {{Propagation of polarized gravitational waves}},\
  }\href {https://doi.org/10.1103/PhysRevD.103.044053} {\bibfield  {journal}
  {\bibinfo  {journal} {Phys. Rev. D}\ }\textbf {\bibinfo {volume} {103}},\
  \bibinfo {pages} {044053} (\bibinfo {year} {2021})}\BibitemShut {NoStop}%
\bibitem [{\citenamefont {Andersson}\ and\ \citenamefont
  {Oancea}(2023)}]{andersson2023spin}%
  \BibitemOpen
  \bibfield  {author} {\bibinfo {author} {\bibfnamefont {L.}~\bibnamefont
  {Andersson}}\ and\ \bibinfo {author} {\bibfnamefont {M.~A.}\ \bibnamefont
  {Oancea}},\ }\bibfield  {title} {\bibinfo {title} {{Spin Hall effects in the
  sky}},\ }\href {https://doi.org/10.1088/1361-6382/ace021} {\bibfield
  {journal} {\bibinfo  {journal} {Class. Quantum Grav.}\ }\textbf {\bibinfo
  {volume} {40}},\ \bibinfo {pages} {154002} (\bibinfo {year}
  {2023})}\BibitemShut {NoStop}%
\bibitem [{\citenamefont {Oancea}\ \emph {et~al.}(2024)\citenamefont {Oancea},
  \citenamefont {Stiskalek},\ and\ \citenamefont
  {Zumalacarregui}}]{oancea2024frequency}%
  \BibitemOpen
  \bibfield  {author} {\bibinfo {author} {\bibfnamefont {M.~A.}\ \bibnamefont
  {Oancea}}, \bibinfo {author} {\bibfnamefont {R.}~\bibnamefont {Stiskalek}},\
  and\ \bibinfo {author} {\bibfnamefont {M.}~\bibnamefont {Zumalacarregui}},\
  }\bibfield  {title} {\bibinfo {title} {{Frequency- and polarization-dependent
  lensing of gravitational waves in strong gravitational fields}},\ }\href
  {https://doi.org/10.1103/PhysRevD.109.124045} {\bibfield  {journal} {\bibinfo
   {journal} {Phys. Rev. D}\ }\textbf {\bibinfo {volume} {109}},\ \bibinfo
  {pages} {124045} (\bibinfo {year} {2024})}\BibitemShut {NoStop}%
\bibitem [{\citenamefont {Kubota}\ \emph {et~al.}(2024)\citenamefont {Kubota},
  \citenamefont {Arai},\ and\ \citenamefont {Mukohyama}}]{kubota2024spin}%
  \BibitemOpen
  \bibfield  {author} {\bibinfo {author} {\bibfnamefont {K.~I.}\ \bibnamefont
  {Kubota}}, \bibinfo {author} {\bibfnamefont {S.}~\bibnamefont {Arai}},\ and\
  \bibinfo {author} {\bibfnamefont {S.}~\bibnamefont {Mukohyama}},\ }\bibfield
  {title} {\bibinfo {title} {{Spin optics for gravitational waves lensed by a
  rotating object}},\ }\href {https://doi.org/10.1103/PhysRevD.109.044027}
  {\bibfield  {journal} {\bibinfo  {journal} {Phys. Rev. D}\ }\textbf {\bibinfo
  {volume} {109}},\ \bibinfo {pages} {044027} (\bibinfo {year}
  {2024})}\BibitemShut {NoStop}%
\bibitem [{\citenamefont {Frolov}\ and\ \citenamefont
  {Koek}(2025)}]{frolov2025spinoptics}%
  \BibitemOpen
  \bibfield  {author} {\bibinfo {author} {\bibfnamefont {V.~P.}\ \bibnamefont
  {Frolov}}\ and\ \bibinfo {author} {\bibfnamefont {A.}~\bibnamefont {Koek}},\
  }\bibfield  {title} {\bibinfo {title} {{Spinoptics in the Kerr Spacetime:
  Polarized Wave Scattering}},\ }\href {https://doi.org/10.1103/trh5-4sgq}
  {\bibfield  {journal} {\bibinfo  {journal} {Phys. Rev. D}\ }\textbf {\bibinfo
  {volume} {111}},\ \bibinfo {pages} {104081} (\bibinfo {year}
  {2025})}\BibitemShut {NoStop}%
\bibitem [{\citenamefont {Ito}\ \emph {et~al.}(2026)\citenamefont {Ito},
  \citenamefont {Mameda},\ and\ \citenamefont {Yamamoto}}]{ito2026spin}%
  \BibitemOpen
  \bibfield  {author} {\bibinfo {author} {\bibfnamefont {R.}~\bibnamefont
  {Ito}}, \bibinfo {author} {\bibfnamefont {K.}~\bibnamefont {Mameda}},\ and\
  \bibinfo {author} {\bibfnamefont {N.}~\bibnamefont {Yamamoto}},\ }\bibfield
  {title} {\bibinfo {title} {{Spin Hall effect and Berry curvature of gravitons
  from quantum field theory}},\ }\href {https://arxiv.org/abs/2605.19817}
  {\bibfield  {journal} {\bibinfo  {journal} {arXiv:2605.19817}\ } (\bibinfo
  {year} {2026})}\BibitemShut {NoStop}%
\bibitem [{\citenamefont {Dyakonov}\ and\ \citenamefont
  {Perel}(1971{\natexlab{a}})}]{dyakonov1971current}%
  \BibitemOpen
  \bibfield  {author} {\bibinfo {author} {\bibfnamefont {M.~I.}\ \bibnamefont
  {Dyakonov}}\ and\ \bibinfo {author} {\bibfnamefont {V.~I.}\ \bibnamefont
  {Perel}},\ }\bibfield  {title} {\bibinfo {title} {{Current-induced spin
  orientation of electrons in semiconductors}},\ }\href
  {https://doi.org/10.1016/0375-9601(71)90196-4} {\bibfield  {journal}
  {\bibinfo  {journal} {Phys. Lett. A}\ }\textbf {\bibinfo {volume} {35}},\
  \bibinfo {pages} {459} (\bibinfo {year} {1971}{\natexlab{a}})}\BibitemShut
  {NoStop}%
\bibitem [{\citenamefont {Dyakonov}\ and\ \citenamefont
  {Perel}(1971{\natexlab{b}})}]{dyakonov1971possibility}%
  \BibitemOpen
  \bibfield  {author} {\bibinfo {author} {\bibfnamefont {M.~I.}\ \bibnamefont
  {Dyakonov}}\ and\ \bibinfo {author} {\bibfnamefont {V.~I.}\ \bibnamefont
  {Perel}},\ }\bibfield  {title} {\bibinfo {title} {{Possibility of orienting
  electron spins with current}},\ }\href
  {https://cir.nii.ac.jp/crid/1370290617742857866} {\bibfield  {journal}
  {\bibinfo  {journal} {JETP Lett.}\ }\textbf {\bibinfo {volume} {13}},\
  \bibinfo {pages} {467} (\bibinfo {year} {1971}{\natexlab{b}})}\BibitemShut
  {NoStop}%
\bibitem [{\citenamefont {Hirsch}(1999)}]{hirsch1999spin}%
  \BibitemOpen
  \bibfield  {author} {\bibinfo {author} {\bibfnamefont {J.~E.}\ \bibnamefont
  {Hirsch}},\ }\bibfield  {title} {\bibinfo {title} {{Spin Hall Effect}},\
  }\href {https://doi.org/10.1103/PhysRevLett.83.1834} {\bibfield  {journal}
  {\bibinfo  {journal} {Phys. Rev. Lett.}\ }\textbf {\bibinfo {volume} {83}},\
  \bibinfo {pages} {1834} (\bibinfo {year} {1999})}\BibitemShut {NoStop}%
\bibitem [{\citenamefont {Zhang}(2000)}]{zhang2000spin}%
  \BibitemOpen
  \bibfield  {author} {\bibinfo {author} {\bibfnamefont {S.}~\bibnamefont
  {Zhang}},\ }\bibfield  {title} {\bibinfo {title} {{Spin Hall Effect in the
  Presence of Spin Diffusion}},\ }\href
  {https://doi.org/10.1103/PhysRevLett.85.393} {\bibfield  {journal} {\bibinfo
  {journal} {Phys. Rev. Lett.}\ }\textbf {\bibinfo {volume} {85}},\ \bibinfo
  {pages} {393} (\bibinfo {year} {2000})}\BibitemShut {NoStop}%
\bibitem [{\citenamefont {Murakami}\ \emph {et~al.}(2003)\citenamefont
  {Murakami}, \citenamefont {Nagaosa},\ and\ \citenamefont
  {Zhang}}]{murakami2003dissipationless}%
  \BibitemOpen
  \bibfield  {author} {\bibinfo {author} {\bibfnamefont {S.}~\bibnamefont
  {Murakami}}, \bibinfo {author} {\bibfnamefont {N.}~\bibnamefont {Nagaosa}},\
  and\ \bibinfo {author} {\bibfnamefont {S.-C.}\ \bibnamefont {Zhang}},\
  }\bibfield  {title} {\bibinfo {title} {{Dissipationless Quantum Spin Current
  at Room Temperature}},\ }\href {https://doi.org/10.1126/science.1087128}
  {\bibfield  {journal} {\bibinfo  {journal} {Science}\ }\textbf {\bibinfo
  {volume} {301}},\ \bibinfo {pages} {1348} (\bibinfo {year}
  {2003})}\BibitemShut {NoStop}%
\bibitem [{\citenamefont {Sinova}\ \emph {et~al.}(2004)\citenamefont {Sinova},
  \citenamefont {Culcer}, \citenamefont {Niu}, \citenamefont {Sinitsyn},
  \citenamefont {Jungwirth},\ and\ \citenamefont
  {MacDonald}}]{sinova2004universal}%
  \BibitemOpen
  \bibfield  {author} {\bibinfo {author} {\bibfnamefont {J.}~\bibnamefont
  {Sinova}}, \bibinfo {author} {\bibfnamefont {D.}~\bibnamefont {Culcer}},
  \bibinfo {author} {\bibfnamefont {Q.}~\bibnamefont {Niu}}, \bibinfo {author}
  {\bibfnamefont {N.~A.}\ \bibnamefont {Sinitsyn}}, \bibinfo {author}
  {\bibfnamefont {T.}~\bibnamefont {Jungwirth}},\ and\ \bibinfo {author}
  {\bibfnamefont {A.~H.}\ \bibnamefont {MacDonald}},\ }\bibfield  {title}
  {\bibinfo {title} {{Universal Intrinsic Spin Hall Effect}},\ }\href
  {https://doi.org/10.1103/PhysRevLett.92.126603} {\bibfield  {journal}
  {\bibinfo  {journal} {Phys. Rev. Lett.}\ }\textbf {\bibinfo {volume} {92}},\
  \bibinfo {pages} {126603} (\bibinfo {year} {2004})}\BibitemShut {NoStop}%
\bibitem [{\citenamefont {Sinova}\ \emph {et~al.}(2015)\citenamefont {Sinova},
  \citenamefont {Valenzuela}, \citenamefont {Wunderlich}, \citenamefont
  {Back},\ and\ \citenamefont {Jungwirth}}]{sinova2015spin}%
  \BibitemOpen
  \bibfield  {author} {\bibinfo {author} {\bibfnamefont {J.}~\bibnamefont
  {Sinova}}, \bibinfo {author} {\bibfnamefont {S.~O.}\ \bibnamefont
  {Valenzuela}}, \bibinfo {author} {\bibfnamefont {J.}~\bibnamefont
  {Wunderlich}}, \bibinfo {author} {\bibfnamefont {C.~H.}\ \bibnamefont
  {Back}},\ and\ \bibinfo {author} {\bibfnamefont {T.}~\bibnamefont
  {Jungwirth}},\ }\bibfield  {title} {\bibinfo {title} {{Spin Hall effects}},\
  }\href {https://doi.org/10.1103/RevModPhys.87.1213} {\bibfield  {journal}
  {\bibinfo  {journal} {Rev. Mod. Phys.}\ }\textbf {\bibinfo {volume} {87}},\
  \bibinfo {pages} {1213} (\bibinfo {year} {2015})}\BibitemShut {NoStop}%
\bibitem [{\citenamefont {Vilenkin}(1978)}]{vilenkin1978parity}%
  \BibitemOpen
  \bibfield  {author} {\bibinfo {author} {\bibfnamefont {A.}~\bibnamefont
  {Vilenkin}},\ }\bibfield  {title} {\bibinfo {title} {{Parity violating
  currents in thermal radiation}},\ }\href
  {https://doi.org/10.1016/0370-2693(78)90330-1} {\bibfield  {journal}
  {\bibinfo  {journal} {Phys. Lett. B}\ }\textbf {\bibinfo {volume} {80}},\
  \bibinfo {pages} {150} (\bibinfo {year} {1978})}\BibitemShut {NoStop}%
\bibitem [{\citenamefont {Vilenkin}(1979)}]{vilenkin1979macroscopic}%
  \BibitemOpen
  \bibfield  {author} {\bibinfo {author} {\bibfnamefont {A.}~\bibnamefont
  {Vilenkin}},\ }\bibfield  {title} {\bibinfo {title} {{Macroscopic
  parity-violating effects: Neutrino fluxes from rotating black holes and in
  rotating thermal radiation}},\ }\href
  {https://doi.org/10.1103/PhysRevD.20.1807} {\bibfield  {journal} {\bibinfo
  {journal} {Phys. Rev. D}\ }\textbf {\bibinfo {volume} {20}},\ \bibinfo
  {pages} {1807} (\bibinfo {year} {1979})}\BibitemShut {NoStop}%
\bibitem [{\citenamefont {Vilenkin}(1980)}]{vilenkin1980quantum}%
  \BibitemOpen
  \bibfield  {author} {\bibinfo {author} {\bibfnamefont {A.}~\bibnamefont
  {Vilenkin}},\ }\bibfield  {title} {\bibinfo {title} {{Quantum Field Theory at
  Finite Temperature in a Rotating System}},\ }\href
  {https://doi.org/10.1103/PhysRevD.21.2260} {\bibfield  {journal} {\bibinfo
  {journal} {Phys. Rev. D}\ }\textbf {\bibinfo {volume} {21}},\ \bibinfo
  {pages} {2260} (\bibinfo {year} {1980})}\BibitemShut {NoStop}%
\bibitem [{\citenamefont {Son}\ and\ \citenamefont
  {Surówka}(2009)}]{son2009hydrodynamics}%
  \BibitemOpen
  \bibfield  {author} {\bibinfo {author} {\bibfnamefont {D.~T.}\ \bibnamefont
  {Son}}\ and\ \bibinfo {author} {\bibfnamefont {P.}~\bibnamefont {Surówka}},\
  }\bibfield  {title} {\bibinfo {title} {{Hydrodynamics with Triangle
  Anomalies}},\ }\href {https://doi.org/10.1103/PhysRevLett.103.191601}
  {\bibfield  {journal} {\bibinfo  {journal} {Phys. Rev. Lett.}\ }\textbf
  {\bibinfo {volume} {103}},\ \bibinfo {pages} {191601} (\bibinfo {year}
  {2009})}\BibitemShut {NoStop}%
\bibitem [{\citenamefont {Erdmenger}\ \emph {et~al.}(2009)\citenamefont
  {Erdmenger}, \citenamefont {Haack}, \citenamefont {Kaminski},\ and\
  \citenamefont {Yarom}}]{erdmenger2009fluid}%
  \BibitemOpen
  \bibfield  {author} {\bibinfo {author} {\bibfnamefont {J.}~\bibnamefont
  {Erdmenger}}, \bibinfo {author} {\bibfnamefont {M.}~\bibnamefont {Haack}},
  \bibinfo {author} {\bibfnamefont {M.}~\bibnamefont {Kaminski}},\ and\
  \bibinfo {author} {\bibfnamefont {A.}~\bibnamefont {Yarom}},\ }\bibfield
  {title} {\bibinfo {title} {{Fluid dynamics of R-charged black holes}},\
  }\href {https://doi.org/10.1088/1126-6708/2009/01/055} {\bibfield  {journal}
  {\bibinfo  {journal} {J. High Energy Phys.}\ }\textbf {\bibinfo {volume}
  {2009}}\bibinfo  {number} { (1)},\ \bibinfo {pages} {055}}\BibitemShut
  {NoStop}%
\bibitem [{\citenamefont {Landsteiner}\ \emph {et~al.}(2011)\citenamefont
  {Landsteiner}, \citenamefont {Megías},\ and\ \citenamefont
  {Pena-Benitez}}]{landsteiner2011gravitational}%
  \BibitemOpen
\bibfield  {number} {  }\bibfield  {author} {\bibinfo {author} {\bibfnamefont
  {K.}~\bibnamefont {Landsteiner}}, \bibinfo {author} {\bibfnamefont
  {E.}~\bibnamefont {Megías}},\ and\ \bibinfo {author} {\bibfnamefont
  {F.}~\bibnamefont {Pena-Benitez}},\ }\bibfield  {title} {\bibinfo {title}
  {{Gravitational Anomaly and Transport Phenomena}},\ }\href
  {https://doi.org/10.1103/PhysRevLett.107.021601} {\bibfield  {journal}
  {\bibinfo  {journal} {Phys. Rev. Lett.}\ }\textbf {\bibinfo {volume} {107}},\
  \bibinfo {pages} {021601} (\bibinfo {year} {2011})}\BibitemShut {NoStop}%
\bibitem [{\citenamefont {Stephanov}\ and\ \citenamefont
  {Yin}(2012)}]{stephanov2012chiral}%
  \BibitemOpen
  \bibfield  {author} {\bibinfo {author} {\bibfnamefont {M.~A.}\ \bibnamefont
  {Stephanov}}\ and\ \bibinfo {author} {\bibfnamefont {Y.}~\bibnamefont
  {Yin}},\ }\bibfield  {title} {\bibinfo {title} {{Chiral Kinetic Theory}},\
  }\href {https://doi.org/10.1103/PhysRevLett.109.162001} {\bibfield  {journal}
  {\bibinfo  {journal} {Phys. Rev. Lett.}\ }\textbf {\bibinfo {volume} {109}},\
  \bibinfo {pages} {162001} (\bibinfo {year} {2012})}\BibitemShut {NoStop}%
\bibitem [{\citenamefont {Landsteiner}(2016)}]{landsteiner2016notes}%
  \BibitemOpen
  \bibfield  {author} {\bibinfo {author} {\bibfnamefont {K.}~\bibnamefont
  {Landsteiner}},\ }\bibfield  {title} {\bibinfo {title} {{Notes on Anomaly
  Induced Transport}},\ }\href {https://doi.org/10.5506/APhysPolB.47.2617}
  {\bibfield  {journal} {\bibinfo  {journal} {Acta Phys. Pol. B}\ }\textbf
  {\bibinfo {volume} {47}},\ \bibinfo {pages} {2617} (\bibinfo {year}
  {2016})}\BibitemShut {NoStop}%
\bibitem [{\citenamefont {Kharzeev}\ \emph {et~al.}(2016)\citenamefont
  {Kharzeev}, \citenamefont {Liao}, \citenamefont {Voloshin},\ and\
  \citenamefont {Wang}}]{kharzeev2016chiral}%
  \BibitemOpen
  \bibfield  {author} {\bibinfo {author} {\bibfnamefont {D.~E.}\ \bibnamefont
  {Kharzeev}}, \bibinfo {author} {\bibfnamefont {J.}~\bibnamefont {Liao}},
  \bibinfo {author} {\bibfnamefont {S.~A.}\ \bibnamefont {Voloshin}},\ and\
  \bibinfo {author} {\bibfnamefont {G.}~\bibnamefont {Wang}},\ }\bibfield
  {title} {\bibinfo {title} {{Chiral magnetic and vortical effects in
  high-energy nuclear collisions—A status report}},\ }\href
  {https://doi.org/10.1016/j.ppnp.2016.01.001} {\bibfield  {journal} {\bibinfo
  {journal} {Prog. Part. Nucl. Phys.}\ }\textbf {\bibinfo {volume} {88}},\
  \bibinfo {pages} {1} (\bibinfo {year} {2016})}\BibitemShut {NoStop}%
\bibitem [{\citenamefont {Yan}\ and\ \citenamefont
  {Felser}(2017)}]{yan2017topological}%
  \BibitemOpen
  \bibfield  {author} {\bibinfo {author} {\bibfnamefont {B.}~\bibnamefont
  {Yan}}\ and\ \bibinfo {author} {\bibfnamefont {C.}~\bibnamefont {Felser}},\
  }\bibfield  {title} {\bibinfo {title} {{Topological Materials: Weyl
  Semimetals}},\ }\href
  {https://doi.org/10.1146/annurev-conmatphys-031016-025458} {\bibfield
  {journal} {\bibinfo  {journal} {Annu. Rev. Condens. Matter Phys.}\ }\textbf
  {\bibinfo {volume} {8}},\ \bibinfo {pages} {337} (\bibinfo {year}
  {2017})}\BibitemShut {NoStop}%
\bibitem [{\citenamefont {Armitage}\ \emph {et~al.}(2018)\citenamefont
  {Armitage}, \citenamefont {Mele},\ and\ \citenamefont
  {Vishwanath}}]{armitage2018weyl}%
  \BibitemOpen
  \bibfield  {author} {\bibinfo {author} {\bibfnamefont {N.~P.}\ \bibnamefont
  {Armitage}}, \bibinfo {author} {\bibfnamefont {E.~J.}\ \bibnamefont {Mele}},\
  and\ \bibinfo {author} {\bibfnamefont {A.}~\bibnamefont {Vishwanath}},\
  }\bibfield  {title} {\bibinfo {title} {{Weyl and Dirac semimetals in
  three-dimensional solids}},\ }\href
  {https://doi.org/10.1103/RevModPhys.90.015001} {\bibfield  {journal}
  {\bibinfo  {journal} {Rev. Mod. Phys.}\ }\textbf {\bibinfo {volume} {90}},\
  \bibinfo {pages} {015001} (\bibinfo {year} {2018})}\BibitemShut {NoStop}%
\bibitem [{\citenamefont {Gorbar}\ \emph {et~al.}(2018)\citenamefont {Gorbar},
  \citenamefont {Miransky}, \citenamefont {Shovkovy},\ and\ \citenamefont
  {Sukhachov}}]{gorbar2018anomalous}%
  \BibitemOpen
  \bibfield  {author} {\bibinfo {author} {\bibfnamefont {E.~V.}\ \bibnamefont
  {Gorbar}}, \bibinfo {author} {\bibfnamefont {V.~A.}\ \bibnamefont
  {Miransky}}, \bibinfo {author} {\bibfnamefont {I.~A.}\ \bibnamefont
  {Shovkovy}},\ and\ \bibinfo {author} {\bibfnamefont {P.~O.}\ \bibnamefont
  {Sukhachov}},\ }\bibfield  {title} {\bibinfo {title} {{Anomalous transport
  properties of Dirac and Weyl semimetals}},\ }\href
  {https://doi.org/10.1063/1.5037551} {\bibfield  {journal} {\bibinfo
  {journal} {Low Temp. Phys.}\ }\textbf {\bibinfo {volume} {44}},\ \bibinfo
  {pages} {487} (\bibinfo {year} {2018})}\BibitemShut {NoStop}%
\bibitem [{\citenamefont {Nagaosa}\ \emph {et~al.}(2020)\citenamefont
  {Nagaosa}, \citenamefont {Morimoto},\ and\ \citenamefont
  {Tokura}}]{nagaosa2020transport}%
  \BibitemOpen
  \bibfield  {author} {\bibinfo {author} {\bibfnamefont {N.}~\bibnamefont
  {Nagaosa}}, \bibinfo {author} {\bibfnamefont {T.}~\bibnamefont {Morimoto}},\
  and\ \bibinfo {author} {\bibfnamefont {Y.}~\bibnamefont {Tokura}},\
  }\bibfield  {title} {\bibinfo {title} {{Transport, magnetic and optical
  properties of Weyl materials}},\ }\href
  {https://doi.org/10.1038/s41578-020-0208-y} {\bibfield  {journal} {\bibinfo
  {journal} {Nat. Rev. Mater.}\ }\textbf {\bibinfo {volume} {5}},\ \bibinfo
  {pages} {621} (\bibinfo {year} {2020})}\BibitemShut {NoStop}%
\bibitem [{\citenamefont {Kamada}\ \emph {et~al.}(2023)\citenamefont {Kamada},
  \citenamefont {Yamamoto},\ and\ \citenamefont {Yang}}]{kamada2023chiral}%
  \BibitemOpen
  \bibfield  {author} {\bibinfo {author} {\bibfnamefont {K.}~\bibnamefont
  {Kamada}}, \bibinfo {author} {\bibfnamefont {N.}~\bibnamefont {Yamamoto}},\
  and\ \bibinfo {author} {\bibfnamefont {D.-L.}\ \bibnamefont {Yang}},\
  }\bibfield  {title} {\bibinfo {title} {{Chiral effects in astrophysics and
  cosmology}},\ }\href {https://doi.org/10.1016/j.ppnp.2022.104016} {\bibfield
  {journal} {\bibinfo  {journal} {Prog. Part. Nucl. Phys.}\ }\textbf {\bibinfo
  {volume} {129}},\ \bibinfo {pages} {104016} (\bibinfo {year}
  {2023})}\BibitemShut {NoStop}%
\bibitem [{Note1()}]{Note1}%
  \BibitemOpen
  \bibinfo {note} {Two recent exceptions to this with quantum metric effects
  are Refs.~\cite {oancea2026quantum} and~\cite {akiba2026quantum}. The former
  focusses on fermion propagation via a subbundle geometry formalism, while the
  latter explores photon transport using a semiclassical action.}\BibitemShut
  {Stop}%
\bibitem [{\citenamefont {T\"orm\"a}(2023)}]{torma2023essay}%
  \BibitemOpen
  \bibfield  {author} {\bibinfo {author} {\bibfnamefont {P.}~\bibnamefont
  {T\"orm\"a}},\ }\bibfield  {title} {\bibinfo {title} {{Essay: Where Can
  Quantum Geometry Lead Us?}},\ }\href
  {https://link.aps.org/doi/10.1103/PhysRevLett.131.240001} {\bibfield
  {journal} {\bibinfo  {journal} {Phys. Rev. Lett.}\ }\textbf {\bibinfo
  {volume} {131}},\ \bibinfo {pages} {240001} (\bibinfo {year}
  {2023})}\BibitemShut {NoStop}%
\bibitem [{\citenamefont {Liu}\ \emph {et~al.}(2024)\citenamefont {Liu},
  \citenamefont {Qiang}, \citenamefont {Lu},\ and\ \citenamefont
  {Xie}}]{liu2024quantum}%
  \BibitemOpen
  \bibfield  {author} {\bibinfo {author} {\bibfnamefont {T.}~\bibnamefont
  {Liu}}, \bibinfo {author} {\bibfnamefont {X.-B.}\ \bibnamefont {Qiang}},
  \bibinfo {author} {\bibfnamefont {H.-Z.}\ \bibnamefont {Lu}},\ and\ \bibinfo
  {author} {\bibfnamefont {X.}~\bibnamefont {Xie}},\ }\bibfield  {title}
  {\bibinfo {title} {{Quantum geometry in condensed matter}},\ }\href
  {https://doi.org/10.1093/nsr/nwae334} {\bibfield  {journal} {\bibinfo
  {journal} {Natl. Sci. Rev.}\ ,\ \bibinfo {pages} {nwae334}} (\bibinfo {year}
  {2024})}\BibitemShut {NoStop}%
\bibitem [{\citenamefont {Chen}(2024)}]{chen2024quantum}%
  \BibitemOpen
  \bibfield  {author} {\bibinfo {author} {\bibfnamefont {W.}~\bibnamefont
  {Chen}},\ }\bibfield  {title} {\bibinfo {title} {Quantum geometrical
  properties of topological materials},\ }\href
  {http://dx.doi.org/10.1088/1361-648X/ad8619} {\bibfield  {journal} {\bibinfo
  {journal} {J. Phys.: Condens. Matter}\ }\textbf {\bibinfo {volume} {37}},\
  \bibinfo {pages} {025605} (\bibinfo {year} {2024})}\BibitemShut {NoStop}%
\bibitem [{\citenamefont {Shim}\ \emph {et~al.}(2025)\citenamefont {Shim},
  \citenamefont {Mehraeen}, \citenamefont {Sklenar}, \citenamefont {Zhang},
  \citenamefont {Hoffmann},\ and\ \citenamefont {Mason}}]{shim2025spin}%
  \BibitemOpen
  \bibfield  {author} {\bibinfo {author} {\bibfnamefont {S.}~\bibnamefont
  {Shim}}, \bibinfo {author} {\bibfnamefont {M.}~\bibnamefont {Mehraeen}},
  \bibinfo {author} {\bibfnamefont {J.}~\bibnamefont {Sklenar}}, \bibinfo
  {author} {\bibfnamefont {S.~S.-L.}\ \bibnamefont {Zhang}}, \bibinfo {author}
  {\bibfnamefont {A.}~\bibnamefont {Hoffmann}},\ and\ \bibinfo {author}
  {\bibfnamefont {N.}~\bibnamefont {Mason}},\ }\bibfield  {title} {\bibinfo
  {title} {{Spin-polarized antiferromagnetic metals}},\ }\href
  {https://doi.org/10.1146/annurev-conmatphys-042924-123620} {\bibfield
  {journal} {\bibinfo  {journal} {Annu. Rev. Condens. Matter Phys.}\ }\textbf
  {\bibinfo {volume} {16}} (\bibinfo {year} {2025})}\BibitemShut {NoStop}%
\bibitem [{\citenamefont {Jiang}\ \emph {et~al.}(2025)\citenamefont {Jiang},
  \citenamefont {Holder},\ and\ \citenamefont {Yan}}]{jiang2025revealing}%
  \BibitemOpen
  \bibfield  {author} {\bibinfo {author} {\bibfnamefont {Y.}~\bibnamefont
  {Jiang}}, \bibinfo {author} {\bibfnamefont {T.}~\bibnamefont {Holder}},\ and\
  \bibinfo {author} {\bibfnamefont {B.}~\bibnamefont {Yan}},\ }\bibfield
  {title} {\bibinfo {title} {{Revealing quantum geometry in nonlinear quantum
  materials}},\ }\href {https://doi.org/10.1088/1361-6633/ade454} {\bibfield
  {journal} {\bibinfo  {journal} {Rep. Prog. Phys.}\ }\textbf {\bibinfo
  {volume} {88}} (\bibinfo {year} {2025})}\BibitemShut {NoStop}%
\bibitem [{\citenamefont {Yu}\ \emph {et~al.}(2025)\citenamefont {Yu},
  \citenamefont {Bernevig}, \citenamefont {Queiroz}, \citenamefont {Rossi},
  \citenamefont {Törmä},\ and\ \citenamefont {Yang}}]{yu2025quantum}%
  \BibitemOpen
  \bibfield  {author} {\bibinfo {author} {\bibfnamefont {J.}~\bibnamefont
  {Yu}}, \bibinfo {author} {\bibfnamefont {B.~A.}\ \bibnamefont {Bernevig}},
  \bibinfo {author} {\bibfnamefont {R.}~\bibnamefont {Queiroz}}, \bibinfo
  {author} {\bibfnamefont {E.}~\bibnamefont {Rossi}}, \bibinfo {author}
  {\bibfnamefont {P.}~\bibnamefont {Törmä}},\ and\ \bibinfo {author}
  {\bibfnamefont {B.-J.}\ \bibnamefont {Yang}},\ }\bibfield  {title} {\bibinfo
  {title} {{Quantum geometry in quantum materials}},\ }\href
  {https://doi.org/10.1038/s41535-025-00801-3} {\bibfield  {journal} {\bibinfo
  {journal} {npj Quantum Mater.}\ }\textbf {\bibinfo {volume} {10}},\ \bibinfo
  {pages} {101} (\bibinfo {year} {2025})}\BibitemShut {NoStop}%
\bibitem [{\citenamefont {Verma}\ \emph {et~al.}(2026)\citenamefont {Verma},
  \citenamefont {Moll}, \citenamefont {Holder},\ and\ \citenamefont
  {Queiroz}}]{verma2026quantum}%
  \BibitemOpen
  \bibfield  {author} {\bibinfo {author} {\bibfnamefont {N.}~\bibnamefont
  {Verma}}, \bibinfo {author} {\bibfnamefont {P.~J.~W.}\ \bibnamefont {Moll}},
  \bibinfo {author} {\bibfnamefont {T.}~\bibnamefont {Holder}},\ and\ \bibinfo
  {author} {\bibfnamefont {R.}~\bibnamefont {Queiroz}},\ }\bibfield  {title}
  {\bibinfo {title} {{Quantum geometry and the hidden scales in materials}},\
  }\href {https://doi.org/10.1038/s42254-026-00923-y} {\bibfield  {journal}
  {\bibinfo  {journal} {Nat. Rev. Phys.}\ }\textbf {\bibinfo {volume} {8}},\
  \bibinfo {pages} {226} (\bibinfo {year} {2026})}\BibitemShut {NoStop}%
\bibitem [{\citenamefont {Gao}\ \emph {et~al.}(2025)\citenamefont {Gao},
  \citenamefont {Nagaosa}, \citenamefont {Ni},\ and\ \citenamefont
  {Xu}}]{gao2025quantum}%
  \BibitemOpen
  \bibfield  {author} {\bibinfo {author} {\bibfnamefont {A.}~\bibnamefont
  {Gao}}, \bibinfo {author} {\bibfnamefont {N.}~\bibnamefont {Nagaosa}},
  \bibinfo {author} {\bibfnamefont {N.}~\bibnamefont {Ni}},\ and\ \bibinfo
  {author} {\bibfnamefont {S.-Y.}\ \bibnamefont {Xu}},\ }\bibfield  {title}
  {\bibinfo {title} {{Quantum Geometry Phenomena in Condensed Matter
  Systems}},\ }\href {https://arxiv.org/abs/2508.00469} {\bibfield  {journal}
  {\bibinfo  {journal} {arXiv:2508.00469}\ } (\bibinfo {year}
  {2025})}\BibitemShut {NoStop}%
\bibitem [{\citenamefont {Provost}\ and\ \citenamefont
  {Vallee}(1980)}]{provost1980riemannian}%
  \BibitemOpen
  \bibfield  {author} {\bibinfo {author} {\bibfnamefont {J.}~\bibnamefont
  {Provost}}\ and\ \bibinfo {author} {\bibfnamefont {G.}~\bibnamefont
  {Vallee}},\ }\bibfield  {title} {\bibinfo {title} {{Riemannian structure on
  manifolds of quantum states}},\ }\href {https://doi.org/10.1007/BF02193559}
  {\bibfield  {journal} {\bibinfo  {journal} {Commun. Math. Phys.}\ }\textbf
  {\bibinfo {volume} {76}},\ \bibinfo {pages} {289} (\bibinfo {year}
  {1980})}\BibitemShut {NoStop}%
\bibitem [{\citenamefont {Xiao}\ \emph {et~al.}(2010)\citenamefont {Xiao},
  \citenamefont {Chang},\ and\ \citenamefont {Niu}}]{xiao2010berry}%
  \BibitemOpen
  \bibfield  {author} {\bibinfo {author} {\bibfnamefont {D.}~\bibnamefont
  {Xiao}}, \bibinfo {author} {\bibfnamefont {M.-C.}\ \bibnamefont {Chang}},\
  and\ \bibinfo {author} {\bibfnamefont {Q.}~\bibnamefont {Niu}},\ }\bibfield
  {title} {\bibinfo {title} {{Berry phase effects on electronic properties}},\
  }\href {https://link.aps.org/doi/10.1103/RevModPhys.82.1959} {\bibfield
  {journal} {\bibinfo  {journal} {Rev. Mod. Phys.}\ }\textbf {\bibinfo {volume}
  {82}},\ \bibinfo {pages} {1959} (\bibinfo {year} {2010})}\BibitemShut
  {NoStop}%
\bibitem [{\citenamefont {Marzari}\ and\ \citenamefont
  {Vanderbilt}(1997)}]{marzari1997maximally}%
  \BibitemOpen
  \bibfield  {author} {\bibinfo {author} {\bibfnamefont {N.}~\bibnamefont
  {Marzari}}\ and\ \bibinfo {author} {\bibfnamefont {D.}~\bibnamefont
  {Vanderbilt}},\ }\bibfield  {title} {\bibinfo {title} {{Maximally localized
  generalized Wannier functions for composite energy bands}},\ }\href
  {https://doi.org/10.1103/PhysRevB.56.12847} {\bibfield  {journal} {\bibinfo
  {journal} {Phys. Rev. B}\ }\textbf {\bibinfo {volume} {56}},\ \bibinfo
  {pages} {12847} (\bibinfo {year} {1997})}\BibitemShut {NoStop}%
\bibitem [{\citenamefont {Souza}\ \emph {et~al.}(2000)\citenamefont {Souza},
  \citenamefont {Wilkens},\ and\ \citenamefont
  {Martin}}]{souza2000polarization}%
  \BibitemOpen
  \bibfield  {author} {\bibinfo {author} {\bibfnamefont {I.}~\bibnamefont
  {Souza}}, \bibinfo {author} {\bibfnamefont {T.}~\bibnamefont {Wilkens}},\
  and\ \bibinfo {author} {\bibfnamefont {R.~M.}\ \bibnamefont {Martin}},\
  }\bibfield  {title} {\bibinfo {title} {{Polarization and localization in
  insulators: Generating function approach}},\ }\href
  {https://link.aps.org/doi/10.1103/PhysRevB.62.1666} {\bibfield  {journal}
  {\bibinfo  {journal} {Phys. Rev. B}\ }\textbf {\bibinfo {volume} {62}},\
  \bibinfo {pages} {1666} (\bibinfo {year} {2000})}\BibitemShut {NoStop}%
\bibitem [{\citenamefont {Campos~Venuti}\ and\ \citenamefont
  {Zanardi}(2007)}]{venuti2007quantum}%
  \BibitemOpen
  \bibfield  {author} {\bibinfo {author} {\bibfnamefont {L.}~\bibnamefont
  {Campos~Venuti}}\ and\ \bibinfo {author} {\bibfnamefont {P.}~\bibnamefont
  {Zanardi}},\ }\bibfield  {title} {\bibinfo {title} {{Quantum Critical Scaling
  of the Geometric Tensors}},\ }\href
  {https://doi.org/10.1103/PhysRevLett.99.095701} {\bibfield  {journal}
  {\bibinfo  {journal} {Phys. Rev. Lett.}\ }\textbf {\bibinfo {volume} {99}},\
  \bibinfo {pages} {095701} (\bibinfo {year} {2007})}\BibitemShut {NoStop}%
\bibitem [{\citenamefont {Ma}\ \emph {et~al.}(2010)\citenamefont {Ma},
  \citenamefont {Chen}, \citenamefont {Fan},\ and\ \citenamefont
  {Liu}}]{ma2010abelian}%
  \BibitemOpen
  \bibfield  {author} {\bibinfo {author} {\bibfnamefont {Y.-Q.}\ \bibnamefont
  {Ma}}, \bibinfo {author} {\bibfnamefont {S.}~\bibnamefont {Chen}}, \bibinfo
  {author} {\bibfnamefont {H.}~\bibnamefont {Fan}},\ and\ \bibinfo {author}
  {\bibfnamefont {W.-M.}\ \bibnamefont {Liu}},\ }\bibfield  {title} {\bibinfo
  {title} {{Abelian and non-Abelian quantum geometric tensor}},\ }\href
  {https://link.aps.org/doi/10.1103/PhysRevB.81.245129} {\bibfield  {journal}
  {\bibinfo  {journal} {Phys. Rev. B}\ }\textbf {\bibinfo {volume} {81}},\
  \bibinfo {pages} {245129} (\bibinfo {year} {2010})}\BibitemShut {NoStop}%
\bibitem [{\citenamefont {Resta}(2011)}]{resta2011insulating}%
  \BibitemOpen
  \bibfield  {author} {\bibinfo {author} {\bibfnamefont {R.}~\bibnamefont
  {Resta}},\ }\bibfield  {title} {\bibinfo {title} {{The insulating state of
  matter: a geometrical theory}},\ }\href
  {https://doi.org/10.1140/epjb/e2010-10874-4} {\bibfield  {journal} {\bibinfo
  {journal} {Eur. Phys. J. B}\ }\textbf {\bibinfo {volume} {79}},\ \bibinfo
  {pages} {121} (\bibinfo {year} {2011})}\BibitemShut {NoStop}%
\bibitem [{\citenamefont {De~Grandi}\ \emph {et~al.}(2011)\citenamefont
  {De~Grandi}, \citenamefont {Polkovnikov},\ and\ \citenamefont
  {Sandvik}}]{degrandi2011universal}%
  \BibitemOpen
  \bibfield  {author} {\bibinfo {author} {\bibfnamefont {C.}~\bibnamefont
  {De~Grandi}}, \bibinfo {author} {\bibfnamefont {A.}~\bibnamefont
  {Polkovnikov}},\ and\ \bibinfo {author} {\bibfnamefont {A.~W.}\ \bibnamefont
  {Sandvik}},\ }\bibfield  {title} {\bibinfo {title} {{Universal nonequilibrium
  quantum dynamics in imaginary time}},\ }\href
  {https://doi.org/10.1103/PhysRevB.84.224303} {\bibfield  {journal} {\bibinfo
  {journal} {Phys. Rev. B}\ }\textbf {\bibinfo {volume} {84}},\ \bibinfo
  {pages} {224303} (\bibinfo {year} {2011})}\BibitemShut {NoStop}%
\bibitem [{\citenamefont {Neupert}\ \emph {et~al.}(2013)\citenamefont
  {Neupert}, \citenamefont {Chamon},\ and\ \citenamefont
  {Mudry}}]{neupert2013measuring}%
  \BibitemOpen
  \bibfield  {author} {\bibinfo {author} {\bibfnamefont {T.}~\bibnamefont
  {Neupert}}, \bibinfo {author} {\bibfnamefont {C.}~\bibnamefont {Chamon}},\
  and\ \bibinfo {author} {\bibfnamefont {C.}~\bibnamefont {Mudry}},\ }\bibfield
   {title} {\bibinfo {title} {{Measuring the quantum geometry of Bloch bands
  with current noise}},\ }\href
  {https://link.aps.org/doi/10.1103/PhysRevB.87.245103} {\bibfield  {journal}
  {\bibinfo  {journal} {Phys. Rev. B}\ }\textbf {\bibinfo {volume} {87}},\
  \bibinfo {pages} {245103} (\bibinfo {year} {2013})}\BibitemShut {NoStop}%
\bibitem [{\citenamefont {Claassen}\ \emph {et~al.}(2015)\citenamefont
  {Claassen}, \citenamefont {Lee}, \citenamefont {Thomale}, \citenamefont
  {Qi},\ and\ \citenamefont {Devereaux}}]{claassen2015position}%
  \BibitemOpen
  \bibfield  {author} {\bibinfo {author} {\bibfnamefont {M.}~\bibnamefont
  {Claassen}}, \bibinfo {author} {\bibfnamefont {C.~H.}\ \bibnamefont {Lee}},
  \bibinfo {author} {\bibfnamefont {R.}~\bibnamefont {Thomale}}, \bibinfo
  {author} {\bibfnamefont {X.-L.}\ \bibnamefont {Qi}},\ and\ \bibinfo {author}
  {\bibfnamefont {T.~P.}\ \bibnamefont {Devereaux}},\ }\bibfield  {title}
  {\bibinfo {title} {{Position-Momentum Duality and Fractional Quantum Hall
  Effect in Chern Insulators}},\ }\href
  {https://link.aps.org/doi/10.1103/PhysRevLett.114.236802} {\bibfield
  {journal} {\bibinfo  {journal} {Phys. Rev. Lett.}\ }\textbf {\bibinfo
  {volume} {114}},\ \bibinfo {pages} {236802} (\bibinfo {year}
  {2015})}\BibitemShut {NoStop}%
\bibitem [{\citenamefont {Gao}\ \emph {et~al.}(2014)\citenamefont {Gao},
  \citenamefont {Yang},\ and\ \citenamefont {Niu}}]{gao2014field}%
  \BibitemOpen
  \bibfield  {author} {\bibinfo {author} {\bibfnamefont {Y.}~\bibnamefont
  {Gao}}, \bibinfo {author} {\bibfnamefont {S.~A.}\ \bibnamefont {Yang}},\ and\
  \bibinfo {author} {\bibfnamefont {Q.}~\bibnamefont {Niu}},\ }\bibfield
  {title} {\bibinfo {title} {{Field Induced Positional Shift of Bloch Electrons
  and Its Dynamical Implications}},\ }\href
  {https://link.aps.org/doi/10.1103/PhysRevLett.112.166601} {\bibfield
  {journal} {\bibinfo  {journal} {Phys. Rev. Lett.}\ }\textbf {\bibinfo
  {volume} {112}},\ \bibinfo {pages} {166601} (\bibinfo {year}
  {2014})}\BibitemShut {NoStop}%
\bibitem [{\citenamefont {Ahn}\ \emph {et~al.}(2022)\citenamefont {Ahn},
  \citenamefont {Guo}, \citenamefont {Nagaosa},\ and\ \citenamefont
  {Vishwanath}}]{ahn2022riemannian}%
  \BibitemOpen
  \bibfield  {author} {\bibinfo {author} {\bibfnamefont {J.}~\bibnamefont
  {Ahn}}, \bibinfo {author} {\bibfnamefont {G.-Y.}\ \bibnamefont {Guo}},
  \bibinfo {author} {\bibfnamefont {N.}~\bibnamefont {Nagaosa}},\ and\ \bibinfo
  {author} {\bibfnamefont {A.}~\bibnamefont {Vishwanath}},\ }\bibfield  {title}
  {\bibinfo {title} {Riemannian geometry of resonant optical responses},\
  }\href {https://doi.org/10.1038/s41567-021-01465-z} {\bibfield  {journal}
  {\bibinfo  {journal} {Nat. Phys.}\ }\textbf {\bibinfo {volume} {18}},\
  \bibinfo {pages} {290} (\bibinfo {year} {2022})}\BibitemShut {NoStop}%
\bibitem [{\citenamefont {Hsu}\ \emph {et~al.}(2023)\citenamefont {Hsu},
  \citenamefont {You}, \citenamefont {Ahn},\ and\ \citenamefont
  {Guo}}]{hsu2023nonlinear}%
  \BibitemOpen
  \bibfield  {author} {\bibinfo {author} {\bibfnamefont {H.-C.}\ \bibnamefont
  {Hsu}}, \bibinfo {author} {\bibfnamefont {J.-S.}\ \bibnamefont {You}},
  \bibinfo {author} {\bibfnamefont {J.}~\bibnamefont {Ahn}},\ and\ \bibinfo
  {author} {\bibfnamefont {G.-Y.}\ \bibnamefont {Guo}},\ }\bibfield  {title}
  {\bibinfo {title} {{Nonlinear photoconductivities and quantum geometry of
  chiral multifold fermions}},\ }\href
  {https://link.aps.org/doi/10.1103/PhysRevB.107.155434} {\bibfield  {journal}
  {\bibinfo  {journal} {Phys. Rev. B}\ }\textbf {\bibinfo {volume} {107}},\
  \bibinfo {pages} {155434} (\bibinfo {year} {2023})}\BibitemShut {NoStop}%
\bibitem [{\citenamefont {Jain}\ \emph
  {et~al.}(2026{\natexlab{a}})\citenamefont {Jain}, \citenamefont {Jankowski},
  \citenamefont {Mehraeen},\ and\ \citenamefont {Slager}}]{jain2026nonlinear}%
  \BibitemOpen
  \bibfield  {author} {\bibinfo {author} {\bibfnamefont {A.}~\bibnamefont
  {Jain}}, \bibinfo {author} {\bibfnamefont {W.~J.}\ \bibnamefont {Jankowski}},
  \bibinfo {author} {\bibfnamefont {M.}~\bibnamefont {Mehraeen}},\ and\
  \bibinfo {author} {\bibfnamefont {R.-J.}\ \bibnamefont {Slager}},\ }\bibfield
   {title} {\bibinfo {title} {{Nonlinear Odd Viscoelastic Effect}},\ }\href
  {https://link.aps.org/doi/10.1103/jg6l-gzfr} {\bibfield  {journal} {\bibinfo
  {journal} {Phys. Rev. Lett.}\ }\textbf {\bibinfo {volume} {137}},\ \bibinfo
  {pages} {036302} (\bibinfo {year} {2026}{\natexlab{a}})}\BibitemShut
  {NoStop}%
\bibitem [{\citenamefont {Jain}\ \emph
  {et~al.}(2026{\natexlab{b}})\citenamefont {Jain}, \citenamefont {Jankowski},
  \citenamefont {Mehraeen},\ and\ \citenamefont
  {Slager}}]{jain2026topological}%
  \BibitemOpen
  \bibfield  {author} {\bibinfo {author} {\bibfnamefont {A.}~\bibnamefont
  {Jain}}, \bibinfo {author} {\bibfnamefont {W.~J.}\ \bibnamefont {Jankowski}},
  \bibinfo {author} {\bibfnamefont {M.}~\bibnamefont {Mehraeen}},\ and\
  \bibinfo {author} {\bibfnamefont {R.-J.}\ \bibnamefont {Slager}},\ }\bibfield
   {title} {\bibinfo {title} {{Topological acoustic diode}},\ }\href
  {https://arxiv.org/abs/2601.20951} {\bibfield  {journal} {\bibinfo  {journal}
  {arXiv:2601.20951}\ } (\bibinfo {year} {2026}{\natexlab{b}})}\BibitemShut
  {NoStop}%
\bibitem [{\citenamefont {Mehraeen}(2025)}]{mehraeen2025quantum}%
  \BibitemOpen
  \bibfield  {author} {\bibinfo {author} {\bibfnamefont {M.}~\bibnamefont
  {Mehraeen}},\ }\bibfield  {title} {\bibinfo {title} {{Quantum Response Theory
  and Momentum-Space Gravity}},\ }\href
  {https://link.aps.org/doi/10.1103/t6nt-qzws} {\bibfield  {journal} {\bibinfo
  {journal} {Phys. Rev. Lett.}\ }\textbf {\bibinfo {volume} {135}},\ \bibinfo
  {pages} {156302} (\bibinfo {year} {2025})}\BibitemShut {NoStop}%
\bibitem [{\citenamefont {Guo}\ \emph {et~al.}(2025)\citenamefont {Guo},
  \citenamefont {Lu}, \citenamefont {Wang},\ and\ \citenamefont
  {Chang}}]{guo2025bicircular}%
  \BibitemOpen
  \bibfield  {author} {\bibinfo {author} {\bibfnamefont {Z.}~\bibnamefont
  {Guo}}, \bibinfo {author} {\bibfnamefont {Z.}~\bibnamefont {Lu}}, \bibinfo
  {author} {\bibfnamefont {H.}~\bibnamefont {Wang}},\ and\ \bibinfo {author}
  {\bibfnamefont {K.}~\bibnamefont {Chang}},\ }\bibfield  {title} {\bibinfo
  {title} {Bicircular light-induced multistate geometric current},\ }\href
  {https://link.aps.org/doi/10.1103/7ytw-vyb7} {\bibfield  {journal} {\bibinfo
  {journal} {Phys. Rev. B}\ }\textbf {\bibinfo {volume} {112}},\ \bibinfo
  {pages} {035162} (\bibinfo {year} {2025})}\BibitemShut {NoStop}%
\bibitem [{\citenamefont {Kharzeev}\ \emph {et~al.}(2008)\citenamefont
  {Kharzeev}, \citenamefont {McLerran},\ and\ \citenamefont
  {Warringa}}]{kharzeev2008effects}%
  \BibitemOpen
  \bibfield  {author} {\bibinfo {author} {\bibfnamefont {D.~E.}\ \bibnamefont
  {Kharzeev}}, \bibinfo {author} {\bibfnamefont {L.~D.}\ \bibnamefont
  {McLerran}},\ and\ \bibinfo {author} {\bibfnamefont {H.~J.}\ \bibnamefont
  {Warringa}},\ }\bibfield  {title} {\bibinfo {title} {{The effects of
  topological charge change in heavy ion collisions: "Event by event P and CP
  violation"}},\ }\href {https://doi.org/10.1016/j.nuclphysa.2008.02.298}
  {\bibfield  {journal} {\bibinfo  {journal} {Nucl. Phys. A}\ }\textbf
  {\bibinfo {volume} {803}},\ \bibinfo {pages} {227} (\bibinfo {year}
  {2008})}\BibitemShut {NoStop}%
\bibitem [{\citenamefont {Fukushima}\ \emph {et~al.}(2008)\citenamefont
  {Fukushima}, \citenamefont {Kharzeev},\ and\ \citenamefont
  {Warringa}}]{fukushima2008chiral}%
  \BibitemOpen
  \bibfield  {author} {\bibinfo {author} {\bibfnamefont {K.}~\bibnamefont
  {Fukushima}}, \bibinfo {author} {\bibfnamefont {D.~E.}\ \bibnamefont
  {Kharzeev}},\ and\ \bibinfo {author} {\bibfnamefont {H.~J.}\ \bibnamefont
  {Warringa}},\ }\bibfield  {title} {\bibinfo {title} {{Chiral magnetic
  effect}},\ }\href {https://doi.org/10.1103/PhysRevD.78.074033} {\bibfield
  {journal} {\bibinfo  {journal} {Phys. Rev. D}\ }\textbf {\bibinfo {volume}
  {78}},\ \bibinfo {pages} {074033} (\bibinfo {year} {2008})}\BibitemShut
  {NoStop}%
\bibitem [{\citenamefont {Kharzeev}\ and\ \citenamefont
  {Son}(2011)}]{kharzeev2011testing}%
  \BibitemOpen
  \bibfield  {author} {\bibinfo {author} {\bibfnamefont {D.~E.}\ \bibnamefont
  {Kharzeev}}\ and\ \bibinfo {author} {\bibfnamefont {D.~T.}\ \bibnamefont
  {Son}},\ }\bibfield  {title} {\bibinfo {title} {{Testing the Chiral Magnetic
  and Chiral Vortical Effects in Heavy Ion Collisions}},\ }\href
  {https://doi.org/10.1103/PhysRevLett.106.062301} {\bibfield  {journal}
  {\bibinfo  {journal} {Phys. Rev. Lett.}\ }\textbf {\bibinfo {volume} {106}},\
  \bibinfo {pages} {062301} (\bibinfo {year} {2011})}\BibitemShut {NoStop}%
\bibitem [{\citenamefont {Tashiro}\ \emph {et~al.}(2012)\citenamefont
  {Tashiro}, \citenamefont {Vachaspati},\ and\ \citenamefont
  {Vilenkin}}]{tashiro2012chiral}%
  \BibitemOpen
  \bibfield  {author} {\bibinfo {author} {\bibfnamefont {H.}~\bibnamefont
  {Tashiro}}, \bibinfo {author} {\bibfnamefont {T.}~\bibnamefont
  {Vachaspati}},\ and\ \bibinfo {author} {\bibfnamefont {A.}~\bibnamefont
  {Vilenkin}},\ }\bibfield  {title} {\bibinfo {title} {{Chiral Effects and
  Cosmic Magnetic Fields}},\ }\href
  {https://doi.org/10.1103/PhysRevD.86.105033} {\bibfield  {journal} {\bibinfo
  {journal} {Phys. Rev. D}\ }\textbf {\bibinfo {volume} {86}},\ \bibinfo
  {pages} {105033} (\bibinfo {year} {2012})}\BibitemShut {NoStop}%
\bibitem [{\citenamefont {Boyarsky}\ \emph {et~al.}(2012)\citenamefont
  {Boyarsky}, \citenamefont {Fröhlich},\ and\ \citenamefont
  {Ruchayskiy}}]{boyarsky2012self}%
  \BibitemOpen
  \bibfield  {author} {\bibinfo {author} {\bibfnamefont {A.}~\bibnamefont
  {Boyarsky}}, \bibinfo {author} {\bibfnamefont {J.}~\bibnamefont
  {Fröhlich}},\ and\ \bibinfo {author} {\bibfnamefont {O.}~\bibnamefont
  {Ruchayskiy}},\ }\bibfield  {title} {\bibinfo {title} {{Self-consistent
  evolution of magnetic fields and chiral asymmetry in the early Universe}},\
  }\href {https://doi.org/10.1103/PhysRevLett.108.031301} {\bibfield  {journal}
  {\bibinfo  {journal} {Phys. Rev. Lett.}\ }\textbf {\bibinfo {volume} {108}},\
  \bibinfo {pages} {031301} (\bibinfo {year} {2012})}\BibitemShut {NoStop}%
\bibitem [{\citenamefont {Long}\ \emph {et~al.}(2014)\citenamefont {Long},
  \citenamefont {Sabancilar},\ and\ \citenamefont
  {Vachaspati}}]{long2014leptogenesis}%
  \BibitemOpen
  \bibfield  {author} {\bibinfo {author} {\bibfnamefont {A.~J.}\ \bibnamefont
  {Long}}, \bibinfo {author} {\bibfnamefont {E.}~\bibnamefont {Sabancilar}},\
  and\ \bibinfo {author} {\bibfnamefont {T.}~\bibnamefont {Vachaspati}},\
  }\bibfield  {title} {\bibinfo {title} {{Leptogenesis and primordial magnetic
  fields}},\ }\href {https://doi.org/10.1088/1475-7516/2014/02/036} {\bibfield
  {journal} {\bibinfo  {journal} {JCAP}\ }\textbf {\bibinfo {volume}
  {2014}}\bibinfo  {number} { (2)},\ \bibinfo {pages} {036}}\BibitemShut
  {NoStop}%
\bibitem [{\citenamefont {Abbaslu}\ \emph {et~al.}(2019)\citenamefont
  {Abbaslu}, \citenamefont {Rostam~Zadeh},\ and\ \citenamefont
  {Gousheh}}]{abbaslu2019contribution}%
  \BibitemOpen
\bibfield  {number} {  }\bibfield  {author} {\bibinfo {author} {\bibfnamefont
  {S.}~\bibnamefont {Abbaslu}}, \bibinfo {author} {\bibfnamefont
  {S.}~\bibnamefont {Rostam~Zadeh}},\ and\ \bibinfo {author} {\bibfnamefont
  {S.~S.}\ \bibnamefont {Gousheh}},\ }\bibfield  {title} {\bibinfo {title}
  {{Contribution of the chiral vortical effect to the evolution of the
  hypermagnetic field and the matter-antimatter asymmetry in the early
  Universe}},\ }\href {https://doi.org/10.1103/PhysRevD.100.116022} {\bibfield
  {journal} {\bibinfo  {journal} {Phys. Rev. D}\ }\textbf {\bibinfo {volume}
  {100}},\ \bibinfo {pages} {116022} (\bibinfo {year} {2019})}\BibitemShut
  {NoStop}%
\bibitem [{\citenamefont {Huang}(2021)}]{huang2021vorticity}%
  \BibitemOpen
  \bibfield  {author} {\bibinfo {author} {\bibfnamefont {X.-G.}\ \bibnamefont
  {Huang}},\ }\bibfield  {title} {\bibinfo {title} {{Vorticity and Spin
  Polarization --- A Theoretical Perspective}},\ }\href
  {https://doi.org/10.1016/j.nuclphysa.2020.121752} {\bibfield  {journal}
  {\bibinfo  {journal} {Nucl. Phys. A}\ }\textbf {\bibinfo {volume} {1005}},\
  \bibinfo {pages} {121752} (\bibinfo {year} {2021})}\BibitemShut {NoStop}%
\bibitem [{\citenamefont {Abbaslu}\ \emph {et~al.}(2021)\citenamefont
  {Abbaslu}, \citenamefont {Rostam~Zadeh}, \citenamefont {Mehraeen},\ and\
  \citenamefont {Gousheh}}]{abbaslu2021generation}%
  \BibitemOpen
  \bibfield  {author} {\bibinfo {author} {\bibfnamefont {S.}~\bibnamefont
  {Abbaslu}}, \bibinfo {author} {\bibfnamefont {S.}~\bibnamefont
  {Rostam~Zadeh}}, \bibinfo {author} {\bibfnamefont {M.}~\bibnamefont
  {Mehraeen}},\ and\ \bibinfo {author} {\bibfnamefont {S.}~\bibnamefont
  {Gousheh}},\ }\bibfield  {title} {\bibinfo {title} {{The generation of
  matter--antimatter asymmetries and hypermagnetic fields by the chiral
  vortical effect of transient fluctuations}},\ }\href
  {https://doi.org/10.1140/epjc/s10052-021-09272-9} {\bibfield  {journal}
  {\bibinfo  {journal} {Eur. Phys. J. C}\ }\textbf {\bibinfo {volume} {81}},\
  \bibinfo {pages} {1} (\bibinfo {year} {2021})}\BibitemShut {NoStop}%
\bibitem [{\citenamefont {Kleiss}\ and\ \citenamefont
  {Stirling}(1985)}]{kleiss1985spinor}%
  \BibitemOpen
  \bibfield  {author} {\bibinfo {author} {\bibfnamefont {R.}~\bibnamefont
  {Kleiss}}\ and\ \bibinfo {author} {\bibfnamefont {W.~J.}\ \bibnamefont
  {Stirling}},\ }\bibfield  {title} {\bibinfo {title} {{Spinor Techniques for
  Calculating $p\bar{p} \to W^{\pm}/Z^0 +$ Jets}},\ }\href
  {https://doi.org/10.1016/0550-3213(85)90285-8} {\bibfield  {journal}
  {\bibinfo  {journal} {Nucl. Phys. B}\ }\textbf {\bibinfo {volume} {262}},\
  \bibinfo {pages} {235} (\bibinfo {year} {1985})}\BibitemShut {NoStop}%
\bibitem [{\citenamefont {Gunion}\ and\ \citenamefont
  {Kunszt}(1985)}]{gunion1985improved}%
  \BibitemOpen
  \bibfield  {author} {\bibinfo {author} {\bibfnamefont {J.~F.}\ \bibnamefont
  {Gunion}}\ and\ \bibinfo {author} {\bibfnamefont {Z.}~\bibnamefont
  {Kunszt}},\ }\bibfield  {title} {\bibinfo {title} {{Improved Analytic
  Techniques for Tree Graph Calculations and the $Ggq\bar{q}$ Lepton Antilepton
  Subprocess}},\ }\href {https://doi.org/10.1016/0370-2693(85)90774-9}
  {\bibfield  {journal} {\bibinfo  {journal} {Phys. Lett. B}\ }\textbf
  {\bibinfo {volume} {161}},\ \bibinfo {pages} {333} (\bibinfo {year}
  {1985})}\BibitemShut {NoStop}%
\bibitem [{\citenamefont {Xu}\ \emph {et~al.}(1987)\citenamefont {Xu},
  \citenamefont {Zhang},\ and\ \citenamefont {Chang}}]{xu1987helicity}%
  \BibitemOpen
  \bibfield  {author} {\bibinfo {author} {\bibfnamefont {Z.}~\bibnamefont
  {Xu}}, \bibinfo {author} {\bibfnamefont {D.-H.}\ \bibnamefont {Zhang}},\ and\
  \bibinfo {author} {\bibfnamefont {L.}~\bibnamefont {Chang}},\ }\bibfield
  {title} {\bibinfo {title} {{Helicity Amplitudes for Multiple Bremsstrahlung
  in Massless Nonabelian Gauge Theories}},\ }\href
  {https://doi.org/10.1016/0550-3213(87)90479-2} {\bibfield  {journal}
  {\bibinfo  {journal} {Nucl. Phys. B}\ }\textbf {\bibinfo {volume} {291}},\
  \bibinfo {pages} {392} (\bibinfo {year} {1987})}\BibitemShut {NoStop}%
\bibitem [{\citenamefont {Peskin}(2011)}]{peskin2011simplifying}%
  \BibitemOpen
  \bibfield  {author} {\bibinfo {author} {\bibfnamefont {M.~E.}\ \bibnamefont
  {Peskin}},\ }\bibfield  {title} {\bibinfo {title} {{Simplifying QCD
  Computations}},\ }\href {https://arxiv.org/abs/1101.2414} {\bibfield
  {journal} {\bibinfo  {journal} {arXiv:1101.2414}\ } (\bibinfo {year}
  {2011})}\BibitemShut {NoStop}%
\bibitem [{\citenamefont {Gupta}(1952)}]{gupta1952quantization}%
  \BibitemOpen
  \bibfield  {author} {\bibinfo {author} {\bibfnamefont {S.~N.}\ \bibnamefont
  {Gupta}},\ }\bibfield  {title} {\bibinfo {title} {{Quantization of Einstein's
  Gravitational Field: General Treatment}},\ }\href
  {https://doi.org/10.1088/0370-1298/65/8/304} {\bibfield  {journal} {\bibinfo
  {journal} {Proc. Phys. Soc. A}\ }\textbf {\bibinfo {volume} {65}},\ \bibinfo
  {pages} {608} (\bibinfo {year} {1952})}\BibitemShut {NoStop}%
\bibitem [{\citenamefont {Weinberg}(1964)}]{weinberg1964photons}%
  \BibitemOpen
  \bibfield  {author} {\bibinfo {author} {\bibfnamefont {S.}~\bibnamefont
  {Weinberg}},\ }\bibfield  {title} {\bibinfo {title} {{Photons and Gravitons
  in S-Matrix Theory: Derivation of Charge Conservation and Equality of
  Gravitational and Inertial Mass}},\ }\href
  {https://doi.org/10.1103/PhysRev.135.B1049} {\bibfield  {journal} {\bibinfo
  {journal} {Phys. Rev.}\ }\textbf {\bibinfo {volume} {135}},\ \bibinfo {pages}
  {B1049} (\bibinfo {year} {1964})}\BibitemShut {NoStop}%
\bibitem [{Note2()}]{Note2}%
  \BibitemOpen
  \bibinfo {note} {For further details, see the Supplemental Material at [url],
  which includes Refs.~\cite {ahn2022riemannian, provost1980riemannian,
  mehraeen2025quantum, hehl1995metric, blauGR, jain2026topological,
  gao2014field, hsu2023nonlinear, mehraeen2025quantum, jain2026nonlinear,
  ito2026spin, bellac2011thermal, gupta1952quantization, weinberg1964photons,
  kleiss1985spinor, gunion1985improved, xu1987helicity, hattori2021wigner,
  fonarev1994wigner, nakahara2018geometry, liu2019chiral, altas2019second,
  mameda2022photonic} and the following sections: S1.~Graviton field operator
  in linearized gravity; S2.~Quantum state geometry; S3.~Graviton Wigner
  functions in flat spacetime; S4.~Phase-space horizontal lift and Wigner
  functions in curved spacetime; S5.~Dressed distribution function;
  S6.~Graviton stress-energy tensor in phase space; S7.~Graviton transport in
  linearized background.}\BibitemShut {Stop}%
\bibitem [{\citenamefont {Bellac}(2011)}]{bellac2011thermal}%
  \BibitemOpen
  \bibfield  {author} {\bibinfo {author} {\bibfnamefont {M.~L.}\ \bibnamefont
  {Bellac}},\ }\href@noop {} {\emph {\bibinfo {title} {{Thermal Field
  Theory}}}},\ Cambridge Monographs on Mathematical Physics\ (\bibinfo
  {publisher} {Cambridge University Press},\ \bibinfo {year}
  {2011})\BibitemShut {NoStop}%
\bibitem [{Note3()}]{Note3}%
  \BibitemOpen
  \bibinfo {note} {Regarding notation and conventions, throughout this work, we
  assume natural units $\hbar =c=k_B=1$, unless to show explicit $\hbar $
  dependence. We work in the metric signature $\protect \text
  {diag}(1,-\protect \boldsymbol {1})$, and the Levi-Civita tensor $\varepsilon
  ^{\mu \nu \rho \sigma }$ is defined as $\varepsilon ^{0123} = (-g)^{-1/2}$.
  Tensor (anti-)symmetrizations are defined as $\protect \mathcal {O}_{(\mu \nu
  )} \equiv (\protect \mathcal {O}_{\mu \nu } + \protect \mathcal {O}_{\nu \mu
  })/2 $ and $\protect \mathcal {O}_{[\mu \nu ]} \equiv (\protect \mathcal
  {O}_{\mu \nu } - \protect \mathcal {O}_{\nu \mu })/2 $.}\BibitemShut {Stop}%
\bibitem [{\citenamefont {Nakahara}(2018)}]{nakahara2018geometry}%
  \BibitemOpen
  \bibfield  {author} {\bibinfo {author} {\bibfnamefont {M.}~\bibnamefont
  {Nakahara}},\ }\href@noop {} {\emph {\bibinfo {title} {Geometry, topology and
  physics}}}\ (\bibinfo  {publisher} {CRC press},\ \bibinfo {year}
  {2018})\BibitemShut {NoStop}%
\bibitem [{\citenamefont {Fonarev}(1994)}]{fonarev1994wigner}%
  \BibitemOpen
  \bibfield  {author} {\bibinfo {author} {\bibfnamefont {O.~A.}\ \bibnamefont
  {Fonarev}},\ }\bibfield  {title} {\bibinfo {title} {{Wigner function and
  quantum kinetic theory in field theory and nuclear physics}},\ }\href
  {https://doi.org/10.1063/1.530542} {\bibfield  {journal} {\bibinfo  {journal}
  {J. Math. Phys.}\ }\textbf {\bibinfo {volume} {35}},\ \bibinfo {pages} {2105}
  (\bibinfo {year} {1994})}\BibitemShut {NoStop}%
\bibitem [{\citenamefont {Liu}\ \emph {et~al.}(2019)\citenamefont {Liu},
  \citenamefont {Gao}, \citenamefont {Mameda},\ and\ \citenamefont
  {Huang}}]{liu2019chiral}%
  \BibitemOpen
  \bibfield  {author} {\bibinfo {author} {\bibfnamefont {Y.-C.}\ \bibnamefont
  {Liu}}, \bibinfo {author} {\bibfnamefont {L.-L.}\ \bibnamefont {Gao}},
  \bibinfo {author} {\bibfnamefont {K.}~\bibnamefont {Mameda}},\ and\ \bibinfo
  {author} {\bibfnamefont {X.-G.}\ \bibnamefont {Huang}},\ }\bibfield  {title}
  {\bibinfo {title} {{Chiral kinetic theory in curved spacetime}},\ }\href
  {https://doi.org/10.1103/PhysRevD.99.085014} {\bibfield  {journal} {\bibinfo
  {journal} {Phys. Rev. D}\ }\textbf {\bibinfo {volume} {99}},\ \bibinfo
  {pages} {085014} (\bibinfo {year} {2019})}\BibitemShut {NoStop}%
\bibitem [{\citenamefont {Sipe}\ and\ \citenamefont
  {Ghahramani}(1993)}]{sipe1993nonlinear}%
  \BibitemOpen
  \bibfield  {author} {\bibinfo {author} {\bibfnamefont {J.~E.}\ \bibnamefont
  {Sipe}}\ and\ \bibinfo {author} {\bibfnamefont {E.}~\bibnamefont
  {Ghahramani}},\ }\bibfield  {title} {\bibinfo {title} {{Nonlinear optical
  response of semiconductors in the independent-particle approximation}},\
  }\href {https://doi.org/10.1103/PhysRevB.48.11705} {\bibfield  {journal}
  {\bibinfo  {journal} {Phys. Rev. B}\ }\textbf {\bibinfo {volume} {48}},\
  \bibinfo {pages} {11705} (\bibinfo {year} {1993})}\BibitemShut {NoStop}%
\bibitem [{\citenamefont {Aversa}\ and\ \citenamefont
  {Sipe}(1995)}]{aversa1995nonlinear}%
  \BibitemOpen
  \bibfield  {author} {\bibinfo {author} {\bibfnamefont {C.}~\bibnamefont
  {Aversa}}\ and\ \bibinfo {author} {\bibfnamefont {J.~E.}\ \bibnamefont
  {Sipe}},\ }\bibfield  {title} {\bibinfo {title} {{Nonlinear optical
  susceptibilities of semiconductors: Results with a length-gauge analysis}},\
  }\href {https://doi.org/10.1103/PhysRevB.52.14636} {\bibfield  {journal}
  {\bibinfo  {journal} {Phys. Rev. B}\ }\textbf {\bibinfo {volume} {52}},\
  \bibinfo {pages} {14636} (\bibinfo {year} {1995})}\BibitemShut {NoStop}%
\bibitem [{\citenamefont {Sipe}\ and\ \citenamefont
  {Shkrebtii}(2000)}]{sipe2000second}%
  \BibitemOpen
  \bibfield  {author} {\bibinfo {author} {\bibfnamefont {J.~E.}\ \bibnamefont
  {Sipe}}\ and\ \bibinfo {author} {\bibfnamefont {A.~I.}\ \bibnamefont
  {Shkrebtii}},\ }\bibfield  {title} {\bibinfo {title} {{Second-order optical
  response in semiconductors}},\ }\href
  {https://doi.org/10.1103/PhysRevB.61.5337} {\bibfield  {journal} {\bibinfo
  {journal} {Phys. Rev. B}\ }\textbf {\bibinfo {volume} {61}},\ \bibinfo
  {pages} {5337} (\bibinfo {year} {2000})}\BibitemShut {NoStop}%
\bibitem [{\citenamefont {Collett}\ and\ \citenamefont
  {Schaefer}(2008)}]{collett2008visualization}%
  \BibitemOpen
  \bibfield  {author} {\bibinfo {author} {\bibfnamefont {E.}~\bibnamefont
  {Collett}}\ and\ \bibinfo {author} {\bibfnamefont {B.}~\bibnamefont
  {Schaefer}},\ }\bibfield  {title} {\bibinfo {title} {{Visualization and
  calculation of polarized light. I. The polarization ellipse, the Poincaré
  sphere and the hybrid polarization sphere}},\ }\href
  {https://doi.org/10.1364/AO.47.004009} {\bibfield  {journal} {\bibinfo
  {journal} {Appl. Opt.}\ }\textbf {\bibinfo {volume} {47}},\ \bibinfo {pages}
  {4009} (\bibinfo {year} {2008})}\BibitemShut {NoStop}%
\bibitem [{\citenamefont {Hattori}\ \emph {et~al.}(2021)\citenamefont
  {Hattori}, \citenamefont {Hidaka}, \citenamefont {Yamamoto},\ and\
  \citenamefont {Yang}}]{hattori2021wigner}%
  \BibitemOpen
  \bibfield  {author} {\bibinfo {author} {\bibfnamefont {K.}~\bibnamefont
  {Hattori}}, \bibinfo {author} {\bibfnamefont {Y.}~\bibnamefont {Hidaka}},
  \bibinfo {author} {\bibfnamefont {N.}~\bibnamefont {Yamamoto}},\ and\
  \bibinfo {author} {\bibfnamefont {D.-L.}\ \bibnamefont {Yang}},\ }\bibfield
  {title} {\bibinfo {title} {{Wigner functions and quantum kinetic theory of
  polarized photons}},\ }\href {https://doi.org/10.1007/JHEP02(2021)001}
  {\bibfield  {journal} {\bibinfo  {journal} {J. High Energy Phys.}\ }\textbf
  {\bibinfo {volume} {2021}}\bibinfo  {number} { (2)},\ \bibinfo {pages}
  {1}}\BibitemShut {NoStop}%
\bibitem [{\citenamefont {Altas}\ and\ \citenamefont
  {Tekin}(2019)}]{altas2019second}%
  \BibitemOpen
\bibfield  {number} {  }\bibfield  {author} {\bibinfo {author} {\bibfnamefont
  {E.}~\bibnamefont {Altas}}\ and\ \bibinfo {author} {\bibfnamefont
  {B.}~\bibnamefont {Tekin}},\ }\bibfield  {title} {\bibinfo {title} {{Second
  order perturbation theory in general relativity: Taub charges as integral
  constraints}},\ }\href {https://link.aps.org/doi/10.1103/PhysRevD.99.104078}
  {\bibfield  {journal} {\bibinfo  {journal} {Phys. Rev. D}\ }\textbf {\bibinfo
  {volume} {99}},\ \bibinfo {pages} {104078} (\bibinfo {year}
  {2019})}\BibitemShut {NoStop}%
\bibitem [{\citenamefont {Mameda}\ \emph {et~al.}(2022)\citenamefont {Mameda},
  \citenamefont {Yamamoto},\ and\ \citenamefont {Yang}}]{mameda2022photonic}%
  \BibitemOpen
  \bibfield  {author} {\bibinfo {author} {\bibfnamefont {K.}~\bibnamefont
  {Mameda}}, \bibinfo {author} {\bibfnamefont {N.}~\bibnamefont {Yamamoto}},\
  and\ \bibinfo {author} {\bibfnamefont {D.-L.}\ \bibnamefont {Yang}},\
  }\bibfield  {title} {\bibinfo {title} {{Photonic spin Hall effect from
  quantum kinetic theory in curved spacetime}},\ }\href
  {https://doi.org/10.1103/PhysRevD.105.096019} {\bibfield  {journal} {\bibinfo
   {journal} {Phys. Rev. D}\ }\textbf {\bibinfo {volume} {105}},\ \bibinfo
  {pages} {096019} (\bibinfo {year} {2022})}\BibitemShut {NoStop}%
\bibitem [{\citenamefont {Hayata}\ \emph {et~al.}(2021)\citenamefont {Hayata},
  \citenamefont {Hidaka},\ and\ \citenamefont {Mameda}}]{hayata2021second}%
  \BibitemOpen
  \bibfield  {author} {\bibinfo {author} {\bibfnamefont {T.}~\bibnamefont
  {Hayata}}, \bibinfo {author} {\bibfnamefont {Y.}~\bibnamefont {Hidaka}},\
  and\ \bibinfo {author} {\bibfnamefont {K.}~\bibnamefont {Mameda}},\
  }\bibfield  {title} {\bibinfo {title} {{Second order chiral kinetic theory
  under gravity and antiparallel charge-energy flow}},\ }\href
  {https://doi.org/10.1007/JHEP05(2021)023} {\bibfield  {journal} {\bibinfo
  {journal} {J. High Energy Phys.}\ }\textbf {\bibinfo {volume} {2021}}\bibinfo
   {number} { (5)},\ \bibinfo {pages} {23}}\BibitemShut {NoStop}%
\bibitem [{\citenamefont {Stern}\ and\ \citenamefont {von
  Oppen}(2026)}]{stern2026how}%
  \BibitemOpen
\bibfield  {number} {  }\bibfield  {author} {\bibinfo {author} {\bibfnamefont
  {A.}~\bibnamefont {Stern}}\ and\ \bibinfo {author} {\bibfnamefont
  {F.}~\bibnamefont {von Oppen}},\ }\bibfield  {title} {\bibinfo {title} {{How
  quantum is quantum geometry?}},\ }\href {https://arxiv.org/abs/2608.26269}
  {\bibfield  {journal} {\bibinfo  {journal} {arXiv:2608.26269}\ } (\bibinfo
  {year} {2026})}\BibitemShut {NoStop}%
\bibitem [{\citenamefont {Atencia}\ \emph {et~al.}(2022)\citenamefont
  {Atencia}, \citenamefont {Niu},\ and\ \citenamefont
  {Culcer}}]{atencia2022semiclassical}%
  \BibitemOpen
  \bibfield  {author} {\bibinfo {author} {\bibfnamefont {R.~B.}\ \bibnamefont
  {Atencia}}, \bibinfo {author} {\bibfnamefont {Q.}~\bibnamefont {Niu}},\ and\
  \bibinfo {author} {\bibfnamefont {D.}~\bibnamefont {Culcer}},\ }\bibfield
  {title} {\bibinfo {title} {Semiclassical response of disordered conductors:
  Extrinsic carrier velocity and spin and field-corrected collision integral},\
  }\href {https://link.aps.org/doi/10.1103/PhysRevResearch.4.013001} {\bibfield
   {journal} {\bibinfo  {journal} {Phys. Rev. Res.}\ }\textbf {\bibinfo
  {volume} {4}},\ \bibinfo {pages} {013001} (\bibinfo {year}
  {2022})}\BibitemShut {NoStop}%
\bibitem [{\citenamefont {Atencia}\ \emph {et~al.}(2023)\citenamefont
  {Atencia}, \citenamefont {Xiao},\ and\ \citenamefont
  {Culcer}}]{atencia2023disorder}%
  \BibitemOpen
  \bibfield  {author} {\bibinfo {author} {\bibfnamefont {R.~B.}\ \bibnamefont
  {Atencia}}, \bibinfo {author} {\bibfnamefont {D.}~\bibnamefont {Xiao}},\ and\
  \bibinfo {author} {\bibfnamefont {D.}~\bibnamefont {Culcer}},\ }\bibfield
  {title} {\bibinfo {title} {{Disorder in the nonlinear anomalous Hall effect
  of $\mathcal{PT}$-symmetric Dirac fermions}},\ }\href
  {https://link.aps.org/doi/10.1103/PhysRevB.108.L201115} {\bibfield  {journal}
  {\bibinfo  {journal} {Phys. Rev. B}\ }\textbf {\bibinfo {volume} {108}},\
  \bibinfo {pages} {L201115} (\bibinfo {year} {2023})}\BibitemShut {NoStop}%
\bibitem [{\citenamefont {Mehraeen}(2024)}]{mehraeen2024quantum}%
  \BibitemOpen
  \bibfield  {author} {\bibinfo {author} {\bibfnamefont {M.}~\bibnamefont
  {Mehraeen}},\ }\bibfield  {title} {\bibinfo {title} {Quantum kinetic theory
  of quadratic responses},\ }\href
  {https://link.aps.org/doi/10.1103/PhysRevB.110.174423} {\bibfield  {journal}
  {\bibinfo  {journal} {Phys. Rev. B}\ }\textbf {\bibinfo {volume} {110}},\
  \bibinfo {pages} {174423} (\bibinfo {year} {2024})}\BibitemShut {NoStop}%
\bibitem [{\citenamefont {Huang}\ \emph {et~al.}(2025)\citenamefont {Huang},
  \citenamefont {Xiao}, \citenamefont {Yang},\ and\ \citenamefont
  {Li}}]{huang2025scaling}%
  \BibitemOpen
  \bibfield  {author} {\bibinfo {author} {\bibfnamefont {Y.-X.}\ \bibnamefont
  {Huang}}, \bibinfo {author} {\bibfnamefont {C.}~\bibnamefont {Xiao}},
  \bibinfo {author} {\bibfnamefont {S.~A.}\ \bibnamefont {Yang}},\ and\
  \bibinfo {author} {\bibfnamefont {X.}~\bibnamefont {Li}},\ }\bibfield
  {title} {\bibinfo {title} {Scaling law and extrinsic mechanisms for
  time-reversal-odd second-order nonlinear transport},\ }\href
  {https://link.aps.org/doi/10.1103/PhysRevB.111.155127} {\bibfield  {journal}
  {\bibinfo  {journal} {Phys. Rev. B}\ }\textbf {\bibinfo {volume} {111}},\
  \bibinfo {pages} {155127} (\bibinfo {year} {2025})}\BibitemShut {NoStop}%
\bibitem [{\citenamefont {Gong}\ \emph {et~al.}(2026)\citenamefont {Gong},
  \citenamefont {Du}, \citenamefont {Sun}, \citenamefont {Lu},\ and\
  \citenamefont {Xie}}]{gong2026scaling}%
  \BibitemOpen
  \bibfield  {author} {\bibinfo {author} {\bibfnamefont {Z.-H.}\ \bibnamefont
  {Gong}}, \bibinfo {author} {\bibfnamefont {Z.-Z.}\ \bibnamefont {Du}},
  \bibinfo {author} {\bibfnamefont {H.-P.}\ \bibnamefont {Sun}}, \bibinfo
  {author} {\bibfnamefont {H.-Z.}\ \bibnamefont {Lu}},\ and\ \bibinfo {author}
  {\bibfnamefont {X.-C.}\ \bibnamefont {Xie}},\ }\bibfield  {title} {\bibinfo
  {title} {{Scaling analysis of quantum geometry in second-order nonlinear
  transport}},\ }\href {https://doi.org/10.1016/j.scib.2026.07.062} {\bibfield
  {journal} {\bibinfo  {journal} {Sci. Bull.}\ } (\bibinfo {year}
  {2026})}\BibitemShut {NoStop}%
\bibitem [{\citenamefont {Ji}\ \emph {et~al.}(2025)\citenamefont {Ji},
  \citenamefont {Palomino}, \citenamefont {Goldman}, \citenamefont {Ozawa},
  \citenamefont {Riseborough}, \citenamefont {Wang},\ and\ \citenamefont
  {Mera}}]{ji2025density}%
  \BibitemOpen
  \bibfield  {author} {\bibinfo {author} {\bibfnamefont {G.}~\bibnamefont
  {Ji}}, \bibinfo {author} {\bibfnamefont {D.~E.}\ \bibnamefont {Palomino}},
  \bibinfo {author} {\bibfnamefont {N.}~\bibnamefont {Goldman}}, \bibinfo
  {author} {\bibfnamefont {T.}~\bibnamefont {Ozawa}}, \bibinfo {author}
  {\bibfnamefont {P.}~\bibnamefont {Riseborough}}, \bibinfo {author}
  {\bibfnamefont {J.}~\bibnamefont {Wang}},\ and\ \bibinfo {author}
  {\bibfnamefont {B.}~\bibnamefont {Mera}},\ }\bibfield  {title} {\bibinfo
  {title} {{Density Matrix Geometry and Sum Rules}},\ }\href
  {https://arxiv.org/abs/2507.14028} {\bibfield  {journal} {\bibinfo  {journal}
  {arXiv:2507.14028}\ } (\bibinfo {year} {2025})}\BibitemShut {NoStop}%
\bibitem [{\citenamefont {Guan}\ and\ \citenamefont
  {Bradlyn}(2026)}]{guan2026exploring}%
  \BibitemOpen
  \bibfield  {author} {\bibinfo {author} {\bibfnamefont {Y.}~\bibnamefont
  {Guan}}\ and\ \bibinfo {author} {\bibfnamefont {B.}~\bibnamefont {Bradlyn}},\
  }\bibfield  {title} {\bibinfo {title} {{Exploring many-body quantum geometry
  beyond the quantum metric with correlation functions: A time-dependent
  perspective}},\ }\href {https://doi.org/10.1103/3xjs-c7v7} {\bibfield
  {journal} {\bibinfo  {journal} {Phys. Rev. Research}\ }\textbf {\bibinfo
  {volume} {8}},\ \bibinfo {pages} {013291} (\bibinfo {year}
  {2026})}\BibitemShut {NoStop}%
\bibitem [{\citenamefont {Bradlyn}(2026)}]{bradlyn2026multistate}%
  \BibitemOpen
  \bibfield  {author} {\bibinfo {author} {\bibfnamefont {B.}~\bibnamefont
  {Bradlyn}},\ }\bibfield  {title} {\bibinfo {title} {{Multi-State Geometry of
  Density Matrices and Rectification Sum Rules}},\ }\href
  {https://arxiv.org/abs/2608.06326} {\bibfield  {journal} {\bibinfo  {journal}
  {arXiv:2608.06326}\ } (\bibinfo {year} {2026})}\BibitemShut {NoStop}%
\bibitem [{\citenamefont {Bradlyn}\ \emph {et~al.}(2012)\citenamefont
  {Bradlyn}, \citenamefont {Goldstein},\ and\ \citenamefont
  {Read}}]{bradlyn2012kubo}%
  \BibitemOpen
  \bibfield  {author} {\bibinfo {author} {\bibfnamefont {B.}~\bibnamefont
  {Bradlyn}}, \bibinfo {author} {\bibfnamefont {M.}~\bibnamefont {Goldstein}},\
  and\ \bibinfo {author} {\bibfnamefont {N.}~\bibnamefont {Read}},\ }\bibfield
  {title} {\bibinfo {title} {{Kubo formulas for viscosity: Hall viscosity, Ward
  identities, and the relation with conductivity}},\ }\href
  {https://doi.org/10.1103/PhysRevB.86.245309} {\bibfield  {journal} {\bibinfo
  {journal} {Phys. Rev. B}\ }\textbf {\bibinfo {volume} {86}},\ \bibinfo
  {pages} {245309} (\bibinfo {year} {2012})}\BibitemShut {NoStop}%
\bibitem [{\citenamefont {Shapourian}\ \emph {et~al.}(2015)\citenamefont
  {Shapourian}, \citenamefont {Hughes},\ and\ \citenamefont
  {Ryu}}]{shapourian2015viscoelastic}%
  \BibitemOpen
  \bibfield  {author} {\bibinfo {author} {\bibfnamefont {H.}~\bibnamefont
  {Shapourian}}, \bibinfo {author} {\bibfnamefont {T.~L.}\ \bibnamefont
  {Hughes}},\ and\ \bibinfo {author} {\bibfnamefont {S.}~\bibnamefont {Ryu}},\
  }\bibfield  {title} {\bibinfo {title} {{Viscoelastic response of topological
  tight-binding models in two and three dimensions}},\ }\href
  {https://doi.org/10.1103/PhysRevB.92.165131} {\bibfield  {journal} {\bibinfo
  {journal} {Phys. Rev. B}\ }\textbf {\bibinfo {volume} {92}},\ \bibinfo
  {pages} {165131} (\bibinfo {year} {2015})}\BibitemShut {NoStop}%
\bibitem [{\citenamefont {Osborne}\ \emph {et~al.}(2026)\citenamefont
  {Osborne}, \citenamefont {Monteiro},\ and\ \citenamefont
  {Bradlyn}}]{osborne2026geometry}%
  \BibitemOpen
  \bibfield  {author} {\bibinfo {author} {\bibfnamefont {I.}~\bibnamefont
  {Osborne}}, \bibinfo {author} {\bibfnamefont {G.}~\bibnamefont {Monteiro}},\
  and\ \bibinfo {author} {\bibfnamefont {B.}~\bibnamefont {Bradlyn}},\
  }\bibfield  {title} {\bibinfo {title} {{Geometry of Contact Terms in Linear
  Response: Applications to Elasticity}},\ }\href
  {https://arxiv.org/abs/2603.10144} {\bibfield  {journal} {\bibinfo  {journal}
  {arXiv:2603.10144}\ } (\bibinfo {year} {2026})}\BibitemShut {NoStop}%
\bibitem [{\citenamefont {Liu}\ \emph {et~al.}(2025)\citenamefont {Liu},
  \citenamefont {Zhang}, \citenamefont {Dong}, \citenamefont {Mason},
  \citenamefont {van~der Zande},\ and\ \citenamefont
  {Johnson}}]{liu2025programmable}%
  \BibitemOpen
  \bibfield  {author} {\bibinfo {author} {\bibfnamefont {Q.}~\bibnamefont
  {Liu}}, \bibinfo {author} {\bibfnamefont {Y.}~\bibnamefont {Zhang}}, \bibinfo
  {author} {\bibfnamefont {H.}~\bibnamefont {Dong}}, \bibinfo {author}
  {\bibfnamefont {N.}~\bibnamefont {Mason}}, \bibinfo {author} {\bibfnamefont
  {A.~M.}\ \bibnamefont {van~der Zande}},\ and\ \bibinfo {author}
  {\bibfnamefont {H.~T.}\ \bibnamefont {Johnson}},\ }\bibfield  {title}
  {\bibinfo {title} {{Programmable Strainscapes in a Two-Dimensional (2D)
  Material Monolayer}},\ }\href {https://doi.org/10.1021/acsnano.5c06381}
  {\bibfield  {journal} {\bibinfo  {journal} {ACS Nano}\ }\textbf {\bibinfo
  {volume} {19}},\ \bibinfo {pages} {30125} (\bibinfo {year}
  {2025})}\BibitemShut {NoStop}%
\bibitem [{\citenamefont {Bukharaev}\ \emph {et~al.}(2018)\citenamefont
  {Bukharaev}, \citenamefont {Zvezdin}, \citenamefont {Pyatakov},\ and\
  \citenamefont {Fetisov}}]{bukharaev2018straintronics}%
  \BibitemOpen
  \bibfield  {author} {\bibinfo {author} {\bibfnamefont {A.~A.}\ \bibnamefont
  {Bukharaev}}, \bibinfo {author} {\bibfnamefont {A.~K.}\ \bibnamefont
  {Zvezdin}}, \bibinfo {author} {\bibfnamefont {A.~P.}\ \bibnamefont
  {Pyatakov}},\ and\ \bibinfo {author} {\bibfnamefont {Y.~K.}\ \bibnamefont
  {Fetisov}},\ }\bibfield  {title} {\bibinfo {title} {{Straintronics: a new
  trend in micro- and nanoelectronics and materials science}},\ }\href
  {https://doi.org/10.3367/UFNe.2018.01.038279} {\bibfield  {journal} {\bibinfo
   {journal} {Phys. Usp.}\ }\textbf {\bibinfo {volume} {61}},\ \bibinfo {pages}
  {1175} (\bibinfo {year} {2018})}\BibitemShut {NoStop}%
\bibitem [{\citenamefont {Berdyugin}\ \emph {et~al.}(2019)\citenamefont
  {Berdyugin}, \citenamefont {Xu}, \citenamefont {Pellegrino}, \citenamefont
  {Krishna~Kumar}, \citenamefont {Principi}, \citenamefont {Torre},
  \citenamefont {Ben~Shalom}, \citenamefont {Taniguchi}, \citenamefont
  {Watanabe}, \citenamefont {Grigorieva}, \citenamefont {Polini}, \citenamefont
  {Geim},\ and\ \citenamefont {Bandurin}}]{berdyugin2019measuring}%
  \BibitemOpen
  \bibfield  {author} {\bibinfo {author} {\bibfnamefont {A.~I.}\ \bibnamefont
  {Berdyugin}}, \bibinfo {author} {\bibfnamefont {S.~G.}\ \bibnamefont {Xu}},
  \bibinfo {author} {\bibfnamefont {F.~M.~D.}\ \bibnamefont {Pellegrino}},
  \bibinfo {author} {\bibfnamefont {R.}~\bibnamefont {Krishna~Kumar}}, \bibinfo
  {author} {\bibfnamefont {A.}~\bibnamefont {Principi}}, \bibinfo {author}
  {\bibfnamefont {I.}~\bibnamefont {Torre}}, \bibinfo {author} {\bibfnamefont
  {M.}~\bibnamefont {Ben~Shalom}}, \bibinfo {author} {\bibfnamefont
  {T.}~\bibnamefont {Taniguchi}}, \bibinfo {author} {\bibfnamefont
  {K.}~\bibnamefont {Watanabe}}, \bibinfo {author} {\bibfnamefont {I.~V.}\
  \bibnamefont {Grigorieva}}, \bibinfo {author} {\bibfnamefont
  {M.}~\bibnamefont {Polini}}, \bibinfo {author} {\bibfnamefont {A.~K.}\
  \bibnamefont {Geim}},\ and\ \bibinfo {author} {\bibfnamefont {D.~A.}\
  \bibnamefont {Bandurin}},\ }\bibfield  {title} {\bibinfo {title} {{Measuring
  Hall viscosity of graphene's electron fluid}},\ }\href
  {https://doi.org/10.1126/science.aau0685} {\bibfield  {journal} {\bibinfo
  {journal} {Science}\ }\textbf {\bibinfo {volume} {364}},\ \bibinfo {pages}
  {162} (\bibinfo {year} {2019})}\BibitemShut {NoStop}%
\bibitem [{\citenamefont {Aharon-Steinberg}\ \emph {et~al.}(2022)\citenamefont
  {Aharon-Steinberg}, \citenamefont {Völkl}, \citenamefont {Kaplan},
  \citenamefont {Pariari}, \citenamefont {Roy}, \citenamefont {Holder},
  \citenamefont {Wolf}, \citenamefont {Meltzer}, \citenamefont {Myasoedov},
  \citenamefont {Huber}, \citenamefont {Yan}, \citenamefont {Falkovich},
  \citenamefont {Levitov}, \citenamefont {Hücker},\ and\ \citenamefont
  {Zeldov}}]{aharon-steinberg2022direct}%
  \BibitemOpen
  \bibfield  {author} {\bibinfo {author} {\bibfnamefont {A.}~\bibnamefont
  {Aharon-Steinberg}}, \bibinfo {author} {\bibfnamefont {T.}~\bibnamefont
  {Völkl}}, \bibinfo {author} {\bibfnamefont {A.}~\bibnamefont {Kaplan}},
  \bibinfo {author} {\bibfnamefont {A.~K.}\ \bibnamefont {Pariari}}, \bibinfo
  {author} {\bibfnamefont {I.}~\bibnamefont {Roy}}, \bibinfo {author}
  {\bibfnamefont {T.}~\bibnamefont {Holder}}, \bibinfo {author} {\bibfnamefont
  {Y.}~\bibnamefont {Wolf}}, \bibinfo {author} {\bibfnamefont {A.~Y.}\
  \bibnamefont {Meltzer}}, \bibinfo {author} {\bibfnamefont {Y.}~\bibnamefont
  {Myasoedov}}, \bibinfo {author} {\bibfnamefont {M.~E.}\ \bibnamefont
  {Huber}}, \bibinfo {author} {\bibfnamefont {B.}~\bibnamefont {Yan}}, \bibinfo
  {author} {\bibfnamefont {G.}~\bibnamefont {Falkovich}}, \bibinfo {author}
  {\bibfnamefont {L.~S.}\ \bibnamefont {Levitov}}, \bibinfo {author}
  {\bibfnamefont {M.}~\bibnamefont {Hücker}},\ and\ \bibinfo {author}
  {\bibfnamefont {E.}~\bibnamefont {Zeldov}},\ }\bibfield  {title} {\bibinfo
  {title} {{Direct observation of vortices in an electron fluid}},\ }\href
  {https://doi.org/10.1038/s41586-022-04794-y} {\bibfield  {journal} {\bibinfo
  {journal} {Nature}\ }\textbf {\bibinfo {volume} {607}},\ \bibinfo {pages}
  {74} (\bibinfo {year} {2022})}\BibitemShut {NoStop}%
\bibitem [{\citenamefont {Mitscherling}\ \emph {et~al.}(2026)\citenamefont
  {Mitscherling}, \citenamefont {Borgnia}, \citenamefont {Ahuja}, \citenamefont
  {Moore},\ and\ \citenamefont {Bulchandani}}]{mitscherling2026orbital}%
  \BibitemOpen
  \bibfield  {author} {\bibinfo {author} {\bibfnamefont {J.}~\bibnamefont
  {Mitscherling}}, \bibinfo {author} {\bibfnamefont {D.~S.}\ \bibnamefont
  {Borgnia}}, \bibinfo {author} {\bibfnamefont {S.}~\bibnamefont {Ahuja}},
  \bibinfo {author} {\bibfnamefont {J.~E.}\ \bibnamefont {Moore}},\ and\
  \bibinfo {author} {\bibfnamefont {V.~B.}\ \bibnamefont {Bulchandani}},\
  }\bibfield  {title} {\bibinfo {title} {{Orbital Wigner functions and quantum
  transport in multiband systems}},\ }\href {https://doi.org/10.1103/tdsv-whd9}
  {\bibfield  {journal} {\bibinfo  {journal} {Phys. Rev. B}\ }\textbf {\bibinfo
  {volume} {114}},\ \bibinfo {pages} {175111} (\bibinfo {year}
  {2026})}\BibitemShut {NoStop}%
\bibitem [{\citenamefont {Arkani-Hamed}\ \emph {et~al.}(2021)\citenamefont
  {Arkani-Hamed}, \citenamefont {Huang},\ and\ \citenamefont
  {Huang}}]{arkani-hamed2021scattering}%
  \BibitemOpen
  \bibfield  {author} {\bibinfo {author} {\bibfnamefont {N.}~\bibnamefont
  {Arkani-Hamed}}, \bibinfo {author} {\bibfnamefont {T.-C.}\ \bibnamefont
  {Huang}},\ and\ \bibinfo {author} {\bibfnamefont {Y.-t.}\ \bibnamefont
  {Huang}},\ }\bibfield  {title} {\bibinfo {title} {{Scattering amplitudes for
  all masses and spins}},\ }\href {https://doi.org/10.1007/JHEP11(2021)070}
  {\bibfield  {journal} {\bibinfo  {journal} {J. High Energy Phys.}\ }\textbf
  {\bibinfo {volume} {2021}}\bibinfo  {number} { (11)},\ \bibinfo {pages}
  {070}}\BibitemShut {NoStop}%
\bibitem [{\citenamefont {Oancea}\ \emph {et~al.}(2026)\citenamefont {Oancea},
  \citenamefont {Mieling},\ and\ \citenamefont {Palumbo}}]{oancea2026quantum}%
  \BibitemOpen
\bibfield  {number} {  }\bibfield  {author} {\bibinfo {author} {\bibfnamefont
  {M.~A.}\ \bibnamefont {Oancea}}, \bibinfo {author} {\bibfnamefont {T.~B.}\
  \bibnamefont {Mieling}},\ and\ \bibinfo {author} {\bibfnamefont
  {G.}~\bibnamefont {Palumbo}},\ }\bibfield  {title} {\bibinfo {title}
  {{Quantum geometric tensors from sub-bundle geometry}},\ }\href
  {https://doi.org/10.22331/q-2026-01-14-1965} {\bibfield  {journal} {\bibinfo
  {journal} {Quantum}\ }\textbf {\bibinfo {volume} {10}},\ \bibinfo {pages}
  {1965} (\bibinfo {year} {2026})}\BibitemShut {NoStop}%
\bibitem [{\citenamefont {Akiba}\ and\ \citenamefont
  {Yamamoto}(2026)}]{akiba2026quantum}%
  \BibitemOpen
  \bibfield  {author} {\bibinfo {author} {\bibfnamefont {K.}~\bibnamefont
  {Akiba}}\ and\ \bibinfo {author} {\bibfnamefont {N.}~\bibnamefont
  {Yamamoto}},\ }\bibfield  {title} {\bibinfo {title} {{Quantum Metric and
  Nonlinear Hall Effect of Photons}},\ }\href
  {https://arxiv.org/abs/2604.27751} {\bibfield  {journal} {\bibinfo  {journal}
  {arXiv:2604.27751}\ } (\bibinfo {year} {2026})}\BibitemShut {NoStop}%
\bibitem [{\citenamefont {Hehl}\ \emph {et~al.}(1995)\citenamefont {Hehl},
  \citenamefont {McCrea}, \citenamefont {Mielke},\ and\ \citenamefont
  {Ne'eman}}]{hehl1995metric}%
  \BibitemOpen
  \bibfield  {author} {\bibinfo {author} {\bibfnamefont {F.~W.}\ \bibnamefont
  {Hehl}}, \bibinfo {author} {\bibfnamefont {J.~D.}\ \bibnamefont {McCrea}},
  \bibinfo {author} {\bibfnamefont {E.~W.}\ \bibnamefont {Mielke}},\ and\
  \bibinfo {author} {\bibfnamefont {Y.}~\bibnamefont {Ne'eman}},\ }\bibfield
  {title} {\bibinfo {title} {{Metric-affine gauge theory of gravity: field
  equations, Noether identities, world spinors, and breaking of dilation
  invariance}},\ }\href {https://doi.org/10.1016/0370-1573(94)00111-F}
  {\bibfield  {journal} {\bibinfo  {journal} {Phys. Rep.}\ }\textbf {\bibinfo
  {volume} {258}},\ \bibinfo {pages} {1} (\bibinfo {year} {1995})}\BibitemShut
  {NoStop}%
\bibitem [{\citenamefont {Blau}()}]{blauGR}%
  \BibitemOpen
  \bibfield  {author} {\bibinfo {author} {\bibfnamefont {M.}~\bibnamefont
  {Blau}},\ }\href@noop {} {\bibinfo {title} {{Lecture Notes on General
  Relativity}}},\ \bibinfo {howpublished}
  {\url{http://blau.itp.unibe.ch/GRLecturenotes.html}}\BibitemShut {NoStop}%
\end{thebibliography}%

\clearpage
\onecolumngrid
\setcounter{equation}{0}
\setcounter{figure}{0}
\setcounter{table}{0}
\setcounter{page}{1}
\makeatletter
\renewcommand*{\thesection}{S\arabic{section}}
\renewcommand{\theequation}{S.\arabic{equation}}
\renewcommand{\thefigure}{S.\arabic{figure}}
\renewcommand*{\theHequation}{S.\arabic{equation}}
\renewcommand*{\theHfigure}{S.\arabic{figure}}
\renewcommand*{\theHtable}{S.\arabic{table}}
\renewcommand*{\theHsection}{S.\arabic{section}}
\makeatother

\begin{center}
\textbf{\large 
\medskip
Supplemental Material for ``Quantum geometry of gravitons"}
\end{center}

\tableofcontents

\section{S1.~Graviton field operator in linearized gravity}

Consider the Einstein-Hilbert action
\begin{equation}
S
=
\frac{1}{16 \pi G} \int d^4x \sqrt{-g} R,
\end{equation}
where $G$ is Newton's contant and $g = \det g_{\mu\nu}$.
Here, 
$R
=
g_{\mu\nu}R^{\mu\nu}$
is the Ricci scalar with
$R_{\mu\nu}
=
R^{\rho}_{\phantom{\rho}\mu\rho\nu}
$ the Ricci tensor,
$R^{\mu}_{\phantom{\mu}\nu\rho\sigma}
=
\partial_{[\sigma}\Gamma^\mu_{\rho]\nu} + \Gamma^\mu_{\lambda[\sigma}\Gamma^\lambda_{\rho]\nu}$
the Riemann tensor, and the spacetime Christoffel symbol given by
$\Gamma^\rho_{\mu\nu} 
=
\frac{1}{2} 
g^{\rho\lambda}(\partial_\mu g_{\nu\lambda} 
+
\partial_\nu g_{\mu\lambda}
-
\partial_\lambda g_{\mu\nu})$. We first consider small perturbations around the Minkowski background,
$g_{\mu\nu}
=
\eta_{\mu\nu}
+
h_{\mu\nu}$, with 
$\eta_{\mu\nu}
=
\text{diag}(1, -\mathbf{1})$, and work within the transverse-traceless (TT) gauge
\begin{equation}
\label{TT_gauge_h}
\partial^\mu h_{\mu\nu} = 0, \quad n^\mu h_{\mu\nu} = 0, \quad \eta^{\mu\nu} h_{\mu\nu} = 0.
\end{equation}
The action reduces to
\begin{equation}
S
=
\frac{1}{64\pi G} 
\int d^4x 
\partial_\rho h_{\mu\nu} \partial^\rho h^{\mu\nu}.
\end{equation}
Imposing canonical quantization conditions and noting the equation of motion 
$\pd^2 h_{\mu\nu}
=
0$, the plane-wave expansion of the metric perturbation  takes the form
\begin{equation}
\label{Seq_field_exp}
h^h_{\mu\nu}(x)
=
\int \frac{\text{d}^3 \mathbf{p}}{(2\pi)^3} 
\sqrt{\frac{16\pi G}{|\mathbf{p}|}}
\left[
a^h_{\mathbf{p}} 
e^h_{\mu\nu}(p)\text{e}^{-\text{i}p\cdot x}
+
a^{h\dagger}_{\mathbf{p}}
e^{h*}_{\mu\nu}(p)\text{e}^{\text{i}p\cdot x}
\right],
\end{equation}
where $e_{\mu\nu}^h$ is the polarization tensor of the quantized gravitational wave, with $h= R,L$ denoting the helicity of the graviton state, and graviton creation and annihilation operators satisfying the canonical commutation relations 
\begin{equation}
\left[ a_{\boldsymbol{p}}^h, a_{\boldsymbol{p}'}^{h'\dagger} \right]
=
(2\pi)^3 \delta^{hh'}
\delta^{(3)}(\boldsymbol{p} - \boldsymbol{p}').
\end{equation}

\section{S2.~Quantum state geometry}

Here, we detail the quantum geometric structure underlying the Wigner function expansion that follows, which is induced by the momentum dependence of the basis states
\begin{equation}
\partial^{q}_{\alpha} c^{a}_{h}
=
-i \sum_{b} \mathcal{A}^{ba}_{\alpha h} c^{b}_{h}.
\end{equation}
Here, 
$\mathcal{A}^{ba}_{\alpha h}
=
i c^{b\dagger}_{h} \partial^{q}_{\alpha} c^{a}_{h}$
is the Berry connection, with $a,b$ labeling basis states and helicity labels retained. We further decompose the Berry connection into diagonal and off-diagonal elements as
$\mathcal{A}^{ba}_{h} = a^{a}_{h} \delta^{ba} + \mathcal{A}^{\prime ba}_{h}$. 
The off-diagonal Berry connections can be regarded as elements of complex-valued veilbeins in the basis of states~\cite{ahn2022riemannian} 
\begin{equation}
\hat{\theta}^{ab}_{\mu h}
=
\mathcal{A}^{\prime ab}_{\mu h} c^{a}_{h} c^{b\dagger}_{h},
\end{equation}
with the Hilbert-Schmidt inner product $\left(A, B\right) =
\mathrm{Tr}\left(A^{\dagger} B\right)$ of the tangent basis vectors and their derivatives inducing gauge-invariant geometric structures on the quantum-state manifold.

The simplest invariant is the two-state quantum
geometric tensor, obtained from the inner product of tangent basis vectors
\begin{equation}
\mathcal{Q}^{ab}_{\mu\nu,h}
\equiv
\left(\hat{\theta}^{ab}_{\nu h}, \hat{\theta}^{ab}_{\mu h}\right)
=
\mathcal{A}^{\prime ab}_{\mu h} 
\mathcal{A}^{\prime ba}_{\nu h}
=
\mathcal{G}^{ab}_{\mu\nu,h} 
-
\frac{i}{2} \Omega^{ab}_{\mu\nu,h},
\end{equation}
the real and imaginary parts of which define the two-state quantum metric
and Berry curvature tensors,
\begin{equation}
\mathcal{G}^{ab}_{\mu\nu,h}
=
\mathcal{A}^{\prime ab}_{(\mu h} \mathcal{A}^{\prime ba}_{\nu h)},
\qquad
\Omega^{ab}_{\mu\nu,h}
=
2i \mathcal{A}^{\prime ab}_{[\mu h} 
\mathcal{A}^{\prime ba}_{\nu h]},
\end{equation}
with the single-state quantum geometric tensor obtained by summing over intermediate states~\cite{provost1980riemannian}. Higher-order geometric quantities are generated by differentiation of
the vielbeins. Specifically, the Hermitian connection components are defined as~\cite{ahn2022riemannian}
\begin{equation}
\mathcal{C}^{ba}_{\mu\nu\rho,h}
\equiv
\left(\hat{\theta}^{ba}_{\mu h},
\partial^{q}_{\nu} \hat{\theta}^{ba}_{\rho h}\right)
= \mathcal{A}^{\prime ab}_{\mu h}
\left(\mathcal{D}^{h}_{\nu} \mathcal{A}^{\prime}_{\rho h}\right)^{ba},
\end{equation}
where
\begin{equation}
\label{Seq_Berry_cov}
\left(\mathcal{D}^{h}_{\alpha} O_{\beta}\right)^{ab}
= \left(\partial^{q}_{\alpha} - i \Delta^{ab}_{\alpha h}\right)
O^{ab}_{\beta},
\end{equation}
is the Berry covariant derivative, with
$\Delta^{ab}_{\alpha h} \equiv a^{a}_{\alpha h} -
a^{b}_{\alpha h}$ the diagonal Berry connection mismatch. For an arbitrary momentum-space rank-2 tensor $\mathcal{O}_{\mu\nu}$, this corresponds to the momentum-space covariant derivative~\cite{mehraeen2025quantum}
\begin{equation}
\nabla_{\rho}^{\mathcal{C}} \mathcal{O}_{\mu\nu}
=
\partial_{\rho}^q \mathcal{O}_{\mu\nu}
-
\mathcal{C}_{\sigma\mu\rho,h}
\tensor{\mathcal{O}}{^{\sigma}_{\nu}}
-
\mathcal{C}_{\sigma\nu\rho,h}
\tensor{\mathcal{O}}{^{\sigma}_{\mu}}.
\end{equation}

Following the standard decomposition of affine connections~\cite{hehl1995metric, blauGR}, the Hermitian connection components can be similarly decomposed as~\cite{jain2026topological}
\begin{equation}
\mathcal{C}^{ba}_{\mu\nu\rho,h}
=
\left(\Gamma_{\mathcal{G}}\right)^{ba}_{\mu\nu\rho,h}
+
\mathcal{K}^{ba}_{\mu\nu\rho,h}
+
\mathcal{L}^{ba}_{\mu\nu\rho,h},
\end{equation}
where 
\begin{equation}
(\Gamma_{\mathcal{G}})_{\mu\nu\rho}^{ba}
=
\frac{1}{2}
\left(
\pd_\rho^q \mathcal{G}_{\mu\nu}^{ba}
+
\pd_\nu^q \mathcal{G}_{\mu\rho}^{ba}
-
\pd_\mu^q \mathcal{G}_{\nu\rho}^{ba}
\right),
\end{equation}
is a component of the Levi-Civita connection~\cite{gao2014field} of the quantum metric tensor,
\begin{equation}
\mathcal{K}_{\mu\nu\rho,h}^{ba}
=
\frac{1}{2} 
(\mathcal{T}_{\mu\nu\rho,h}^{ba}
-
\mathcal{T}_{\nu\mu\rho,h}^{ba}
-
\mathcal{T}_{\rho\mu\nu,h}^{ba}),
\end{equation}
is the quantum contorsion tensor~\cite{hsu2023nonlinear, mehraeen2025quantum}, capturing the quantum torsion tensor defined, acting on a scalar field $\phi$, as~\cite{jain2026topological}
\begin{equation}
\mathcal{T}_{\mu\nu\rho,h} \partial^\mu \phi 
\equiv
[\nabla_\nu^\mathcal{C}, \nabla_\rho^\mathcal{C}]\phi
=
2 \, \mathcal{C}_{\mu[\nu\rho],h} \pd_q^\mu \phi,
\end{equation}
with state indices implied, and
\begin{equation}
\mathcal{L}^{ba}_{\mu\nu\rho,h}
= \frac{1}{2}
\left(\mathcal{N}^{ba}_{\mu\nu\rho,h} 
+
\mathcal{N}^{ba}_{\rho\mu\nu,h}
-
\mathcal{N}^{ba}_{\nu\rho\mu,h}\right),
\end{equation}
is the quantum disformation tensor, composed of the quantum nonmetricity tensor~\cite{jain2026nonlinear, jain2026topological}
\begin{equation}
\mathcal{N}^{ba}_{\mu\nu\rho,h} = -\nabla^{\mathcal{C}}_{\rho} \mathcal{G}^{ba}_{\mu\nu,h},
\end{equation}
which measures the failure of the Hermitian connection to
parallel-transport the quantum metric in the momentum-space manifold. As we show below, in the derivation of Wigner functions of massless gravitons, the quantum Levi-Civita connection and nonmetricity tensor can, in principle, contribute to the graviton polarization tensor, while the quantum torsion tensor does not. This is due to the fact that the existence of a torsion tensor requires the eigenbasis to consist of at least three states, which is not the case for the massless Weyl spinors comprising the polarization vectors of massless spin-1 fields. We also note that the quantum Levi-Civita connection is purely real, as it is comprised solely of the quantum metric tensor, while the quantum nonmetricity tensor is purely imaginary, as is shown in Refs.~\cite{mehraeen2025quantum, jain2026topological}.

\section{S3.~Graviton Wigner functions in flat spacetime}

\subsection{Wigner function expansion}

We first consider Wigner functions in the flat Minkowski background. Later on, we generalize this to curved backgrounds. The helicity-dependent lesser propagator for metric perturbations is defined as~\cite{bellac2011thermal, ito2026spin}
\begin{equation}
\label{Seq_Wigner_flat}
{W^h}_{\mu\nu\rho\sigma}(x, q)
=
\int d^4y \, e^{-\frac{i}{\hbar} q \cdot y} 
\left\langle h^h_{\rho\sigma}\left(x + \frac{y}{2}\right) h^h_{\mu\nu}\left(x - \frac{y}{2}\right) \right\rangle.
\end{equation}
In order to capture the quantum geometry of the Wigner functions, including the quantum geometric tensor and quantum connection, it is necessary to consider the expansion of the lesser propagator to third order in spacetime gradients. Inserting Eq.~(\ref{Seq_field_exp}), the lesser propagator reads
\begin{equation}
\begin{split}
{W^h}_{\mu\nu\rho\sigma}(x, q)
=&
\frac{32\pi^2 G}{|\mathbf{q}|} 
\int \frac{d^3\mathbf{p}_-}{(2\pi)^3}
e^{-\frac{i}{\hbar} x \cdot p_-}
\left[
\mathfrak{P}_{\mu\nu\rho\sigma}^{h(+)} 
\braket{ a_{\mathbf{q} - \frac{\mathbf{p}_-}{2}}^{h\dagger} a_{\mathbf{q} + \frac{\mathbf{p}_-}{2}}^h}
+
\mathfrak{P}_{\mu\nu\rho\sigma}^{h(-)} 
\braket{ a_{-\mathbf{q} + \frac{\mathbf{p}_-}{2}}^h 
a_{-\mathbf{q} - \frac{\mathbf{p}_-}{2}}^{h\dagger}}
\right]
\\
=&
\frac{32\pi^2 G}{|\mathbf{q}|}
\left[
\theta(q_0) \mathfrak{P}_{\mu\nu\rho\sigma}^{h(+)}
-
\theta(- q_0) \mathfrak{P}_{\mu\nu\rho\sigma}^{h(-)}
\right]
\check{f}^h(x,q),
\end{split}
\end{equation}
where the bare distribution function is given by
\begin{equation}
\label{Seq_bare_dist}
\check{f}^h(x,q)
=
\begin{cases}
\check{f}^h(x, \mathbf{q}), & q_0 = |\mathbf{q}|, 
\\
- [1 + \check{f}^h(x, \mathbf{-q})],  & q_0 = - |\mathbf{q}|,
\end{cases}
\end{equation}
\begin{equation}
\check{f}^h(x, \mathbf{q})
=
\int \frac{d^3\mathbf{p}_-}{(2\pi)^3}
e^{-\frac{i}{\hbar} x \cdot p_-}
\braket{ a_{\mathbf{q} - \frac{\mathbf{p}_-}{2}}^{h\dagger} a_{\mathbf{q} + \frac{\mathbf{p}_-}{2}}^h},
\end{equation} 
and the tensors $\mathfrak{P}_{\mu\nu\rho\sigma}^{h(\pm)}$ capture the polarization product expansion terms on the positive and negative mass shells. Specifically, for the product expansion
\begin{equation}
\label{Seq_prod_exp}
\begin{split}
\mathcal{P}_{\mu\nu\rho\sigma}^h (q ; p_-)
=&
e^h_{\mu\nu} \left(q + \frac{p_-}{2} \right)
e^{h*}_{\rho\sigma} \left(q - \frac{p_-}{2} \right)
\\
=&
\Pi _{\mu\nu\rho\sigma}^h (q)
+
p_-^{\alpha}
\Sigma _{\mu\nu\rho\sigma\alpha}^h (q)
+
p_-^{\alpha} p_-^{\beta} 
\Theta _{\mu\nu\rho\sigma\alpha\beta}^h (q)
+
p_-^{\alpha} p_-^{\beta} p_-^{\gamma} 
\Upsilon _{\mu\nu\rho\sigma\alpha\beta\gamma}^h (q)
+
\mathcal{O}(p_-^4),
\end{split}
\end{equation}
this reads
\begin{equation}
\mathfrak{P}^{h (\pm)}_{\mu\nu\rho\sigma}
=
\sum_n \mathfrak{P}^{h (\pm, n)}_{\mu\nu\rho\sigma}
\end{equation}
with
\begin{equation}
\mathfrak{P}^{h (+ , 0)}_{\mu\nu\rho\sigma} 
=
\delta_+ \Pi^h_{\mu\nu\rho\sigma}(q),
\qquad
\mathfrak{P}^{h (- , 0)}_{\mu\nu\rho\sigma}
=
\delta_- \Pi^h_{\rho\sigma\mu\nu}(-q),
\end{equation}
\begin{equation}
\mathfrak{P}^{h (+ , 1)}_{\mu\nu\rho\sigma} 
=
\delta_{+} 
p_-^\alpha \Sigma^h_{\mu\nu\rho\sigma\alpha}(q),
\qquad
\mathfrak{P}^{h (- , 1)}_{\mu\nu\rho\sigma} 
=
\delta_{-} 
p_-^\alpha \Sigma^h_{\rho\sigma\mu\nu\alpha}(-q),
\end{equation}
\begin{subequations}
\begin{align}
\mathfrak{P}^{h (+ , 2)}_{\mu\nu\rho\sigma} 
=&
\delta_+ 
\left[
p_-^\alpha p_-^\beta \Theta^h_{\mu\nu\rho\sigma\alpha\beta}(q) 
-
K_A \Pi^h_{\mu\nu\rho\sigma}(q)
\right]
-
\delta'_+ \left[ K_B \Pi^h_{\mu\nu\rho\sigma}(q) \right],
\\
\mathfrak{P}^{h (- , 2)}_{\mu\nu\rho\sigma} 
=&
\delta_- 
\left[
p_-^\alpha p_-^\beta \Theta^h_{\rho\sigma\mu\nu\alpha\beta}(-q) 
-
K_A \Pi^h_{\rho\sigma\mu\nu}(-q)
\right]
-
\delta'_- \left[ K_B \Pi^h_{\rho\sigma\mu\nu}(-q) \right],
\end{align}
\end{subequations}

\begin{subequations}
\begin{align}
\mathfrak{P}^{h (+ , 3)}_{\mu\nu\rho\sigma} 
=&
\delta_+
\left[
p_-^\alpha p_-^\beta p_-^\gamma \Upsilon^h_{\mu\nu\rho\sigma\alpha\beta\gamma}(q)
-
K_A p_-^\alpha \Sigma^h_{\mu\nu\rho\sigma\alpha}(q)
\right]
-
\delta'_+ 
\left[
K_B p_-^\alpha \Sigma^h_{\mu\nu\rho\sigma\alpha}(q)
\right],
\\
\mathfrak{P}^{h (- , 3)}_{\mu\nu\rho\sigma} 
=&
\delta_-
\left[
p_-^\alpha p_-^\beta p_-^\gamma \Upsilon^h_{\rho\sigma\mu\nu\alpha\beta\gamma}(-q)
-
K_A p_-^\alpha \Sigma^h_{\rho\sigma\mu\nu\alpha}(-q)
\right]
-
\delta'_- 
\left[
K_B p_-^\alpha \Sigma^h_{\rho\sigma\mu\nu\alpha}(-q)
\right],
\end{align}
\end{subequations}
where
$\delta_{\pm} \equiv \delta(|\mathbf{q}| \mp q_0)$ and
\begin{equation}
\label{eq:KAB}
K_A(\mathbf{q}, \mathbf{p}_-)
=
\frac{1}{8|\mathbf{q}|^2}\left[ \mathbf{p}_-^2 - \frac{2(\mathbf{q} \cdot \mathbf{p}_-)^2}{|\mathbf{q}|^2} \right],
\qquad
K_B(\mathbf{q}, \mathbf{p}_-)
=
\frac{1}{8|\mathbf{q}|}\left[ \mathbf{p}_-^2 - \frac{(\mathbf{q} \cdot \mathbf{p}_-)^2}{|\mathbf{q}|^2} \right],
\end{equation}
or covariantly as
\begin{equation}
K_A = -\frac{\hbar^2}{2} \mathcal{G}_{\lambda\kappa}^{-+} \pd^\lambda \pd^\kappa + \frac{\hbar^2}{8\vert{}\boldsymbol{q}\vert{}^2} (\hat{q}_\perp \cdot \pd)^2,
\qquad
K_B = -\frac{\hbar^2\vert{}\boldsymbol{q}\vert{}}{2} \mathcal{G}_{\lambda\kappa}^{-+} \pd^\lambda \pd^\kappa.
\end{equation}

In terms of graviton polarization tensors, the expansion elements in Eq.~(\ref{Seq_prod_exp}) are explicitly given by
\begin{equation}
\Pi^h_{\mu\nu\rho\sigma}(q) 
=
e^h_{\mu\nu} e^{h*}_{\rho\sigma},
\end{equation}
\begin{equation}
\begin{aligned} 
\Sigma^h_{\mu\nu\rho\sigma\alpha}(q) 
&= 
\frac{1}{2} \left[ (\partial_\alpha^q e^h_{\mu\nu}) e^{h*}_{\rho\sigma}
-
e^h_{\mu\nu} (\partial_\alpha^q 
e^{h*}_{\rho\sigma}) \right],
\end{aligned}
\end{equation}
\begin{equation}
\begin{aligned} 
\Theta^h_{\mu\nu\rho\sigma\alpha\beta}(q)
&=
\frac{1}{8} \left[ 
(\partial_\alpha^q \partial_\beta^q 
e^h_{\mu\nu}) e^{h*}_{\rho\sigma} 
-
2 (\partial_\alpha^q e^h_{\mu\nu})
(\partial_\beta^q 
e^{h*}_{\rho\sigma}) 
+
e^h_{\mu\nu} (\partial_\alpha^q \partial_\beta^q e^{h*}_{\rho\sigma})
\right],
\end{aligned}
\end{equation}
\begin{equation}
\begin{aligned}\Upsilon^h_{\mu\nu\rho\sigma\alpha\beta\gamma}(q) 
&= 
\frac{1}{48} \left[ 
(\partial_\alpha^q \partial_\beta^q \partial_\gamma^q e^h_{\mu\nu}) e^{h*}_{\rho\sigma} 
-
3 (\partial_\alpha^q \partial_\beta^q 
e^h_{\mu\nu})(\partial_\gamma^q 
e^{h*}_{\rho\sigma}) 
+
3(\partial_\alpha^q e^h_{\mu\nu})
(\partial_\beta^q \partial_\gamma^q 
e^{h*}_{\rho\sigma}) 
-
e^h_{\mu\nu} 
(\partial_\alpha^q \partial_\beta^q \partial_\gamma^q e^{h*}_{\rho\sigma})
\right].
\end{aligned}
\end{equation}

\subsection{Spin-1 decomposition of polarization tensor}

To make tensor manipulations more tractable, and to set up the implementation of the spinor-helicity formalism, we make use of the fact that the graviton polarization tensor can be expressed as a local tensor product of spin-1 polarization vectors~\cite{gupta1952quantization, weinberg1964photons},
$e^h_{\mu\nu}(q) = \epsilon^h_\mu(q) \epsilon^h_\nu(q)$. This results in the tensor product
\begin{equation}
\begin{split}
\mathcal{P}_{\mu\nu\rho\sigma}^h (q ; p_-)
=&
\tilde{\mathcal{P}}_{\mu\rho}^h (q ; p_-)
\tilde{\mathcal{P}}_{\nu\sigma}^h (q ; p_-),
\end{split}
\end{equation}
with the analogous expansion
\begin{equation}
\begin{split}
\tilde{\mathcal{P}}_{\mu\rho}^h (q ; p_-)
=&
\eps^h_{\mu} \left(q + \frac{p_-}{2} \right)
\eps^{h*}_{\rho} \left(q - \frac{p_-}{2} \right)
\\
=&
\tilde{\Pi} _{\mu\rho}^h (q)
+
p_-^{\alpha}
\tilde{\Sigma} _{\mu\rho\alpha}^h (q)
+
p_-^{\alpha} p_-^{\beta} 
\tilde{\Theta} _{\mu\rho\alpha\beta}^h (q)
+
p_-^{\alpha} p_-^{\beta} p_-^{\gamma} 
\tilde{\Upsilon}_{\mu\rho\alpha\beta\gamma}^h (q)
+
\mathcal{O}(p_-^4),
\end{split}
\end{equation}
and corresponding elements given by
\begin{equation}
\label{Seq_tilde_Pi}
\tilde{\Pi}_{\mu\nu}^h
=
\epsilon^h_\mu \epsilon^{h*}_\nu,
\end{equation}
\begin{equation}
\label{Seq_tilde_Sigma}
\tilde{\Sigma}_{\mu\nu\alpha}^h
=
\frac{1}{2}
\left[
(\partial_\alpha^q \epsilon^h_\mu)\epsilon^{h*}_\nu - \epsilon^h_\mu(\partial_\alpha^q \epsilon^{h*}_\nu)
\right],
\end{equation}
\begin{equation}
\label{Seq_tilde_Theta}
\tilde{\Theta}_{\mu\nu\alpha\beta}^h
=
\frac{1}{8} \left[ (\partial_\alpha^q\partial_\beta^q \epsilon^h_\mu) \epsilon^{h*}_\nu
-
2(\partial_\alpha^q \epsilon^h_\mu)(\partial_\beta^q \epsilon^{h*}_\nu)
+
\epsilon^h_\mu (\partial_\alpha^q\partial_\beta^q \epsilon^{h*}_\nu) \right],
\end{equation}
\begin{equation}
\label{Seq_tilde_Upsilon}
\tilde{\Upsilon}_{\mu\nu\alpha\beta\gamma}^h
=
\frac{1}{48} \left[ (\partial_\alpha^q\partial_\beta^q\partial_\gamma^q \epsilon^h_\mu) \epsilon^{h*}_\nu
-
3(\partial_\alpha^q\partial_\beta^q \epsilon^h_\mu)(\partial_\gamma^q \epsilon^{h*}_\nu)
+
3(\partial_\alpha^q \epsilon^h_\mu)(\partial_\beta^q\partial_\gamma^q \epsilon^{h*}_\nu)
-
\epsilon^h_\mu (\partial_\alpha^q\partial_\beta^q\partial_\gamma^q
\epsilon^{h*}_\nu) \right].
\end{equation}
These two sets of tensors are related as
\begin{equation}
\Pi_{\mu\nu\rho\sigma}^h(q)
=
\tilde{\Pi}_{\mu\rho}^h \tilde{\Pi}_{\nu\sigma}^h,
\end{equation}
\begin{equation}
\Sigma^h_{\mu\nu\rho\sigma\alpha}(q)
=
\tilde{\Sigma}_{\mu\rho\alpha}^h
\tilde{\Pi}_{\nu\sigma}^h
+
(\mu\rho \leftrightarrow \nu\sigma),
\end{equation}
\begin{equation}
\label{Seq_Theta_spin1}
\Theta^h_{\mu\nu\rho\sigma\alpha\beta}(q)
=
\tilde{\Theta}_{\mu\rho\alpha\beta}^h
\tilde{\Pi}_{\nu\sigma}^h
+
\frac{1}{2}
\tilde{\Sigma}_{\mu\rho\alpha}^h 
\tilde{\Sigma}_{\nu\sigma\beta}^h
+
(\mu\rho \leftrightarrow \nu\sigma),
\end{equation}
\begin{equation}
\label{Seq_Upsilon_spin1}
\Upsilon^h_{\mu\nu\rho\sigma\alpha\beta\gamma}(q)
=
\tilde{\Upsilon}_{\mu\rho\alpha\beta\gamma}^h
\tilde{\Pi}_{\nu\sigma}^h
+
\tilde{\Theta}_{\mu\rho\alpha\beta}^h
\tilde{\Sigma}_{\nu\sigma\gamma}^h
+
(\mu\rho \leftrightarrow \nu\sigma).
\end{equation}

\subsection{Spinor-helicity formalism and spin-1/2 representation}

Within the spinor-helicity formalism, spin-1 polarization vectors can be expressed as~\cite{kleiss1985spinor, gunion1985improved, xu1987helicity, peskin2011simplifying}
\begin{equation}
\epsilon^{h}_{\mu}(q;p)
=
\frac{1}{\sqrt{4p\cdot q}}
\bar{u}^{h}(p)\gamma_{\mu}u^{h}(q),
\end{equation}
where $p^{\mu}$ is an auxiliary reference momentum, with 
$q\cdot p \neq 0$
and $p^{2}=0$.
The massless
helicity eigenspinors satisfy
\begin{equation}
\slashed{q}\left(1+\gamma_{5}\right)u^{\mathrm{R}}(q)
= \slashed{q}\left(1-\gamma_{5}\right)u^{\mathrm{L}}(q) = 0,
\end{equation}
together with the completeness relation
\begin{equation}
u^{\mathrm{R}}(q)\bar{u}^{\mathrm{R}}(q)
+ u^{\mathrm{L}}(q)\bar{u}^{\mathrm{L}}(q) = \slashed{q}.
\end{equation}
Working within the Weyl representation
\begin{equation}
\gamma^{\mu}
=
\begin{pmatrix}
0 & \sigma^{\mu}
\\
\bar{\sigma}^{\mu} & 0
\end{pmatrix},
\qquad
\gamma^{5}
=
\begin{pmatrix}
-I & 0 
\\
0 & I
\end{pmatrix},
\qquad
\sigma^{\mu}
=
\left(1,\bs{\sigma}\right), 
\qquad
\bar{\sigma}^{\mu}
=
\left(1,-\bs{\sigma}\right),
\end{equation}
and introducing two-component spinors $c_h(q)$ through
\begin{equation}
u_{\mathrm{L}}(q) = \sqrt{2\left|\boldsymbol{q}\right|}
\begin{pmatrix}
c_{\mathrm{L}}(q) \\
0
\end{pmatrix}, \qquad
u_{\mathrm{R}}(q) = \sqrt{2\left|\boldsymbol{q}\right|}
\begin{pmatrix}
0 \\
c_{\mathrm{R}}(q)
\end{pmatrix},
\end{equation}
the polarization vector reads
\begin{equation}
\epsilon^h_{\mu}(q;p)
=
\frac{\sqrt{\left|\boldsymbol{q}\right|\left|\boldsymbol{p}\right|}}{\sqrt{q\cdot p}}
c^{\dagger}_h (p) \sigma_{\mu}^h c_h (q),
\end{equation}
where we introduce helicity-dependent Pauli matrix 4-vectors as
$\sigma^{\mathrm{R}} = \sigma$, $\sigma^{\mathrm{L}} = \bar{\sigma}$
and correspondingly
$\bar{\sigma}^{\mathrm{R}} = \bar{\sigma}$,
$\bar{\sigma}^{\mathrm{L}} = \sigma$,
such that
$\bar{\sigma}^{\mathrm{R,L}} = \sigma^{\mathrm{L,R}}$. 

To eliminate the auxiliary vector $p$, it is convenient to express it as
\begin{equation}
p
=
p \cdot n (n - \hat{q}_\perp),
\end{equation}
where
$n$ is the frame vector,
$q_{\perp} = q - q \cdot n \, n$
and
$\hat{q}_\perp
=
q_\perp/\sqrt{|q_\perp^2|}$, so that the spatial part of $p$ is anti-parallel to $\boldsymbol{q}$ in the $n^\mu=(1,\bs{0})$ frame and only the overall scale $p\cdot n$ remains arbitrary. Furthermore, introducing
\begin{equation}
\bar{q}^{\mu} \equiv \left(q\cdot n\right)n^{\mu} - q^{\mu}_{\perp}
= 2\left(q\cdot n\right)n^{\mu} - q^{\mu},
\end{equation}
one has $p = \frac{p\cdot n}{q\cdot n}\bar{q}$ together with
$q\cdot p = 2\left(q\cdot n\right)\left(p\cdot n\right)$. We thus find
$\sqrt{2 | \boldsymbol{q} | |\boldsymbol{p} |}
=
\sqrt{q\cdot p}$,
which eliminates $p$ from the polarization amplitude. With this choice, $c_h(q)$ and $c_h(p)$ correspond to positive- and negative-energy eigenstates of the helicity-$h$ Weyl Hamiltonian
$H_{h} = \iota^{h}\bs{\sigma}\cdot\boldsymbol{q}$, with 
$\iota^{R,L} = \pm 1$.
We thus make the identification 
$c_{h}(q) \to c^{+}_{h}$, $c_{h}(p) \to c^{-}_{h}$, and the polarization vector reads
\begin{equation}
\epsilon_\mu^{h}(q)
=
\frac{1}{\sqrt{2}} c_h^{- \dagger}(q) \sigma_\mu^{h} c_h^+(q),
\end{equation}
with the spinors satisfying
\begin{equation}
c^{+}_{h}c^{+\dagger}_{h}
=
\frac{1}{2\left|\boldsymbol{q}\right|}q\cdot\bar{\sigma}_{h}, \qquad
c^{-}_{h}c^{-\dagger}_{h}
= \frac{1}{2\left|\boldsymbol{q}\right|}\bar{q}\cdot\bar{\sigma}_{h},
\end{equation}
and the completeness relation
\begin{equation}
c^{+}_{h}c^{+\dagger}_{h}
+
c^{-}_{h}c^{-\dagger}_{h}
=
I.
\end{equation}
Finally, using
$\{ \sigma_{\mu} , \bar{\sigma}_\nu \} = 2 \eta_{\mu\nu}$, we also note the useful relations
\begin{equation}
\label{Seq_ccdag_ids}
c_h^{+} c_h^{-\dagger}
=
\frac{1}{\sqrt{2}} \eps^h \cdot \bar{\sigma}
^h,
\qquad
c_h^{-} c_h^{+\dagger}
=
\frac{1}{\sqrt{2}} \eps^{h*} \cdot \bar{\sigma}
^h.
\end{equation}

\subsection{Quantum geometric structure of Wigner functions}

To uncover the quantum geometric structure underlying the Wigner functions, we rexpress the off-diagonal Berry connection in the spinor basis as
\begin{equation}
\begin{aligned}
\mathcal{A}^{\prime-+}_{\mu h}
&=
ic^{-\dagger}_{h}\partial^{q}_{\mu}c^{+}_{h}
=
ic^{-\dagger}_{h}\left[\partial^{q}_{\mu}\left(c^{+}_{h}c^{+\dagger}_{h}\right)\right]c^{+}_{h}
=
ic^{-\dagger}_{h}\left[\partial^{q}_{\mu}\left(\frac{q\cdot\bar{\sigma}^{h}}{2|\boldsymbol{q}|}\right)\right]c^{+}_{h}
=
-\frac{i}{2|\boldsymbol{q}|}
c^{-\dagger}_{h}\sigma^{h}_{\mu}c^{+}_{h},
\end{aligned}
\end{equation}
which establishes the useful relation between the polarization vector and Berry connection
\begin{equation}
\epsilon_\mu^h
=
i\sqrt{2}\vert{}\boldsymbol{q}\vert{}\mathcal{A}^{\prime - +}_{\mu h}.
\end{equation}
Inserting this into the mode expansion elements reveals the geometric structure. At leading order, this implies
\begin{equation}
\tilde{\Pi}^{h}_{\mu\nu}
=
2 |\boldsymbol{q} |^{2} \mathcal{Q}^{-+}_{\mu\nu,h}.
\end{equation}
It is worth mentioning that $\tilde{\Pi}^{h}_{\mu\nu}$ can also be decomposed as~\cite{hattori2021wigner}
\begin{equation}
\tilde{\Pi}^{h}_{\mu\nu}
=
\frac{1}{2}
\left(
P_{\mu\nu}
-
\frac{i}{2} S_{\mu\nu}^h
\right),
\end{equation}
where $P_{\mu\nu}$ is the standard spin-1 polarization tensor in the Coulomb gauge
\begin{equation}
P_{\mu\nu}
=
\frac{p_\mu q_\nu + p_\nu q_\mu}{q \cdot p} - \eta_{\mu\nu} 
=
n_{\mu}n_{\nu} - \eta_{\mu\nu} - \hat{q}^\perp_{\mu}\hat{q}^\perp_{\nu},
\end{equation}
which projects 4-vectors onto the 2d polarization plane transverse to $q$, and
\begin{equation}
S_{\mu\nu}^h
=
2 \iota^h \epsilon_{\mu\nu\alpha\beta} \frac{q^\alpha n^\beta}{q \cdot n},
\end{equation}
is the spin tensor, which generates rotations within the transverse subspace of polarizations. This readily implies the useful reinterpretation of these quantities with quantum geometric tensors
\begin{equation}
P_{\mu\nu}
=4|\boldsymbol{q}|^{2}\mathcal{G}^{-+}_{\mu\nu},
\qquad
S^{h}_{\mu\nu}
=4|\boldsymbol{q}|^{2}\Omega^{-+}_{\mu\nu,h}.
\end{equation}

To evaluate $\tilde{\Sigma}$, using Eq.~(\ref{Seq_Berry_cov}), we have
\begin{align}
\tilde{\Sigma}^{h}_{\mu\nu\alpha}
&=
\frac{1}{2}\left[(\partial^{q}_{\alpha}\epsilon^{h}_{\mu})\epsilon^{h*}_{\nu}-\epsilon^{h}_{\mu}(\partial^{q}_{\alpha}\epsilon^{h*}_{\nu})\right]
\nonumber
\\
&=
\frac{1}{2}\Big[\partial^{q}_{\alpha}\left(i\sqrt{2}|\boldsymbol{q}|\mathcal{A}^{\prime-+}_{\mu h}\right)\left(-i\sqrt{2}|\boldsymbol{q}|\mathcal{A}^{\prime+-}_{\nu h}\right)
-\left(i\sqrt{2}|\boldsymbol{q}|\mathcal{A}^{\prime-+}_{\mu h}\right)\partial^{q}_{\alpha}\left(-i\sqrt{2}|\boldsymbol{q}|\mathcal{A}^{\prime+-}_{\nu h}\right)\Big]
\nonumber
\\
&=
|\boldsymbol{q}|^{2}\left[\left(\partial^{q}_{\alpha}\mathcal{A}^{\prime-+}_{\mu h}\right)\mathcal{A}^{\prime+-}_{\nu h}
-\mathcal{A}^{\prime-+}_{\mu h}\left(\partial^{q}_{\alpha}\mathcal{A}^{\prime+-}_{\nu h}\right)\right]
\nonumber
\\
&=
|\boldsymbol{q}|^{2}\left[2i\Delta^{-+}_{\alpha h}\mathcal{Q}^{-+}_{\mu\nu,h}
+\mathcal{C}^{-+}_{\nu\alpha\mu,h}
-\mathcal{C}^{+-}_{\mu\alpha\nu,h}\right],
\end{align}
revealing the appearance of the quantum connection at linear order in the mode expansion.

Similarly, the second-order contribution reads
\begin{align}
\tilde{\Theta}^{h}_{\mu\nu\alpha\beta}
=&
\frac{|\boldsymbol{q}|}{4}\left(\hat{q}^{\perp}_{\beta}
\mathcal{C}^{-+}_{\nu\alpha\mu,h}
+
\hat{q}^{\perp}_{\beta}\mathcal{C}^{+-}_{\mu\alpha\nu,h}
+
\hat{q}^{\perp}_{\alpha}\mathcal{C}^{-+}_{\nu\beta\mu,h}
+
\hat{q}^{\perp}_{\alpha}\mathcal{C}^{+-}_{\mu\beta\nu,h}\right)
-
\hat{q}^{\perp}_{\alpha}\hat{q}^{\perp}_{\beta}
\mathcal{Q}^{-+}_{\mu\nu,h}
-
\frac{1}{2}\hat{q}^{\perp}_{\mu}\hat{q}^{\perp}_{\nu}
\mathcal{Q}^{-+}_{\alpha\beta,h}
\nonumber
\\
&-
|\boldsymbol{q}|^{2}\left(\mathcal{G}^{-+}_{\beta\mu,h}\mathcal{Q}^{-+}_{\alpha\nu,h}
+
\mathcal{G}^{-+}_{\beta\nu,h}\mathcal{Q}^{-+}_{\mu\alpha,h}\right)
-
|\boldsymbol{q}|^{2}\Delta^{-+}_{\alpha h}\Delta^{-+}_{\beta h}\mathcal{Q}^{-+}_{\mu\nu,h}
\nonumber
\\
&+
\frac{i|\boldsymbol{q}|^{2}}{4}\left[\Delta^{-+}_{\beta h}\left(3\mathcal{C}^{-+}_{\nu\alpha\mu,h}
-
\mathcal{C}^{+-}_{\mu\alpha\nu,h}\right)
+
\Delta^{-+}_{\alpha h}
\left(\mathcal{C}^{-+}_{\nu\beta\mu,h}
-
3\mathcal{C}^{+-}_{\mu\beta\nu,h}\right)\right].
\end{align}
Inserting this into Eq.~(\ref{Seq_Theta_spin1}), we find
\begin{align}
\Theta^h_{\mu\nu\rho\sigma\alpha\beta}(q)
={}& - 4|\mathbf{q}|^2\,\hat{q}^\perp_\alpha \hat{q}^\perp_\beta\,
\mathcal{Q}^{-+}_{\mu\rho,h}\,\mathcal{Q}^{-+}_{\nu\sigma,h}
- 8|\mathbf{q}|^4\,\Delta^{-+}_{\alpha h}\Delta^{-+}_{\beta h}\,
\mathcal{Q}^{-+}_{\mu\rho,h}\,\mathcal{Q}^{-+}_{\nu\sigma,h}
\nonumber\\
&+ \bigg\{
\frac{|\mathbf{q}|^3}{2}\,\mathcal{Q}^{-+}_{\nu\sigma,h}
\Big[ \hat{q}^\perp_\beta \big( \mathcal{C}^{-+}_{\rho\alpha\mu,h} + \mathcal{C}^{+-}_{\mu\alpha\rho,h} \big)
+ \hat{q}^\perp_\alpha \big( \mathcal{C}^{-+}_{\rho\beta\mu,h} + \mathcal{C}^{+-}_{\mu\beta\rho,h} \big) \Big]
\nonumber\\
&\qquad
- |\mathbf{q}|^2\,\hat{q}^\perp_\mu \hat{q}^\perp_\rho\,
\mathcal{Q}^{-+}_{\alpha\beta,h}\,\mathcal{Q}^{-+}_{\nu\sigma,h}
- 2|\mathbf{q}|^4 \big( \mathcal{G}^{-+}_{\beta\mu,h}\,\mathcal{Q}^{-+}_{\alpha\rho,h}
+ \mathcal{G}^{-+}_{\beta\rho,h}\,\mathcal{Q}^{-+}_{\mu\alpha,h} \big)\,
\mathcal{Q}^{-+}_{\nu\sigma,h}
\nonumber\\
&\qquad
+ \frac{i|\mathbf{q}|^4}{2}\,\mathcal{Q}^{-+}_{\nu\sigma,h}
\Big[ \Delta^{-+}_{\beta h} \big( 5\,\mathcal{C}^{-+}_{\rho\alpha\mu,h} - 3\,\mathcal{C}^{+-}_{\mu\alpha\rho,h} \big)
+ \Delta^{-+}_{\alpha h} \big( 3\,\mathcal{C}^{-+}_{\rho\beta\mu,h} - 5\,\mathcal{C}^{+-}_{\mu\beta\rho,h} \big) \Big]
\nonumber\\
&\qquad
+ \frac{|\mathbf{q}|^4}{2}
\big( \mathcal{C}^{-+}_{\rho\alpha\mu,h} - \mathcal{C}^{+-}_{\mu\alpha\rho,h} \big)
\big( \mathcal{C}^{-+}_{\sigma\beta\nu,h} - \mathcal{C}^{+-}_{\nu\beta\sigma,h} \big)
+ (\mu\rho \leftrightarrow \nu\sigma) \bigg\}
\end{align}
In particular, for the purpose of evaluating stress tensor contributions later on, we note the contraction
\begin{equation}
{\Theta^h}_{\alpha\beta\phantom{\alpha\beta}\lambda\kappa}^{\phantom{\alpha\beta}\alpha\beta}
=
-2 \Delta_{\lambda h}^{-+} 
\Delta_{\kappa h}^{-+} 
-
2\mathcal{G}^{-+}_{\lambda\kappa,h}
+
\frac{i}{2} \Omega^{-+}_{\lambda\kappa,h},
\end{equation}
which helps explain the origin of the quantum metric contributions  to the higher-order stress tensor.

The third-order terms turn out to be fairly lengthy. We thus introduce the decomposition
\begin{equation}
\tilde{\Upsilon}^h_{\mu\nu\alpha\beta\gamma}
= 
\sum_{n=1}^4 \mathfrak{U}_n,
\end{equation}
where each term evaluates as
\begin{equation}
\begin{aligned}
48\, \mathfrak{U}_1
=&
(\partial_\alpha^q \partial_\beta^q \partial_\gamma^q \epsilon_\mu^h)\epsilon_\nu^{h*}
\\
=& 2\vert{}\boldsymbol{q}\vert{}^2 \left[ i\partial_\alpha^q \partial_\beta^q \Delta_{\gamma h}^{-+} - \Delta_{\alpha h}^{-+} \partial_\beta^q \Delta_{\gamma h}^{-+} - \Delta_{\beta h}^{-+} \partial_\alpha^q \Delta_{\gamma h}^{-+} - \Delta_{\gamma h}^{-+} \partial_\alpha^q \Delta_{\beta h}^{-+} - i\Delta_{\alpha h}^{-+} \Delta_{\beta h}^{-+} \Delta_{\gamma h}^{-+} \right] \mathcal{Q}_{\mu\nu,h}^{-+} \\ & + 2i\vert{}\boldsymbol{q}\vert{}\hat{q}_\mu^\perp \left[ \left(\partial_\alpha^q \Delta_{\beta h}^{-+}\right)\mathcal{Q}_{\gamma\nu,h}^{-+} + \left(\partial_\alpha^q \Delta_{\gamma h}^{-+}\right)\mathcal{Q}_{\beta\nu,h}^{-+} + \left(\partial_\beta^q \Delta_{\gamma h}^{-+}\right)\mathcal{Q}_{\alpha\nu,h}^{-+} \right] \\ & - 2\vert{}\boldsymbol{q}\vert{}\hat{q}_\mu^\perp \left[ \Delta_{\alpha h}^{-+} \Delta_{\beta h}^{-+} \mathcal{Q}_{\gamma\nu,h}^{-+} + \Delta_{\alpha h}^{-+} \Delta_{\gamma h}^{-+} \mathcal{Q}_{\beta\nu,h}^{-+} + \Delta_{\beta h}^{-+} \Delta_{\gamma h}^{-+} \mathcal{Q}_{\alpha\nu,h}^{-+} \right] \\ & - 8i\vert{}\boldsymbol{q}\vert{}^2 \Big[ \Delta_{\alpha h}^{-+} \mathcal{G}_{\beta\mu,h}^{-+} \mathcal{Q}_{\gamma\nu,h}^{-+} + \Delta_{\beta h}^{-+} \mathcal{G}_{\alpha\mu,h}^{-+} \mathcal{Q}_{\gamma\nu,h}^{-+} + \Delta_{\gamma h}^{-+} \mathcal{G}_{\alpha\mu,h}^{-+} \mathcal{Q}_{\beta\nu,h}^{-+} \\ &\qquad\qquad + \Delta_{\gamma h}^{-+} \mathcal{G}_{\beta\mu,h}^{-+} \mathcal{Q}_{\alpha\nu,h}^{-+} + \Delta_{\beta h}^{-+} \mathcal{G}_{\gamma\mu,h}^{-+} \mathcal{Q}_{\alpha\nu,h}^{-+} + \Delta_{\alpha h}^{-+} \mathcal{G}_{\gamma\mu,h}^{-+} \mathcal{Q}_{\beta\nu,h}^{-+} \Big] \\ & - 8\vert{}\boldsymbol{q}\vert{} \Big[ \hat{q}_\alpha^\perp \mathcal{G}_{\beta\mu,h}^{-+} \mathcal{Q}_{\gamma\nu,h}^{-+} + \hat{q}_\beta^\perp \mathcal{G}_{\alpha\mu,h}^{-+} \mathcal{Q}_{\gamma\nu,h}^{-+} + \hat{q}_\gamma^\perp \mathcal{G}_{\alpha\mu,h}^{-+} \mathcal{Q}_{\beta\nu,h}^{-+} \\ &\qquad\qquad + \hat{q}_\gamma^\perp \mathcal{G}_{\beta\mu,h}^{-+} \mathcal{Q}_{\alpha\nu,h}^{-+} + \hat{q}_\beta^\perp \mathcal{G}_{\gamma\mu,h}^{-+} \mathcal{Q}_{\alpha\nu,h}^{-+} + \hat{q}_\alpha^\perp \mathcal{G}_{\gamma\mu,h}^{-+} \mathcal{Q}_{\beta\nu,h}^{-+} \Big] \\ & + 2i\hat{q}_\mu^\perp \Big[ \Delta_{\alpha h}^{-+} \left(\hat{q}_\beta^\perp \mathcal{Q}_{\gamma\nu,h}^{-+} + \hat{q}_\gamma^\perp \mathcal{Q}_{\beta\nu,h}^{-+}\right) + \Delta_{\beta h}^{-+} \left(\hat{q}_\alpha^\perp \mathcal{Q}_{\gamma\nu,h}^{-+} + \hat{q}_\gamma^\perp \mathcal{Q}_{\alpha\nu,h}^{-+}\right) \\ &\qquad\qquad + \Delta_{\gamma h}^{-+} \left(\hat{q}_\alpha^\perp \mathcal{Q}_{\beta\nu,h}^{-+} + \hat{q}_\beta^\perp \mathcal{Q}_{\alpha\nu,h}^{-+}\right) \Big]
+
2\hat{q}_\mu^\perp \left[ \hat{q}_\alpha^\perp \hat{q}_\beta^\perp \mathcal{Q}_{\gamma\nu,h}^{-+} + \hat{q}_\alpha^\perp \hat{q}_\gamma^\perp \mathcal{Q}_{\beta\nu,h}^{-+} + \hat{q}_\beta^\perp \hat{q}_\gamma^\perp \mathcal{Q}_{\alpha\nu,h}^{-+}
\right],
\end{aligned}
\end{equation}
\begin{equation}
\begin{aligned}
48\, \mathfrak{U}_2
=&
-3(\partial_\alpha^q \partial_\beta^q \epsilon_\mu^h)(\partial_\gamma^q \epsilon_\nu^{h*})
\\
=&
-6 \vert{}\boldsymbol{q}\vert{}^2 \left[ (\partial_\alpha^q \Delta_{\beta h}^{-+}) \Delta_{\gamma h}^{-+} + i\Delta_{\alpha h}^{-+} \Delta_{\beta h}^{-+} \Delta_{\gamma h}^{-+} \right] \mathcal{Q}_{\mu\nu,h}^{-+}
-
6\vert{}\boldsymbol{q}\vert{}\hat{q}_\nu^\perp \left[ i(\partial_\alpha^q \Delta_{\beta h}^{-+}) - \Delta_{\alpha h}^{-+} \Delta_{\beta h}^{-+} \right] \mathcal{Q}_{\mu\gamma,h}^{-+}
\\
&-
6\vert{}\boldsymbol{q}\vert{}\hat{q}_\mu^\perp \Delta_{\gamma h}^{-+} \left[ \Delta_{\alpha h}^{-+} \mathcal{Q}_{\beta\nu,h}^{-+} + \Delta_{\beta h}^{-+} \mathcal{Q}_{\alpha\nu,h}^{-+} \right]
-
6i\hat{q}_\mu^\perp \Delta_{\gamma h}^{-+} \left[ \hat{q}_\alpha^\perp \mathcal{Q}_{\beta\nu,h}^{-+} + \hat{q}_\beta^\perp \mathcal{Q}_{\alpha\nu,h}^{-+} \right]
\\
&-
12i\vert{}\boldsymbol{q}\vert{}^2 \Delta_{\gamma h}^{-+} \left[ \mathcal{G}_{\alpha\mu,h}^{-+} \mathcal{Q}_{\beta\nu,h}^{-+} + \mathcal{G}_{\beta\mu,h}^{-+} \mathcal{Q}_{\alpha\nu,h}^{-+} \right]
-
6i\hat{q}_\mu^\perp \hat{q}_\nu^\perp \left[ \Delta_{\alpha h}^{-+} \mathcal{Q}_{\beta\gamma,h}^{-+} + \Delta_{\beta h}^{-+} \mathcal{Q}_{\alpha\gamma,h}^{-+} \right]
\\
&-
6\hat{q}_\mu^\perp \hat{q}_\nu^\perp \left[ \hat{q}_\alpha^\perp \mathcal{Q}_{\beta\gamma,h}^{-+} + \hat{q}_\beta^\perp \mathcal{Q}_{\alpha\gamma,h}^{-+} \right] + 12\vert{}\boldsymbol{q}\vert{}\hat{q}_\nu^\perp \left[ \mathcal{G}_{\alpha\mu,h}^{-+} \mathcal{Q}_{\beta\gamma,h}^{-+} + \mathcal{G}_{\beta\mu,h}^{-+} \mathcal{Q}_{\alpha\gamma,h}^{-+} \right],
\end{aligned}
\end{equation}
\begin{equation}
\begin{aligned}
48\, \mathfrak{U}_3
=&
3(\partial_\alpha^q \epsilon_\mu^h)(\partial_\beta^q \partial_\gamma^q \epsilon_\nu^{h*})
\\
=&
 6\vert{}\boldsymbol{q}\vert{}^2 \left[ (\partial_\beta^q \Delta_{\gamma h}^{-+}) \Delta_{\alpha h}^{-+} - i\Delta_{\alpha h}^{-+} \Delta_{\beta h}^{-+} \Delta_{\gamma h}^{-+} \right] \mathcal{Q}_{\mu\nu,h}^{-+}
 - 6\vert{}\boldsymbol{q}\vert{}\hat{q}_\mu^\perp \left[ i(\partial_\beta^q \Delta_{\gamma h}^{-+}) + \Delta_{\beta h}^{-+} \Delta_{\gamma h}^{-+} \right] \mathcal{Q}_{\alpha\nu,h}^{-+}
\\
&+
 6\vert{}\boldsymbol{q}\vert{}\hat{q}_\nu^\perp \Delta_{\alpha h}^{-+} \left[ \Delta_{\beta h}^{-+} \mathcal{Q}_{\mu\gamma,h}^{-+} + \Delta_{\gamma h}^{-+} \mathcal{Q}_{\mu\beta,h}^{-+} \right]
 -6 i \hat{q}_\nu^\perp \Delta_{\alpha h}^{-+} \left[ \hat{q}_\beta^\perp \mathcal{Q}_{\mu\gamma,h}^{-+} + \hat{q}_\gamma^\perp \mathcal{Q}_{\mu\beta,h}^{-+} \right]
\\
&+
 12i\vert{}\boldsymbol{q}\vert{}^2 \Delta_{\alpha h}^{-+} \left[ \mathcal{G}_{\beta\nu,h}^{-+} \mathcal{Q}_{\mu\gamma,h}^{-+} + \mathcal{G}_{\gamma\nu,h}^{-+} \mathcal{Q}_{\mu\beta,h}^{-+} \right]
+
6i\hat{q}_\mu^\perp \hat{q}_\nu^\perp \left[ \Delta_{\beta h}^{-+} \mathcal{Q}_{\alpha\gamma,h}^{-+} + \Delta_{\gamma h}^{-+} \mathcal{Q}_{\alpha\beta,h}^{-+} \right]
\\
&-
6\hat{q}_\mu^\perp \hat{q}_\nu^\perp \left[ \hat{q}_\beta^\perp \mathcal{Q}_{\alpha\gamma,h}^{-+} + \hat{q}_\gamma^\perp \mathcal{Q}_{\alpha\beta,h}^{-+} \right]
+
12\vert{}\boldsymbol{q}\vert{}\hat{q}_\mu^\perp \left[ \mathcal{G}_{\beta\nu,h}^{-+} \mathcal{Q}_{\alpha\gamma,h}^{-+} + \mathcal{G}_{\gamma\nu,h}^{-+} \mathcal{Q}_{\alpha\beta,h}^{-+}
\right],
\end{aligned}
\end{equation}
\begin{equation}
\begin{aligned}
48\, \mathfrak{U}_4
=&
-\epsilon_\mu^h(\partial_\alpha^q \partial_\beta^q \partial_\gamma^q \epsilon_\nu^{h*})
\\
=&
2\vert{}\boldsymbol{q}\vert{}^2 \left[ i\partial_\alpha^q \partial_\beta^q \Delta_{\gamma h}^{-+} + \Delta_{\alpha h}^{-+} \partial_\beta^q \Delta_{\gamma h}^{-+} + \Delta_{\beta h}^{-+} \partial_\alpha^q \Delta_{\gamma h}^{-+} + \Delta_{\gamma h}^{-+} \partial_\alpha^q \Delta_{\beta h}^{-+} - i\Delta_{\alpha h}^{-+} \Delta_{\beta h}^{-+} \Delta_{\gamma h}^{-+} \right] \mathcal{Q}_{\mu\nu,h}^{-+} \\ & + 2i\vert{}\boldsymbol{q}\vert{}\hat{q}_\nu^\perp \left[ (\partial_\alpha^q \Delta_{\beta h}^{-+}) \mathcal{Q}_{\mu\gamma,h}^{-+} + (\partial_\alpha^q \Delta_{\gamma h}^{-+}) \mathcal{Q}_{\mu\beta,h}^{-+} + (\partial_\beta^q \Delta_{\gamma h}^{-+}) \mathcal{Q}_{\mu\alpha,h}^{-+} \right] \\ & + 2\vert{}\boldsymbol{q}\vert{}\hat{q}_\nu^\perp \left[ \Delta_{\alpha h}^{-+} \Delta_{\beta h}^{-+} \mathcal{Q}_{\mu\gamma,h}^{-+} + \Delta_{\alpha h}^{-+} \Delta_{\gamma h}^{-+} \mathcal{Q}_{\mu\beta,h}^{-+} + \Delta_{\beta h}^{-+} \Delta_{\gamma h}^{-+} \mathcal{Q}_{\mu\alpha,h}^{-+} \right] \\ & + 8i\vert{}\boldsymbol{q}\vert{}^2 \Big[ \Delta_{\alpha h}^{-+} \mathcal{G}_{\beta\nu,h}^{-+} \mathcal{Q}_{\mu\gamma,h}^{-+} + \Delta_{\beta h}^{-+} \mathcal{G}_{\alpha\nu,h}^{-+} \mathcal{Q}_{\mu\gamma,h}^{-+} + \Delta_{\gamma h}^{-+} \mathcal{G}_{\alpha\nu,h}^{-+} \mathcal{Q}_{\mu\beta,h}^{-+} \\ &\qquad\qquad + \Delta_{\gamma h}^{-+} \mathcal{G}_{\beta\nu,h}^{-+} \mathcal{Q}_{\mu\alpha,h}^{-+} + \Delta_{\beta h}^{-+} \mathcal{G}_{\gamma\nu,h}^{-+} \mathcal{Q}_{\mu\alpha,h}^{-+} + \Delta_{\alpha h}^{-+} \mathcal{G}_{\gamma\nu,h}^{-+} \mathcal{Q}_{\mu\beta,h}^{-+} \Big] \\ & - 8\vert{}\boldsymbol{q}\vert{} \Big[ \hat{q}_\alpha^\perp \mathcal{G}_{\beta\nu,h}^{-+} \mathcal{Q}_{\mu\gamma,h}^{-+} + \hat{q}_\beta^\perp \mathcal{G}_{\alpha\nu,h}^{-+} \mathcal{Q}_{\mu\gamma,h}^{-+} + \hat{q}_\gamma^\perp \mathcal{G}_{\alpha\nu,h}^{-+} \mathcal{Q}_{\mu\beta,h}^{-+} \\ &\qquad\qquad + \hat{q}_\gamma^\perp \mathcal{G}_{\beta\nu,h}^{-+} \mathcal{Q}_{\mu\alpha,h}^{-+} + \hat{q}_\beta^\perp \mathcal{G}_{\gamma\nu,h}^{-+} \mathcal{Q}_{\mu\alpha,h}^{-+} + \hat{q}_\alpha^\perp \mathcal{G}_{\gamma\nu,h}^{-+} \mathcal{Q}_{\mu\beta,h}^{-+} \Big] \\ & - 2i\hat{q}_\nu^\perp \Big[ \Delta_{\alpha h}^{-+} \left(\hat{q}_\beta^\perp \mathcal{Q}_{\mu\gamma,h}^{-+} + \hat{q}_\gamma^\perp \mathcal{Q}_{\mu\beta,h}^{-+}\right) + \Delta_{\beta h}^{-+} \left(\hat{q}_\alpha^\perp \mathcal{Q}_{\mu\gamma,h}^{-+} + \hat{q}_\gamma^\perp \mathcal{Q}_{\mu\alpha,h}^{-+}\right) \\ &\qquad\qquad + \Delta_{\gamma h}^{-+} \left(\hat{q}_\alpha^\perp \mathcal{Q}_{\mu\beta,h}^{-+} + \hat{q}_\beta^\perp \mathcal{Q}_{\mu\alpha,h}^{-+}\right) \Big]
+ 2\hat{q}_\nu^\perp \left[ \hat{q}_\alpha^\perp \hat{q}_\beta^\perp \mathcal{Q}_{\mu\gamma,h}^{-+} + \hat{q}_\alpha^\perp \hat{q}_\gamma^\perp \mathcal{Q}_{\mu\beta,h}^{-+} + \hat{q}_\beta^\perp \hat{q}_\gamma^\perp \mathcal{Q}_{\mu\alpha,h}^{-+}
\right].
\end{aligned}
\end{equation}
Similarly, inserting into Eq.~(\ref{Seq_Upsilon_spin1}) and contracting, we find
\begin{equation}
\begin{split}
\Upsilon^{h\phantom{\alpha\beta}\alpha\beta}_{\phantom{h}\alpha\beta\phantom{\alpha\beta}\lambda\kappa\sigma}
=&
\frac{i}{12} \partial^{q}_{(\kappa} \partial^{q}_{\sigma} \Delta^{-+}_{\lambda)h}
+
\frac{1}{8} \left[
\Delta^{-+}_{\lambda h}
\partial^{q}_{(\kappa} \Delta^{-+}_{\sigma)h}
-
\Delta^{-+}_{\sigma h} \partial^{q}_{(\kappa} \Delta^{-+}_{\lambda)h}
\right]
-
\frac{4i}{3} \Delta^{-+}_{\lambda h} \Delta^{-+}_{\kappa h} \Delta^{-+}_{\sigma h}
\\
&-
\frac{2i}{3} \Delta^{-+}_{\lambda h}
\mathcal{G}^{-+}_{\kappa\sigma}
-
\frac{8i}{3} \Delta^{-+}_{\sigma h}
\mathcal{G}^{-+}_{\lambda\kappa}
-
\frac{2i}{3} \Delta^{-+}_{\kappa h}
\mathcal{G}^{-+}_{\lambda\sigma}
+
\frac{1}{2}
(\Gamma_{\mathcal{G}})_{[\lambda\sigma]\kappa}
\\
&+
\frac{1}{24} \Delta^{-+}_{\lambda h}
\Omega^{-+}_{\kappa\sigma,h}
-
\frac{11}{24} \Delta^{-+}_{\sigma h}
\Omega^{-+}_{\lambda\kappa,h}
-
\frac{5}{12} \Delta^{-+}_{\kappa h}
\Omega^{-+}_{\lambda\sigma,h}
-
\frac{1}{2} \mathcal{L}^{-+}_{(\lambda\sigma)\kappa,h},
\end{split}
\end{equation}
revealing the appearance of the quantum Levi-Civita connection and quantum nonmetricity tensor in the higher-order stress tensors as well.

\section{S4.~Phase-space horizontal lift and Wigner functions in curved spacetime}

In curved spacetime, the definition of the phase space requires care, as a global notion of momentum is not generally available. Instead, to each point $x$ of the spacetime manifold, one attaches a separate momentum space, and the phase space is the resulting fiber bundle. A natural choice of fiber, reproducing the usual momentum space for $\bar{g}_{\mu\nu}=\eta_{\mu\nu}$, is the cotangent space: the momentum variable $q_{\mu}$ is a covariant vector at
$x$, the phase space is the cotangent bundle, and the separation
$y^{\mu}$ conjugate to $q_{\mu}$ is a contravariant vector in
the tangent space at $x$, the set of which builds the tangent bundle. Correspondingly, a more suitable covariant derivative to work with in phase space is the horizontal lift of $\nabla$ to the tangent and cotangent bundles, defined as~\cite{fonarev1994wigner, nakahara2018geometry, liu2019chiral}
\begin{subequations}
\label{Seq_hlift}
\begin{align}
D_{\mu} \Phi(x,y)
=&
\left( \nabla_{\mu} - \Gamma^{\lambda}_{\mu\nu} y^{\nu} \partial_{\lambda}^{y} \right) \Phi(x,y),
\\
D_{\mu} \Phi(x,q)
=&
\left( \nabla_{\mu} + \Gamma^{\lambda}_{\mu\nu} q_{\lambda} \partial_{q}^{\nu} \right) \Phi(x,q).
\end{align}
\end{subequations}
The introduction of the horizontal lift helps simplify the analysis in curved spacetime, as
\begin{equation}
\label{Seq_Dproperties}
\left[D_{\mu},y^{\nu}\right]=0,
\qquad
\left[D_{\mu},q_{\nu}\right]=0,
\qquad
D_{\mu}\bar{g}_{\nu\lambda}=0.
\end{equation}
In addition, the helicity decomposition underlying the graviton Wigner function is defined with respect to the frame vector $n_{\mu}$. To fix the associated frame ambiguity, one can impose the parallel transport condition
$\nabla_{\mu}n_{\nu}=0$ on the frame vector. Together, with the vansihing commutators in Eq.~(\ref{Seq_Dproperties}), this implies that the lifted covariant derivatives acting on Wigner functions pass through all intermediate terms and act only on the distribution function. 

Furthermore, owing to Eq.~\eqref{Seq_Dproperties}, the covariant Taylor expansion of a
field about $x$ exponentiates into a translation generated by the lift,
\begin{equation}
h_{\mu\nu}(x,y)
\equiv
\left(1+y^{\lambda}\nabla_{\lambda}
+\frac{1}{2}y^{\lambda}y^{\kappa}\nabla_{\lambda}\nabla_{\kappa}
+\cdots\right)h_{\mu\nu}(x)
=
e^{y\cdot D}h_{\mu\nu}(x),
\label{eq:lifted}
\end{equation}
which defines the graviton field at covariant separation $y$ from the base point $x$. The graviton Wigner function in curved spacetime then reads
\begin{equation}
\label{Seq_Wigner_curved}
W^{h}_{\phantom{h}\mu\nu\rho\sigma}(x,q)
=
\int_{y}e^{-i\frac{q\cdot y}{\hbar}}
\left\langle
h^{h}_{\rho\sigma}\left(x,\frac{y}{2}\right)
h^{h}_{\mu\nu}\left(x,-\frac{y}{2}\right)
\right\rangle,
\end{equation}
with 
$\int_{y} = \int d^{4}y\sqrt{-\bar{g}}$
and
$\bar{g} = \det(\bar{g}_{\mu\nu})$, 
generalizing the flat-spacetime definition, Eq.~(\ref{Seq_Wigner_flat}), covariantly under diffeomorphisms.

The practical consequence of
Eqs.~\eqref{Seq_Dproperties}--\eqref{Seq_Wigner_curved} is a substitution rule.
In flat spacetime the mode expansion organizes
$W^{h}_{\phantom{h}\mu\nu\rho\sigma}$ as a series of momentum-space kernels
multiplied by powers of the momentum difference $p_{-}$ acting on the
distribution function. Since $p_{-}$ is conjugate to the base point coordinate $x$ in Eq.~\eqref{Seq_Wigner_curved}, it acts on the distribution
function as a spacetime derivative, and covariance under
Eq.~\eqref{Seq_hlift} promotes this derivative to
the phase-space horizontal lift. As a result, the curved-spacetime mode expansion is readily obtained from the flat spacetime results through the substitutions
\begin{equation}
\label{Seq_substitution}
\eta_{\mu\nu}\rightarrow\bar{g}_{\mu\nu},
\qquad
p_{-}\rightarrow i\hbar D.
\end{equation}
In particular, these substitutions (and their inverses) also apply to the derivation of the stress tensor elements that follow when mapping between flat and curved backgrounds.

Finally, we note the TT gauge conditions the phase-space Wigner functions satisfy in curved spacetime follow from Eq.~(\ref{TT_gauge_h}) as
\begin{equation}
\label{Seq_TT_gauge_W}
\left(q^{\mu}+\frac{i\hbar}{2}D^{\mu}\right)W_{\mu\nu\rho\sigma}
=
0,
\qquad
\bar{g}^{\mu\nu}W_{\mu\nu\rho\sigma}
=
0,
\qquad
n^{\mu}W_{\mu\nu\rho\sigma}
=
0.
\end{equation}

\section{S5.~Dressed distribution function}

In this section, we first derive the transformation of the bare distribution function $\check{f}^h(x,q)$ under a momentum-dependent phase rotation of the helicity spinors, and then construct from it the dressed distribution function $f^h(x,q)$, which is
invariant under such rotations through $O(\hbar^3)$. This is an essential step in the analysis, as it allows one to eliminate  gauge redundancies in the stress tensor, as we show below.

The helicity spinors are defined only up to a momentum-dependent phase. Consider the local phase rotation
\begin{equation}
c^{\pm}_h(q) \to e^{i\theta_{\pm}(q)}\,c^{\pm}_h(q),
\label{eq:rot}
\end{equation}
which implies the polarization vector and tensor transform as
\begin{equation}
\epsilon^h_\mu(q) \to e^{i\chi(q)}\,\epsilon^h_\mu(q),
\qquad
e^h_{\mu\nu}(q) \to e^{2i\chi(q)}\,e^h_{\mu\nu}(q),
\label{eq:poltrans}
\end{equation}
with $\chi(q)\equiv\theta_{+}(q)-\theta_{-}(q)$ and the doubling of the phase reflecting the helicity $2$ of the graviton.
Invariance of the field operator $h^h_{\mu\nu}(x)$ in Eq.~(\ref{Seq_field_exp}) then requires
\begin{equation}
a^h_{p} \to e^{-2i\chi(p)} a^h_{p},
\qquad
a^{h\dagger}_{p} \to e^{+2i\chi(p)} a^{h\dagger}_{p}.
\label{eq:optrans}
\end{equation}
The bilinear form of the operators enters the bare distribution function, Eq.~(\ref{Seq_bare_dist}), in which
$q\pm\tfrac{p_-}{2}$ here denote the three-momenta
$\mathbf{q}\pm\tfrac{\mathbf{p}_-}{2}$, while temporal components are fixed by the
on-shell condition. The field expectation value therefore acquires the phase
\begin{equation}
\big\langle a^{h\dagger}_{q-\frac{p_-}{2}}\,a^{h}_{q+\frac{p_-}{2}}\big\rangle
\to
e^{-2i\left[\chi\left(q+\frac{p_-}{2}\right)-\chi\left(q-\frac{p_-}{2}\right)\right]}
\big\langle a^{h\dagger}_{q-\frac{p_-}{2}}\,a^{h}_{q+\frac{p_-}{2}}\big\rangle .
\label{eq:bilinear}
\end{equation}
Expanding the exponent
\begin{equation}
\chi\!\left(q+\tfrac{p_-}{2}\right)-\chi\!\left(q-\tfrac{p_-}{2}\right)
=
p_-^{\alpha}\,\partial^{q}_{\alpha}\chi
+
\tfrac{1}{24}\,p_-^{\alpha}p_-^{\beta}p_-^{\gamma}
\partial^{q}_{\alpha}\partial^{q}_{\beta}\partial^{q}_{\gamma}\chi
+
O(p_-^{5}),
\label{eq:sinh}
\end{equation}
implies that the bare distribution function transforms as
\begin{align}
\check{f}^h(x,q)
\to
&\Big[
1
+
2\hbar (\partial^{q}_{\alpha}\chi)\,D^{\alpha}
+
2\hbar^{2} (\partial^{q}_{\alpha}\chi)(\partial^{q}_{\beta}\chi)
D^{\alpha}D^{\beta}
\nonumber
+
\frac{4\hbar^{3}}{3}
(\partial^{q}_{\alpha}\chi)(\partial^{q}_{\beta}\chi)(\partial^{q}_{\gamma}\chi)
D^{\alpha}D^{\beta}D^{\gamma}
\\
&-
\frac{\hbar^{3}}{12}
(\partial^{q}_{\alpha}\partial^{q}_{\beta}\partial^{q}_{\gamma}\chi)
D^{\alpha}D^{\beta}D^{\gamma}
\Big]
\check{f}^h(x,q)
+
O(\hbar^{4}).
\label{eq:fbartrans}
\end{align}

Based on this, one can construct the gauge-invariant dressed distribution function. Under the rotation \eqref{eq:rot} the diagonal Berry connections transform as
$a^{a}_{\alpha h}\to a^{a}_{\alpha h}-\partial^{q}_{\alpha}\theta_{a}$, so that
\begin{equation}
\Delta^{+-}_{\alpha h}
\to
\Delta^{+-}_{\alpha h}-\partial^{q}_{\alpha}\chi.
\label{eq:contrans}
\end{equation}
Comparing Eqs.~\eqref{eq:fbartrans} and \eqref{eq:contrans}, the shift of
$\Delta^{+-}_{\alpha h}$ is precisely the gradient appearing in the transformation of $\check{f}^h$. We therefore define the dressed distribution function by rotating the phase back with the Berry connection mismatch
\begin{align}
f^h(x,q)
\equiv&
\Big[
1
+
2 \hbar \Delta^{+-}_{\alpha h} D^{\alpha}
+
2 \hbar^{2} \Delta^{+-}_{\alpha h} \Delta^{+-}_{\beta h}
D^{\alpha}D^{\beta}
\nonumber
+
\frac{4\hbar^{3}}{3}
\Delta^{+-}_{\alpha h} \Delta^{+-}_{\beta h}
\Delta^{+-}_{\gamma h}
D^{\alpha}D^{\beta}D^{\gamma}
\\
&-
\frac{\hbar^{3}}{12}
\big(\partial^{q}_{\alpha}\partial^{q}_{\beta}
\Delta^{+-}_{\gamma h}\big)
D^{\alpha}D^{\beta}D^{\gamma}
\Big]
\check{f}^h(x,q)
+
O(\hbar^{4}),
\label{eq:dressed}
\end{align}
which is then invariant under phase rotations and can thus represent a physical distribution function that couples with gauge invariant structures in the stress tensor.

\section{S6.~Graviton stress-energy tensor in phase space}

The stress-energy tensor of the metric perturbation is identified from the expansion of the Einstein tensor in powers of $h_{\mu\nu}$ about the background,
$G_{\mu\nu}
=
G^{[0]}_{\mu\nu}
+
G^{[1]}_{\mu\nu}
+
G^{[2]}_{\mu\nu}$.
One can reorganize the Einstein equation as
$G^{[0]}_{\mu\nu} 
=
- (G^{[1]}_{\mu\nu} + G^{[2]}_{\mu\nu})$, 
so that the perturbation sources the gravitational field, with stress-energy tensor
\begin{align}
T_{\mu\nu}
=
-\frac{1}{8\pi G}\left(G^{[1]}_{\mu\nu}
+
G^{[2]}_{\mu\nu}\right).
\end{align}
The TT gauge conditions eliminate the linear term,
$G^{[1]}_{\mu\nu} = 0$, and reduce
the quadratic term to~\cite{altas2019second}
\begin{equation}
G^{[2]}_{\mu\nu}
=
R^{[2]}_{\mu\nu}
-
\frac{1}{2}\bar{g}_{\mu\nu}R^{[2]},
\end{equation}
with
\begin{align}
R^{[2]}_{\mu\nu}
=&
-\frac{1}{2}h^{\sigma\beta}\nabla_{\sigma}
\left(\nabla_{\nu}h_{\mu\beta}
+
\nabla_{\mu}h_{\nu\beta}
-
\nabla_{\beta}h_{\mu\nu}\right)
+
\frac{1}{4}\nabla_{\nu}\nabla_{\mu}
\left(h_{\alpha\beta}
h^{\alpha\beta}\right) - \frac{1}{4}\nabla_{\nu}h^{\alpha\beta}
\nabla_{\mu}h_{\alpha\beta} 
\nonumber
\\
&-
\frac{1}{2}\nabla^{\sigma}h_{\mu\alpha}\nabla^{\alpha}h_{\nu\sigma}
+
\frac{1}{2}\nabla^{\sigma}h_{\mu\alpha}\nabla_{\sigma}h^{\alpha}_{\nu},
\\
R^{[2]}
=&
\nabla_{\sigma}\nabla_{\lambda}\left(\frac{3}{8}
\bar{g}^{\sigma\lambda}h_{\alpha\beta}h^{\alpha\beta}
- \frac{1}{2}h^{\lambda}_{\rho}h^{\rho\sigma}\right).
\end{align}

To derive the phase-space representation of the stress tensor, we promote the spacetime covariant derivative to the phase-space horizontal lift and apply the Wigner transformation prescription, which symmetrizes the arguments of the gravitational field pairs appearing in the stress tensor, as~\cite{ito2026spin}
\begin{align}
h_{\rho\sigma}\nabla_{\mu}\nabla_{\nu}h_{\lambda\eta}
\to
\frac{1}{2}\left[\left\{D_{\mu}D_{\nu}h_{\lambda\eta}
\left(x, \frac{y}{2}\right)\right\}h_{\rho\sigma}
\left(x, -\frac{y}{2}\right)
+
h_{\rho\sigma}\left(x, \frac{y}{2}\right)
\left\{D_{\mu}D_{\nu}h_{\lambda\eta}
\left(x, -\frac{y}{2}\right)\right\}\right].
\end{align}
Taking the expectation value and Fourier transforming in $y$ as in Eq.~(\ref{Seq_Wigner_curved}), each derivative acting on the field at $(x , \pm y/2)$ produces
$D/2 \pm iq/\hbar$. Applying this to each term, we arrive at the phase-space stress tensor form
\begin{equation}
\label{Seq_stress_phase}
\begin{split}
\mathcal{T}_{\mu\nu}^h(x,q) 
=&
- \frac{1}{32 \pi\hbar^2 G} 
\Bigg\{
\left[ i\hbar q_\nu D_\sigma 
-
\frac{\hbar^2}{2} 
D_{(\nu} D_{\sigma)} \right] 
{W^h}_{\mu\beta}^{\phantom{\mu\beta}\sigma\beta}
-
\left[ i\hbar q_\nu D_\sigma + \frac{\hbar^2}{2}
D_{(\nu} D_{\sigma)}\right] 
{W^h}^{\sigma\beta}_{\phantom{\sigma\beta}\mu\beta}
\\
&+
\left[ i\hbar q_\mu D_\sigma - \frac{\hbar^2}{2}
D_{(\mu} D_{\sigma)} \right] 
{W^h}_{\nu\beta}^{\phantom{\nu\beta}\sigma\beta}
- 
\left[ i\hbar q_\mu D_\sigma + \frac{\hbar^2}{2} (
D_{(\mu} D_{\sigma)} \right] 
{W^h}^{\sigma\beta}_{\phantom{\sigma\beta}\nu\beta}
+
\hbar^2 D_{(\beta} D_{\sigma)} 
\left( {W^h}_{\mu\nu}^{\phantom{\mu\nu}\sigma\beta} 
+ 
{W^h}^{\sigma\beta}_{\phantom{\sigma\beta}\mu\nu} \right) \\
&-
\left[ q_\mu q_\nu 
- 
\frac{3 \hbar^2}{4} D_{(\mu} D_{\nu)}
+
\frac{3 \hbar^2}{8} \bar{g}_{\mu\nu} D_{\sigma} D^{\sigma}
\right]
{W^h}_{\alpha\beta}^{\phantom{\alpha\beta}\alpha\beta}
-
\hbar^2 D_{(\alpha} D_{\sigma)}
{W^h}_{\mu\phantom{\alpha}\nu}^{\phantom{\mu}\alpha\phantom{\nu}\sigma} 
+
\frac{\hbar^2}{4} D_\sigma D^\sigma 
\left( {W^h}_{\nu\alpha\phantom{\alpha}\mu}^{\phantom{\nu\alpha}\alpha\phantom{\mu}}
+ 
{W^h}_{\mu\alpha\phantom{\alpha}\nu}^{\phantom{\mu\alpha}\alpha\phantom{\nu}} \right)
\\
&-
\frac{\hbar^2}{2} \bar{g}_{\mu\nu} 
D_{(\lambda} D_{\sigma)}
\left( {W^h}_{\phantom{\rho\sigma\lambda}\rho}^{\rho\sigma\lambda} 
+ 
{W^h}_{\phantom{\lambda}\rho}^{\lambda\phantom{\rho}\rho\sigma} \right) \Bigg\}.
\end{split}
\end{equation}

To evaluate Eq.~(\ref{Seq_stress_phase}), we decompose in powers of $\hbar$ as
\begin{equation}
\mathcal{T}_{\mu\nu}^h(x,q) 
=
\sum_n
\mathcal{T}_{\mu\nu}^{h(n)}(x,q) ,
\qquad
{W^h}_{\mu\nu\rho\sigma}(x,q) 
=
\sum_n
{W^h_n}_{\mu\nu\rho\sigma}(x,q),
\end{equation}
and make note of the useful identities
\begin{equation}
\delta(q^2)
=
\frac{1}{2|\mathbf{q}|} (\delta_+ + \delta_-),
\qquad
\delta(q^2) \text{sgn}(q_0)
=
\frac{1}{2|\mathbf{q}|} 
(\delta_+ - \delta_-),
\end{equation}
\begin{equation}
\delta'(q^2)
=
\frac{1}{4\vert{}\boldsymbol{q}\vert{}^2} 
( \delta'_+ - \delta'_- )
+
\frac{1}{4\vert{}\boldsymbol{q}\vert{}^3} ( \delta_+ + \delta_- ),
\qquad
\delta'(q^2) \operatorname{sgn}(q_0) = \frac{1}{4\vert{}\boldsymbol{q}\vert{}^2} ( \delta'_+ + \delta'_- )
+
\frac{1}{4\vert{}\boldsymbol{q}\vert{}^3} ( \delta_+ - \delta_- ).
\end{equation}
At zeroth order, this yields the phase-space form of the ideal-fluid stress tensor
\begin{equation}
\label{Seq_mathcalT^0}
\begin{split}
\mathcal{T}_{\mu\nu}^{h(0)} (x,q)
=&
\frac{1}{32 \pi\hbar^2 G} q_\mu q_\nu 
{W_0^h}_{\alpha\beta}^{\phantom{\alpha\beta}\alpha\beta}
\\
=&
\frac{\pi}{\hbar^{2}|\mathbf{q}|} q_{\mu}q_{\nu} \left[ \theta(q_0) \delta_{+} - \theta(- q_0) \delta_{-} \right] f^h(x,q)
\\
=&
\frac{2\pi}{\hbar^{2}} q_{\mu}q_{\nu} \delta(q^2)
\text{sgn}(q_0) f^h(x,q),
\end{split}
\end{equation}
and at first order, we recover the Berry curvature term that contributes to the linear chiral vortical and spin Hall effects~\cite{ito2026spin}
\begin{equation}
\begin{split}
\mathcal{T}_{\mu\nu}^{h(1)} (x,q)
=&
\frac{1}{32 \pi\hbar^2 G}
\left[
q_\mu q_\nu 
{W_1^h}_{\alpha\beta}^{\phantom{\alpha\beta}\alpha\beta}
-
i \hbar q_\mu D_\sigma
 \left(
{W_0^h}_{\nu\beta}^{\phantom{\nu\beta}\sigma\beta}
-
 {W_0^h}^{\sigma\beta}_{\phantom{\sigma\beta}\nu\beta} \right)
-
i \hbar q_\nu D_\sigma 
\left(
{W_0^h}_{\mu\beta}^{\phantom{\mu\beta}\sigma\beta} 
-
{W_0^h}^{\sigma\beta}_{\phantom{\sigma\beta}\mu\beta} \right)
\right]
\\
=&
\frac{8\pi}{\hbar} \delta(q^2) \operatorname{sgn}(q_0) \vert{}\boldsymbol{q}\vert{}^2 q_{(\mu} \Omega_{\nu)\alpha,h}^{-+} D^\alpha f^h(x, q).
\end{split}
\end{equation}

Moving beyond the linear regime, the expansion elements become noticeably more complicated. However, these are also captured by quantum geometric constituents. At second order, we find 
\begin{equation}
\begin{split}
\mathcal{T}_{\mu\nu}^{h(2)} (x,q)
=&
\frac{1}{32 \pi \hbar^2 G} 
\Big\{
q_\mu q_\nu 
{W^h_2}_{\alpha\beta}^{\phantom{\alpha\beta}\alpha\beta}
-
i \hbar q_\mu D_\sigma
 \left(
{W_1^h}_{\nu\beta}^{\phantom{\nu\beta}\sigma\beta}
-
{W_1^h}^{\sigma\beta}_{\phantom{\sigma\beta}\nu\beta} \right)
-
i \hbar q_\nu D_\sigma 
\left(
{W_1^h}_{\mu\beta}^{\phantom{\mu\beta}\sigma\beta} 
-
{W_1^h}^{\sigma\beta}_{\phantom{\sigma\beta}\mu\beta} \right) 
\\
&-
\frac{3}{4} D_{(\mu} D_{\nu)} 
{W_0^h}_{\alpha\beta}^{\phantom{\alpha\beta}\alpha\beta}
+
\frac{1}{2} \left[
D_{(\nu} D_{\sigma)} 
\left(
{W_0^h}_{\mu\beta}^{\phantom{\mu\beta}\sigma\beta}
+
{W_0^h}^{\sigma\beta}_{\phantom{\sigma\beta}\mu\beta} \right) 
+
D_{(\mu} D_{\sigma)}
\left( 
{W_0^h}_{\nu\beta}^{\phantom{\nu\beta}\sigma\beta} 
+
{W_0^h}^{\sigma\beta}_{\phantom{\sigma\beta}\nu\beta} \right) 
\right]
\\
&-
D_{(\beta} D_{\sigma)}
\left(
{W_0^h}_{\mu\nu}^{\phantom{\mu\nu}\sigma\beta}
+
{W_0^h}^{\sigma\beta}_{\phantom{\sigma\beta}\mu\nu} 
- 
{W_0^h}_{\mu\phantom{\beta}\nu\phantom{\sigma}}^{\phantom{\mu}\beta\phantom{\nu}\sigma} 
\right)
+
\frac{1}{2} \bar{g}_{\mu\nu} D_{(\lambda} D_{\sigma)} 
\left( 
{W_0^h}_{\phantom{\rho\sigma\lambda}\rho}^{\rho\sigma\lambda}
+ 
{W_0^h}_{\phantom{\lambda}\rho}^{\lambda\phantom{\rho}\rho\sigma} \right)
\\
&-
\frac{1}{4} D_\sigma D^\sigma
\left(
{W_0^h}_{\nu\alpha\phantom{\alpha}\mu}^{\phantom{\nu\alpha}\alpha\phantom{\mu}}
+
{W_0^h}_{\mu\alpha\phantom{\alpha}\nu}^{\phantom{\mu\alpha}\alpha\phantom{\nu}}
-
\frac{3}{2} \bar{g}_{\mu\nu} 
{W_0^h}_{\alpha\beta}^{\phantom{\alpha\beta}\alpha\beta}
\right)
\bigg\},
\end{split}
\end{equation}
which evaluates to
\begin{equation}
\label{Seq_mathcalT_2_qg}
\begin{split}
\mathcal{T}_{\mu\nu}^{h(2)}(x, q)
=&
\pi \delta(q^2) \operatorname{sgn}(q_0)
\left\{ q_\mu q_\nu 
\left( 5 \mathcal{G}_{\lambda\kappa,h}^{-+} 
- \frac{1}{4\vert \boldsymbol{q} \vert^2} 
\hat{q}_\lambda^\perp \hat{q}_\kappa^\perp \right) 
D^\lambda D^\kappa \right. 
\\
&-
8 \vert \boldsymbol{q} \vert^2 \left[ (\Gamma_\mathcal{G})_{\sigma\lambda(\nu,h}^{-+} q_{\mu)} - q_{(\mu} (\Gamma_\mathcal{G})_{\nu)\sigma\lambda,h}^{-+} \right] D^\sigma D^\lambda
-
\frac{3}{2} D_{(\mu} D_{\nu)}
-
8 \vert \boldsymbol{q} \vert^2 \mathcal{G}_{\sigma(\mu,h}^{-+} D_{\nu)} D^\sigma
\\
&-
8\vert{}\boldsymbol{q}\vert{}^4 
\left[ 2 \mathcal{G}_{\mu\sigma,h}^{-+} 
\mathcal{G}_{\nu\beta,h}^{-+} 
-
\frac{1}{2} \Omega_{\mu\sigma,h}^{-+} \Omega_{\nu\beta,h}^{-+} - \mathcal{G}_{\mu\nu,h}^{-+} \mathcal{G}_{\sigma\beta,h}^{-+} \right] D^\sigma D^\beta
\\
&+
\left. \left( 2\vert \boldsymbol{q} \vert{}^2 
\mathcal{G}_{\mu\nu,h}^{-+} 
+ \frac{3}{4} \bar{g}_{\mu\nu} \right) D_\sigma D^\sigma 
-
4\vert{}\boldsymbol{q}\vert{}^2 \bar{g}_{\mu\nu} 
\mathcal{G}_{\lambda\sigma,h}^{-+} D^\lambda D^\sigma 
\right\} f^h(x, q)
\\
&+
2\pi \vert{}\boldsymbol{q}\vert{} q_0 \delta'(q^2) \operatorname{sgn}(q_0) q_\mu q_\nu 
\mathcal{G}_{\lambda\kappa,h}^{-+} D^\lambda D^\kappa 
f^h(x, q),
\end{split}
\end{equation}
revealing the appearance of the quantum metric and Levi-Civita connection in the expansion. And at third order, the expansion element has a similar formal structure
\begin{equation}
\begin{split}
\mathcal{T}^{h(3)}_{\mu\nu}(x,q)
=&
\frac{1}{32\pi\hbar^{2}G}
\bigg\{
q_\mu q_\nu
{W_3^h}_{\alpha\beta}^{\phantom{\alpha\beta}\alpha\beta}
-
i \hbar q_\mu D_\sigma
\left({W_2^h}^{\sigma\beta}_{\phantom{\sigma\beta}\nu\beta}
-
{W_2^h}_{\nu\beta}^{\phantom{\nu\beta}\sigma\beta}\right)
-
i \hbar q_\nu D_\sigma
\left({W_2^h}^{\sigma\beta}_{\phantom{\sigma\beta}\mu\beta}
-
{W_2^h}_{\mu\beta}^{\phantom{\mu\beta}\sigma\beta}\right)
\\
&-
\frac{3\hbar^{2}}{4}D_{(\mu}D_{\nu)}
{W_1^h}_{\alpha\beta}^{\phantom{\alpha\beta}\alpha\beta}
+
\frac{\hbar^{2}}{2}\left[
D_{(\nu}D_{\sigma)}
\left({W_1^h}^{\sigma\beta}_{\phantom{\sigma\beta}\mu\beta}
+
{W_1^h}_{\mu\beta}^{\phantom{\mu\beta}\sigma\beta}\right)
+
D_{(\mu}D_{\sigma)}
\left({W_1^h}^{\sigma\beta}_{\phantom{\sigma\beta}\nu\beta}
+
{W_1^h}_{\nu\beta}^{\phantom{\nu\beta}\sigma\beta}\right)\right]
\\
&-
\hbar^{2}D_{(\beta}D_{\sigma)}
\left({W_1^h}^{\sigma\beta}_{\phantom{\sigma\beta}\mu\nu}
+
{W_1^h}_{\mu\nu}^{\phantom{\mu\nu}\sigma\beta}
-
{W_1^h}_{\mu\phantom{\beta\sigma}\nu}^{\phantom{\mu}\beta\sigma}\right)
+
\frac{\hbar^{2}}{2}\bar g_{\mu\nu}
D_{(\lambda}D_{\sigma)}
\left({W_1^h}_{\rho}^{\phantom{\rho}\rho\sigma\lambda}
+
{W_1^h}^{\lambda\phantom{\rho}\rho\sigma}_{\phantom{\lambda}\rho}\right)
\\
&-
\frac{\hbar^{2}}{4}D_\sigma D^{\sigma}
\left({W_1^h}_{\nu\alpha\phantom{\alpha}\mu}^{\phantom{\nu\alpha}\alpha}
+
{W_1^h}_{\mu\alpha\phantom{\alpha}\nu}^{\phantom{\mu\alpha}\alpha}
-
\frac{3}{2}\bar g_{\mu\nu}
{W_1^h}_{\alpha\beta}^{\phantom{\alpha\beta}\alpha\beta}\right)
\bigg\},
\end{split}
\end{equation}
resulting in the quantum geometric form
\begin{equation}
\label{Seq_mathcalT3_qg}
\begin{split}
\mathcal{T}^{h(3)}_{\mu\nu}(x,q)
=&
-8\pi\hbar\delta(q^2)\mathrm{sgn}(q_0)
\bigg\{
\frac{3}{8}q_{(\mu}\Omega^{-+}_{\nu)\tau,h}\hat q^\perp_\lambda\hat q^\perp_\kappa
+ \frac{3|\mathbf{q}|^2}{2}q_{(\mu}\Omega^{-+}_{\nu)\tau,h}\mathcal{G}^{-+}_{\lambda\kappa,h}
\\
&+
\frac{|\mathbf{q}|^2}{4}\big( q_{(\mu}\mathcal{G}^{-+}_{\nu)\kappa,h}\Omega^{-+}_{\lambda\tau,h}
+ q_{(\mu}\mathcal{G}^{-+}_{\nu)\lambda,h}\Omega^{-+}_{\kappa\tau,h}
+ q_{(\mu}\Omega^{-+}_{\nu)\lambda,h}\mathcal{G}^{-+}_{\kappa\tau,h}
+ q_{(\mu}\Omega^{-+}_{\nu)\kappa,h}\mathcal{G}^{-+}_{\lambda\tau,h} \big)
\\
&-
\frac{i|\mathbf{q}|}{4}\big(
\big[ \mathcal{N}^{-+}_{\lambda\tau(\nu,h}q_{\mu)} - q_{(\mu}\mathcal{N}^{-+}_{\nu)\lambda\tau,h} \big]\hat q^\perp_\kappa
+ \big[ \mathcal{N}^{-+}_{\kappa\tau(\nu,h}q_{\mu)} - q_{(\mu}\mathcal{N}^{-+}_{\nu)\kappa\tau,h} \big]\hat q^\perp_\lambda
\big)\bigg\} D^\tau D^\lambda D^\kappa f^h
\\
&+ 4\pi\hbar|\mathbf{q}|^2
\Big[ \delta(q^2) - 2|\mathbf{q}|^2\delta'(q^2) \Big]
q_{(\mu}\Omega^{-+}_{\nu)\tau,h}\mathcal{G}^{-+}_{\lambda\kappa,h}
D^\tau D^\lambda D^\kappa f^h
\\
&- 2i\pi\hbar|\mathbf{q}|^2\delta(q^2)\mathrm{sgn}(q_0)
\left[ \mathcal{N}^{-+}_{\rho\mu\alpha,h} D_{(\nu}D^{\rho)}
+ \mathcal{N}^{-+}_{\rho\nu\alpha,h} D_{(\mu}D^{\rho)} 
\right]
D^\alpha f^h
\\
&- 8\pi\hbar|\mathbf{q}|^4\delta(q^2)\mathrm{sgn}(q_0)
\Big[ i\mathcal{N}^{-+}_{\rho\mu\alpha,h}\mathcal{G}^{-+}_{\nu\tau,h}
+ \partial^q_{[\mu}\mathcal{G}^{-+}_{\rho]\alpha,h}\Omega^{-+}_{\nu\tau,h}
+ i\mathcal{N}^{-+}_{\tau\nu\alpha,h}\mathcal{G}^{-+}_{\mu\rho,h}
+ \partial^q_{[\nu}\mathcal{G}^{-+}_{\tau]\alpha,h}\Omega^{-+}_{\mu\rho,h} \Big]
D^{(\rho}D^{\tau)} D^\alpha f^h
\\
&+ 4i\pi\hbar|\mathbf{q}|^4\delta(q^2)\mathrm{sgn}(q_0)
\Big[ \big( \mathcal{C}^{-+}_{\rho\alpha\mu,h} - \mathcal{C}^{+-}_{\mu\alpha\rho,h} \big)
\mathcal{Q}^{-+}_{\sigma\nu,h}
+ \big( \mathcal{C}^{-+}_{\nu\alpha\sigma,h} - \mathcal{C}^{+-}_{\sigma\alpha\nu,h} \big)
\mathcal{Q}^{-+}_{\mu\rho,h} \Big]
D^{(\rho}D^{\sigma)} D^\alpha f^h
\\
&- i\pi\hbar|\mathbf{q}|^2\delta(q^2)\mathrm{sgn}(q_0)
\bar g_{\mu\nu}\mathcal{N}^{-+}_{\sigma\lambda\alpha,h}
D^\lambda D^\sigma D^\alpha f^h
+ i\pi\hbar|\mathbf{q}|^2\delta(q^2)\mathrm{sgn}(q_0)
\mathcal{N}^{-+}_{\mu\nu\alpha,h} D_\sigma D^\sigma D^\alpha f^h.
\end{split}
\end{equation}
At this order, the quantum nonmetricity also appears in the phase-space stress tensor.

The stress-energy tensor in spacetime is then obtained, upon restoring $\hbar$, as
\begin{equation}
\label{Seq_T_from_mathcalT}
T^{h}_{\mu\nu} (x)
=
\hbar^2
\int_{q} \mathcal{T}^{h}_{\mu\nu}(x,q),
\qquad
\int_{q} = \int\frac{d^{4}q}{(2\pi)^{4}\sqrt{-\bar g}}.
\end{equation}
In the next section, we apply this to evaluate graviton transport in a linearized background.

\section{S7.~Graviton transport in linearized background}

\subsection{Setup and useful relations}

We consider the background metric
\begin{equation}
\bar g_{\mu\nu}=\eta_{\mu\nu}+\gamma_{\mu\nu},
\qquad
\gamma_{0i}=\gamma_{0i}(\boldsymbol{x}),
\quad
\gamma_{00}=\gamma_{ij}=0,
\quad |\gamma_{\mu\nu}|\ll1,
\label{eq:metricclass}
\end{equation}
and work to linear order in $\gamma_{\mu\nu}$. In the presence of spin-vorticity coupling, the equilibrium distribution function reads~\cite{liu2019chiral, mameda2022photonic}
\begin{equation}
f^{h}(x,q)
=
N (g^{h}),
\qquad
g^{h}
=
q\cdot U + \frac{\hbar}{2} S^{\alpha\beta}_h
\nabla_{\alpha}U_{\beta},
\label{eq:eqdist}
\end{equation}
where
$N(x)=(e^{x}-1)^{-1}$, $U^{\mu}=\beta u^{\mu}$ with $\beta=1/T$ and $u^{\mu}$ the fluid velocity. We take
$u^{\mu}=n^{\mu}=(1,\boldsymbol{0})$, 
so that 
$u_{\mu} = (u_0, - \bs{u}) =(1,\gamma_{0i})$ 
and $q\cdot U=\beta q_{0}$. This corresponds to the Killing condition
$\nabla_{(\mu}U_{\nu)}=0$. To linear order in the background perturbation $\gamma$, the distribution function reads
\begin{equation}
f^{h}
=
N(q\cdot U)
+
\frac{\hbar}{2} S_h^{\alpha\beta}\nabla_{\alpha}U_{\beta}
N'(q\cdot U).
\label{Seq_fxq_expansion}
\end{equation}
We henceforth apply the shorthand notation 
$N \equiv N(q \cdot U)$.

The Killing condition renders $\nabla_{\alpha}U_{\beta}$ antisymmetric, which is captured by the covariant fluid vorticity
\begin{equation}
\omega^{\mu}
=
(\omega^0, \bs{\omega})
=
\frac{1}{2}\epsilon^{\mu\nu\rho\sigma}u_{\nu}
\nabla_{\rho}u_{\sigma},
\qquad
\omega_0 = O(\gamma^2),
\qquad
\bs{\omega}
=
\frac{1}{2} \bs{\pd} \times \bs{u},
\end{equation}
so that one may write
\begin{equation}
\nabla_{\alpha}U_{\beta}
=-\beta\,\epsilon_{\alpha\beta\mu\nu}\,\omega^{\mu}u^{\nu}.
\label{eq:nablaU}
\end{equation}
The second covariant derivative is captured by the Riemann tensor through the Killing identity
\begin{equation}
\nabla_{\alpha}\nabla^{\beta}U_{\rho}
=
- R^{\beta}_{\phantom{\beta}\rho\alpha\sigma} U^{\sigma},
\end{equation}
which establishes a useful gravitomagnetic relation between the Ricci tensor and vorticity curl
\begin{equation}
R_{0i}
=
- \varepsilon_{ijk} \partial^{j}\omega^{k},
\qquad
R_{00},R_{ij},R=O(\gamma^{2}),
\label{eq:Ricci}
\end{equation}
with
\begin{equation}
\veps_{ijk}
=
\veps_{ijk0},
\qquad
\veps_{ijk}
=
- \veps^{ijk},
\qquad
\veps_{123}
=
-1.
\end{equation}
Accordingly, successive horizontal lifts acting on $N$ yield
\begin{equation}
D_{j}D_{k}N
=
-\beta N'q^{l}\varepsilon_{klm}\partial_{j}\omega^{m},
\qquad
D_{i}D_{j}D_{k}N
=
-\beta N'q^{l}\varepsilon_{klm}\partial_{i}\partial_{j}\omega^{m}.
\end{equation}

To evaluate stress tensor contributions below, we note useful integral identities of the Bose-Einstein distribution
\begin{equation}
\begin{aligned}
\int_0^\infty d|\boldsymbol{q}|\,|\boldsymbol{q}|^3 N &= \frac{\pi^4 T^4}{15}, \qquad
\int_0^\infty d|\boldsymbol{q}|\,|\boldsymbol{q}|^4 N' = -\frac{4\pi^4 T^5}{15}, \qquad
\int_0^\infty d|\boldsymbol{q}|\,|\boldsymbol{q}|^3 N' = -6\zeta(3)\,T^4, \\
\int_0^\infty d|\boldsymbol{q}|\,|\boldsymbol{q}|^2 N' &= -\frac{\pi^2 T^3}{3}, \qquad
\int_0^\infty d|\boldsymbol{q}|\,|\boldsymbol{q}|^3 N'' = \pi^2 T^4,
\end{aligned}
\end{equation}
with $N$ and its derivatives evaluated at $\beta|\boldsymbol{q}|$, and $\zeta(n)$ the Riemann zeta function. We also note the commonly encountered infrared-sensitive integrals
\begin{equation}
\int_{q_{\min}}^\infty d|\boldsymbol{q}|\,|\boldsymbol{q}|\,N' = -T^2\left[\ln\frac{T}{q_{\min}} + 1\right], \qquad
\int_{q_{\min}}^\infty d|\boldsymbol{q}|\,|\boldsymbol{q}|^2 N'' = T^3\left[2\ln\frac{T}{q_{\min}} + 3\right],
\end{equation}
which appear at higher orders and for which one can introduce the cutoff $q_{\min}$ on $|\boldsymbol{q}|$.

Finally, the relevant quantum geometric structures are expressed explicitly as
\begin{equation}
\mathcal{G}^{-+}_{jk}=\frac{1}{4|\boldsymbol{q}|^{2}}
\big(\delta_{jk}-\hat q^{\perp}_{j}\hat q^{\perp}_{k}\big),
\qquad
\Omega^{-+}_{jk,h}=\frac{\iota^{h}}{2|\boldsymbol{q}|^{2}}
\varepsilon_{jkl}\hat q_{\perp}^{l},
\label{eq:GOexplicit}
\end{equation}
\begin{equation}
\mathcal{N}^{-+}_{ijk,h}
=
\frac{i\iota^{h}}{2|\boldsymbol{q}|^{3}}
\hat q^{\perp}_{(i}\varepsilon_{j)kl}\hat q^{\perp l},
\qquad
(\Gamma_{\mathcal{G}})^{-+}_{ijk}
=
\frac{1}{2|\boldsymbol{q}|^{3}}
\left[
\hat q^{\perp}_{(j}\delta_{k)i}
-\hat q^{\perp}_{i}\hat q^{\perp}_{j}\hat q^{\perp}_{k}\right].
\end{equation}

\subsection{Stress tensor in linearized background}

Following the linearized setup presented above, we now present explicit evaluations of the graviton stress-energy tensor.

At zeroth order, we insert Eq.~(\ref{Seq_mathcalT^0}) into Eq.~(\ref{Seq_T_from_mathcalT}) and recover the ideal-fluid stress tensor
\begin{equation}
\begin{split}
T^{h(0)}_{\mu\nu}
=&
2\pi \int_{q}
q_{\mu}q_{\nu}\delta(q^{2})\mathrm{sgn}(q_{0})N(\beta q_{0})
\\
=&
\int\frac{d^{3}\bs{q}}{(2\pi)^{3}}
\frac{q_{\mu}q_{\nu}}{|\boldsymbol{q}|}\bigg|_{q_{0}=|\boldsymbol{q}|-\gamma_{0i}q_{i}}
N \left(\beta(|\boldsymbol{q}|-\gamma_{0i}q_{i})\right)
\\
=&
\frac{\rho_{h}}{3}\left(4u_{\mu}u_{\nu} -\bar g_{\mu\nu}\right)+O(\gamma^{2}),
\qquad
\rho_{h}=\frac{\pi^{2}T^{4}}{30},
\end{split}
\end{equation}
with $\rho_h$ the energy density per helicity channel of the thermal graviton gas.

And at first order, taking into account Eq.~(\ref{Seq_fxq_expansion}), there are two contributions to the stress tensor, namely from $\mathcal{T}_{\mu\nu}^{h(1)}$ acting on $N$, and from $\mathcal{T}_{\mu\nu}^{h(0)}$ acting on the $O(\hbar)$ distribution term. We express this as
\begin{equation}
\begin{split}
T^{h(1)}_{\mu\nu}
=&
T^{h(1,0)}_{\mu\nu}+T^{h(0,1)}_{\mu\nu}
=
2 T^{h(1,0)}_{\mu\nu}
\\
=&
16\pi \hbar \int \frac{d^4 q}{(2\pi)^4\sqrt{-\bar{g}}} \delta(q^2) \operatorname{sgn}(q_0) \vert{}\bs{q}\vert{}^2 q_{(\mu} \Omega_{\nu)\alpha,h}^{-+} D^\alpha N(\beta q_{0})
\\
=&
\frac{4\iota^{h} \hbar \zeta(3)}{\pi^{2}} T^3 u_{(\mu}\omega_{\nu)}+O(\gamma^{2}),
\qquad
T^{h(1)}_{0i}
=\frac{2\iota^{h} \hbar \zeta(3)}{\pi^{2}} T^{3} \omega_{i},
\end{split}
\end{equation}
which recovers the graviton CVE~\cite{ito2026spin}.

Moving on to second order, the contributions are
\begin{equation}
T^{h(2)}_{\mu\nu}=T^{h(2,0)}_{\mu\nu}+T^{h(1,1)}_{\mu\nu}.
\end{equation}
For the first term, we identify three nonvanishing contributions from Eq.~(\ref{Seq_mathcalT_2_qg}), which arise from the quantum metric and Levi-Civita connection
\begin{equation}
T_{\mu\nu}^{h(2,0)}
=
T_{\mu\nu}^{h(2,0)}\Big\vert_{\text{I}}
+
T_{\mu\nu}^{h(2,0)}\Big\vert_{\text{II}}
+
T_{\mu\nu}^{h(2,0)}\Big\vert_{\text{III}},
\end{equation}
\begin{align}
T_{\mu\nu}^{h(2,0)}\Big\vert_{\text{I}}
=& 
\pi \hbar^2 \int \frac{d^4 q}{(2\pi)^4\sqrt{-\bar{g}}} \delta(q^2) \operatorname{sgn}(q_0) q_\mu q_\nu \left( 5 \mathcal{G}_{\lambda\kappa}^{-+} - \frac{1}{4\vert{}\boldsymbol{q}\vert{}^2} \hat{q}_\lambda^\perp \hat{q}_\kappa^\perp \right) D^\lambda D^\kappa N(q \cdot U),
\\
T_{\mu\nu}^{h(2,0)}\Big\vert_{\text{II}}
=& 
- 8 \pi \hbar^2 \int \frac{d^4 q}{(2\pi)^4\sqrt{-\bar{g}}} \delta(q^2) \operatorname{sgn}(q_0) \vert{}\mathbf{q}\vert{}^2 \left[ (\Gamma_{\mathcal{G}})_{\sigma\lambda(\nu}^{-+} q_{\mu)} - q_{(\mu} (\Gamma_{\mathcal{G}})_{\nu)\sigma\lambda}^{-+} \right] D^{(\sigma} D^{\lambda)} N,
\\
T_{\mu\nu}^{h(2,0)}\Big\vert_{\text{III}}
=& 
2\pi \hbar^2 \int \frac{d^4 q}{(2\pi)^4\sqrt{-\bar{g}}} \vert{}\mathbf{q}\vert{} q_0 \delta'(q^2) \operatorname{sgn}(q_0) q_\mu q_\nu \mathcal{G}_{\lambda\kappa}^{-+} D^\lambda D^\kappa N.
\end{align}
Upon evaluation to $O(\gamma)$, all three terms are proportional to the Ricci 3-vector as
\begin{equation}
T_{0i}^{h(2,0)}\Big\vert_{\text{I}} 
=
\frac{5\hbar^2T^2}{288} R_{0i},
\qquad
T_{0i}^{h(2,0)}\Big\vert_{\text{II}}
=
- \frac{\hbar^2T^2}{36} R_{0i},
\qquad
T_{0i}^{h(2,0)}\Big\vert_{\text{III}}
=
\frac{\hbar^2T^2}{144} R_{0i}.
\end{equation}
The remaining term arises from the Berry curvature acting on the spin-vorticity coupling distribution function and reads
\begin{equation}
T_{\mu\nu}^{h(1,1)} 
=
8\pi \hbar \int \frac{d^4 q}{(2\pi)^4\sqrt{-\bar{g}}} \delta(q^2) \operatorname{sgn}(q_0) \vert{}\mathbf{q}\vert{}^2 q_{(\mu} \Omega_{\nu)\alpha,h}^{-+} D^\alpha f^{h(1)},
\qquad
f^{h(1)} = 2\iota^h \hbar \beta \frac{q \cdot \omega}{q \cdot n} N',
\end{equation}
which similarly yields
\begin{equation}
T_{0i}^{h(1,1)}
=
- \frac{\hbar^2 T^2}{18} R_{0i},
\end{equation}
and thus the total second-order contribution
\begin{equation}
T^{h(2)}_{\mu\nu}
=-\frac{17 \hbar^2 T^{2}}{144} u_{(\mu}R_{\nu)\lambda}u^{\lambda}+O(\gamma^{2}),
\qquad
T^{h(2)}_{0i}
=
-\frac{17 \hbar^2 T^{2}}{288}R_{0i}.
\label{eq:T2result}
\end{equation}

Finally, the evaluation of the third-order stress tensor requires considering
\begin{equation}
T^{h(3)}_{\mu\nu}=T^{h(3,0)}_{\mu\nu}+T^{h(2,1)}_{\mu\nu},
\end{equation}
where the surviving terms in Eq.~(\ref{Seq_mathcalT3_qg}) are denoted as
\begin{equation}
T_{\mu\nu}^{h(3,0)}
=
T_{\mu\nu}^{h(3,0)}\Big\vert_{\text{I}}
+
T_{\mu\nu}^{h(3,0)}\Big\vert_{\text{II}}
+
T_{\mu\nu}^{h(3,0)}\Big\vert_{\text{III}}
+
T_{\mu\nu}^{h(3,0)}\Big\vert_{\text{IV}},
\end{equation}
with contributions from the Berry curvature, quantum metric and quantum nonmetricity tensors
\begin{align}
T^{h(3,0)}_{\mu\nu}\Big\vert_{\text{I}}
=&
12\pi\hbar^3 \int\frac{d^{4}q}{(2\pi)^{4}\sqrt{-\bar g}}
\delta(q^{2})\,\mathrm{sgn}(q_{0})
\left[\frac{i|\boldsymbol{q}|}{8}
\mathcal{N}^{-+}_{\lambda\tau(\nu,h} q_{\mu)} \hat q^{\perp}_{\kappa}
- |\boldsymbol{q}|^{2}
q_{(\mu}\Omega^{-+}_{\nu)\tau,h} 
\mathcal{G}^{-+}_{\lambda\kappa}\right]
D^{\tau}D^{\lambda}D^{\kappa}N,
\\
T^{h(3,0)}_{\mu\nu}\Big\vert_{\text{II}}
=&
-4\pi i \hbar^3 \int\frac{d^{4}q}{(2\pi)^{4}\sqrt{-\bar g}}
\delta(q^{2})\mathrm{sgn}(q_{0})\big(|\boldsymbol{q}|^{2}\big)
q_{(\mu}\Big[\mathcal{Q}^{-+}_{\nu)(\kappa,h}\mathcal{Q}^{-+}_{\lambda)\tau,h}
-\mathcal{Q}^{+-}_{\nu)(\kappa,h}\mathcal{Q}^{+-}_{\lambda)\tau,h}\Big]
D^{\tau}D^{\lambda}D^{\kappa}N,
\\
T^{h(3,0)}_{\mu\nu} \Big\vert_{\text{III}}
=&
2 \pi i \hbar^3
\int\frac{d^{4}q}{(2\pi)^{4}\sqrt{-\bar g}}
\delta(q^{2})\mathrm{sgn}(q_{0}) |\boldsymbol{q}|
\\
& \hspace{2cm} \times 
\left(
\left[
\mathcal{N}^{-+}_{\lambda\tau(\nu,h}q_{\mu)}
-q_{(\mu}\mathcal{N}^{-+}_{\nu)\lambda\tau,h}
\right]
\hat q^{\perp}_{\kappa}
+\big[\mathcal{N}^{-+}_{\kappa\tau(\nu,h}q_{\mu)}
-q_{(\mu}\mathcal{N}^{-+}_{\nu)\kappa\tau,h}\big]\hat q^{\perp}_{\lambda}
\right)
D^{\tau}D^{\lambda}D^{\kappa} N,
\\
T_{\mu\nu}^{h(3,0)} \Big\vert_{\text{IV}}
=&
4\pi\hbar^3 \int \frac{d^4 q}{(2\pi)^4\sqrt{-\bar{g}}} \vert{}\mathbf{q}\vert{}^2 \left[ \delta(q^2) - 2\vert{}\mathbf{q}\vert{}^2 \delta'(q^2) \right] q_{(\mu} \Omega_{\nu)\tau,h}^{-+} \mathcal{G}_{\lambda\kappa}^{-+} D^{(\tau} D^\lambda D^{\kappa)} N.
\end{align}
These evaluate to
\begin{equation}
T^{h(3,0)}_{0i} \Big\vert_{\text{I}}
=
\frac{\iota^{h}\hbar^3 T}{24\pi^{2}}
\left(\ln\frac{T}{q_{\min}}+1\right) \bs{\pd}^{2}\omega_{i},
\qquad
T^{h(3,0)}_{0i} \Big\vert_{\text{II}}
=
\frac{\iota^{h}\hbar^3 T}{90\pi^{2}}
\left(\ln\frac{T}{q_{\min}}+1\right) \bs{\pd}^{2} \omega_{i},
\end{equation}
\begin{equation}
T^{h(3,0)}_{0i} \Big\vert_{\text{III}}
=
- \frac{\iota^{h}\hbar^3 T}{60\pi^{2}}
\left(\ln\frac{T}{q_{\min}}+1\right) \bs{\pd}^{2} \omega_{i},
\qquad
T^{h(3,0)}_{0i} \Big\vert_{\text{IV}}
=
\frac{\iota^{h}\hbar^3 T}{90\pi^{2}}
\left( \ln\frac{T}{q_{\min}}+2 \right) \bs{\pd}^{2}\omega_{i},
\end{equation}
\begin{equation}
T^{h(3,0)}_{0i}
=
\frac{\iota^{h}\hbar T}{360\pi^{2}}
\left( 17 \ln\frac{T}{q_{\min}} + 21\right) 
\bs{\pd}^{2}\omega_{i}.
\end{equation}
And the contributions from Eq.~(\ref{Seq_mathcalT_2_qg}) are given by
\begin{equation}
T_{\mu\nu}^{h(2,1)}
=
T_{\mu\nu}^{h(2,1)}\Big\vert_{\text{I}}
+
T_{\mu\nu}^{h(2,1)}\Big\vert_{\text{II}}
+
T_{\mu\nu}^{h(2,1)}\Big\vert_{\text{III}},
\end{equation}
where
\begin{align}
T^{h(2,1)}_{\mu\nu} \Big\vert_{\text{I}}
=&
\pi \hbar^2 \int\frac{d^{4}q}{(2\pi)^{4}\sqrt{-\bar g}}
\delta(q^{2})\mathrm{sgn}(q_{0})
q_{\mu}q_{\nu}
\Big(5\mathcal{G}^{-+}_{\lambda\kappa}
-\frac{1}{4|\boldsymbol{q}|^{2}}\hat q^{\perp}_{\lambda}\hat q^{\perp}_{\kappa}\Big)
D^{\lambda}D^{\kappa}f^{h(1)},
\\
T^{h(2,1)}_{\mu\nu} \Big\vert_{\text{II}}
=&
- 8 \pi \hbar^2
\int\frac{d^{4}q}{(2\pi)^{4}\sqrt{-\bar g}}
\delta(q^{2})\mathrm{sgn}(q_{0}) |\boldsymbol{q}|^{2}
\left
[(\Gamma_{\mathcal{G}})^{-+}_{\sigma\lambda(\nu}q_{\mu)}
-q_{(\mu}(\Gamma_{\mathcal{G}})^{-+}_{\nu)\sigma\lambda}\right]
D^{\sigma}D^{\lambda}f^{h(1)},
\\
T^{h(2,1)}_{\mu\nu} \Big\vert_{\text{III}}
=&
2 \pi \hbar^2
\int\frac{d^{4}q}{(2\pi)^{4}\sqrt{-\bar g}}
|\boldsymbol{q}| q_0 \delta'(q^{2})\mathrm{sgn}(q_0)
q_{\mu}q_{\nu}\,\mathcal{G}^{-+}_{\lambda\kappa}
D^{\lambda}D^{\kappa}f^{h(1)},
\end{align}
yielding
\begin{equation}
T^{h(2,1)}_{0i} \Big\vert_{\text{I}}
=
- \frac{19 \iota^{h} \hbar^3 T}{240\pi^{2}}
\left( \ln\frac{T}{q_{\min}}+1\right)\bs{\pd}^{2}\omega_{i},
\qquad
T^{h(2,1)}_{0i} \Big\vert_{\text{II}}
=
\frac{\iota^{h} \hbar^3 T}{12\pi^{2}}
\left( \ln\frac{T}{q_{\min}}+1\right)\bs{\pd}^{2}\omega_{i},
\end{equation}
\begin{equation}
T^{h(2,1)}_{0i} \Big\vert_{\text{III}}
=
- \frac{\iota^{h} \hbar^3 T}{60\pi^{2}}
\left( 2 \ln\frac{T}{q_{\min}} + 3 \right)\bs{\pd}^{2}\omega_{i},
\qquad
T^{h(2,1)}_{0i}
=
- \frac{\iota^{h} \hbar^3 T}{240\pi^{2}}
\left( 7 \ln\frac{T}{q_{\min}}+11\right)\bs{\pd}^{2}\omega_{i}.
\end{equation}
Adding up all terms, we arrive at the total third-order contribution
\begin{equation}
T^{h(3)}_{\mu\nu}
=
\frac{\iota^{h} \hbar^3 T}{360\pi^{2}}
\left( 13 \ln\frac{T}{q_{\min}}+9\right)
u_{(\mu} \nabla^2 \omega_{\nu)},
\qquad
T^{h(3)}_{0i}
=
\frac{\iota^{h} \hbar^3 T}{720\pi^{2}}
\left( 13 \ln\frac{T}{q_{\min}}+9\right)\bs{\pd}^{2}\omega_{i},
\end{equation}
which captures the inhomogeneous correction to the chiral vortical current.

\end{document}